\documentclass[openacc]{rsproca_new}

\begin{document}

\title{Variational construction of tubular and toroidal streamsurfaces for flow visualization}

\author{
Mingwu Li$^{1}$, B\'alint Kasz\'as$^{2}$ and George Haller$^{2}$}

\address{$^{1}$ Department of Mechanics and Aerospace Engineering, Southern University of Science and Technology, 518055 Shenzhen, China\\
$^{2}$ Institute for Mechanical Systems, ETH Z\"{u}rich Leonhardstrasse 21, 8092 Z\"{u}rich, Switzerland}

\subject{Applied mathematics, Ocean engineering, flow visualization}


\corres{Mingwu Li\\
\email{limw@sustech.edu.cn}}

\begin{abstract}
Approximate streamsurfaces of a 3D velocity field have recently been constructed as isosurfaces of the closest first integral of the velocity field. Such approximate streamsurfaces enable effective and efficient visualization of vortical regions in 3D flows. Here we propose a variational construction of these approximate streamsurfaces to remove the limitation of Fourier series representation of the first integral in earlier work. Specifically, we use finite-element methods to solve a partial-differential equation that describes the best approximate first integral for a given velocity field. We use several examples to demonstrate the power of our approach for 3D flows in domains with arbitrary geometries and boundary conditions. These include generalized axisymmetric flows in the domains of a sphere (spherical vortex), a cylinder (cylindrical vortex), and a hollow cylinder (Taylor-Couette flow) as benchmark studies for various computational domains, non-integrable periodic flows (ABC and Euler flows), and Rayleigh-Bénard convection flows. We also illustrate the use of the variational construction in extracting momentum barriers {in} Rayleigh-Bénard convection.
\end{abstract}


\begin{fmtext}

\section{Introduction}


Streamlines provide a powerful tool for the visualization of 2D flows but have limited usefulness for 3D flows~\cite{van1993implicit}. Indeed, a streamline passes through every point of a flow and hence one needs to select a few illustrative streamlines to provide an efficient visualization~\cite{born2010illustrative,schulze2014sets}.







\end{fmtext}


\maketitle

As an alternative, streamsurfaces are well known techniques for the visualization of 3D flows. As in the case of streamlines, one has to find a {select set} of special streamsurfaces that efficiently convey information about the range of different fluid behaviors in the flow domain. This is not an easy task, given that infinitely many streamsurfaces pass through each point of the flow domain.


Hultquist~\cite{hultquist1992constructing} proposed an advancing front method to construct streamsurfaces. With a properly chosen curve, it is discretized with a set of particles and then advanced downstream. In particular, the spacing between particles at the front and the number of these particles are adaptively changed such that the distance between two adjacent particles are kept the same. This method highly depends on the initial curve and requires a careful implementation~\cite{van1993implicit}.

The stream function $\phi$ of any 2D, incompressible flow is guaranteed to exist and can be used to visualize the streamlines for the 2D flow because the contour lines of this function $\phi$ are the streamlines. Motivated by this observation, van Wijk~\cite{van1993implicit} seeks a scalar function $f$ such that $f(\mathbf{x})=C$ represents a one-parameter family of streamsurfaces for a given 3D flow under the variations in $C$. To solve for the function $f$, a convection equation is used and then simulated with prescribed values of $f$ at boundaries~\cite{van1993implicit}. This simulation is performed until a steady state is reached. A similar approach is proposed to define vortex surfaces where the local vorticity vector is tangent at every point on such a surface~\cite{yang2010lagrangian}. Similarly, one can simulate the convection equation to solve for the scalar function $f$ whose contour plots give the vortex surfaces~\cite{yang2011evolution,xiong2017boundary}.

The aforementioned simulations of the convection equation depend on the choice of the initial distribution of $f$, which is not trivial. In addition, long time simulations are needed to obtain converged solutions~\cite{yang2011evolution}. Therefore, simulating the associated convection equation remains challenging and computationally expensive.

By definition, the scalar function $f$ is a first integral for 3D steady flows or unsteady flows that are instantaneously frozen. However, an exact first integral does not exist for generic 3D flows. Some exceptions include the Bernoulli function, which gives a non-degenerate first integral for steady Euler flows that do not satisfy the Beltrami property~\cite{arnold2021topological}. Analytic first integrals were also constructed for incompressible flows with a volume-preserving symmetry group~\cite{haller1998reduction} and for highly symmetric flows~\cite{he2016construction}.

A level surface of a first integral, $f=C$, is also an invariant manifold. Such manifolds have been broadly used to illustrate local velocity geometry near stationary points~\cite{peikert2009topologically}. However, these invariant manifolds generally stretch and fold globally, which makes them unsuitable for global flow visualization. Exceptions to this general rule are invariant manifolds that are level surfaces of a smooth function.


Motivated by the above observations, Katsanoulis et al.~\cite{stergios}  seek influential streamsurfaces as level sets of approximate first integrals. In particular, they constructed a scalar function $f$ to minimize $|\nabla f\cdot\mathbf{v}|$ at a collection of grid points~\cite{stergios}. Here $\mathbf{v}$ can be a general vector field related to the fluid motion, such as the velocity, vorticity or even a barrier field used for detecting barriers to material transport \cite{haller2020objective}. These barrier fields have been introduced to define material sets that prohibit the transport of active quantities in a frame-indifferent way. For example, the method was used to extract objective momentum barriers as invariant manifolds of the barrier vector fields defined in~\cite{haller2020objective,aksamit2022objective}.

Katsanoulis et al.~\cite{stergios} use a Fourier series to represent the unknown scalar function $f$. This approach works well for spatially periodic flows but has limitations for generic flows that are not periodic in all three directions. Although {such} spatially aperiodic flows have also been successfully treated via a proper choice of smaller subdomains, the selection of such subdomains is problem-dependent and hence requires careful implementation.

To extend the approximate first integral approach of Katsanoulis et al.~\cite{stergios} to generic 3D flows, here we develop a variational construction of approximate first integrals for 3D velocity fields given by either analytic expressions or data sets. This variational approach works for arbitrary geometries and boundary conditions of the computational domain. The approximate first integrals here are obtained as eigenfunctions of a set of linear partial-differential equations (PDEs) obtained as the Euler-Lagrange equations of the variational principle minimizing $|\nabla f\cdot\mathbf{v}|$. We use finite-element methods to discretize the PDEs and then solve for the eigenvectors corresponding to the smallest eigenvalues. These eigenvectors provide the approximate first integrals of the 3D flow.

The rest of this paper is organized as follows. We start with a formulation of an optimization problem whose solution gives the approximate first integrals. We derive the first-order necessary conditions of the optimization problem in Sect.~\ref{sec:formulation}, which are the aforementioned PDEs. Then we establish the weak form of the PDEs in Sect.~\ref{sec:weak-form}, which {leads to an} eigenvalue problem. In Sect.~\ref{sec:eig-to-min}, we discuss the relation between the solutions of this eigenvalue problem and the minimum solution of the optimization problem. The solution of the weak form via finite-element methods is then discussed in Sect.~\ref{sec:fenics}, followed by benchmark studies on 3D flow in domains with arbitrary geometries in Sect.~\ref{sec:benchmark}. We further consider periodic flows in Sect.~\ref{sec:periodicflows} and non-periodic Rayleigh-Bénard convection flows in Sect.~\ref{sec:rbc} to illustrate the broad applicability of our method.

\section{Formulation}
\label{sec:formulation}
Consider a vector field $\mathbf{u}: \Omega \to \mathbb{R}^3$ defined over a spatial domain $\Omega \subset \mathbb{R}^3$. We define the function space of admissible first integrals as 
\begin{equation}
    \mathcal{H}=\left\{H\in H^1(\Omega), \int_\Omega H^2dV=1\right\}
\end{equation}
and consider the optimization problem
\begin{equation}
\label{eq:optH}
    H^\ast = \underset{H\in\mathcal{H}}{\mathrm{argmin}}\,\,\int_{\partial\Omega}||\nabla H\cdot \mathbf{u}||^2dV. 
\end{equation}
{We have introduced the normalization constraint $||H||_{{L}^2}=1$ to exclude the multitude of trivial solutions $H=C$. Indeed, these are minimal solutions for any constant $C$. With the imposed constraint, the only constant solution remains $H = 1/\text{Vol}(\Omega)$, where $\text{Vol}(\Omega)= \int_\Omega dV$. }

We introduce a Lagrange multiplier to enforce the constraint and define the Lagrangian as
\begin{equation}
 \mathcal{L}(H)=\int_{\partial\Omega}||\nabla H\cdot \mathbf{u}||^2dV+\lambda \left(\int_\Omega H^2dV-1\right),
\end{equation}
where $\lambda$ is a Lagrange-multiplier. To express the Lagrangian in terms of the components of $\mathbf{u}$ and $\nabla H$, we make use of the implied summation over repeated indices. The components of the gradient vector are denoted as $(\nabla H)_i = \partial_i H$, which allows us to write the Lagrangian as 
\begin{equation}
\label{eq:lag-sum}
 \mathcal{L}(H)=\int_{\partial\Omega}(\partial_i H u_i)^2dV+\lambda \left(\int_\Omega H^2dV-1\right).
\end{equation}
Note that the variation of the first term in~\eqref{eq:lag-sum} is
\begin{equation}
    \delta\left(\int_{\Omega}(\partial_iHu_i)^2dV\right)=2\int_{\partial\Omega} \partial_iHu_iu_jn_j\delta HdS-2\int_{\Omega}\partial_j(\partial_iHu_iu_j)\delta HdV.
\end{equation}
Since the variation of the second term in~\eqref{eq:lag-sum} is simply $2\lambda H$, we obtain the following first-order necessary conditions for the minimum solution:
\begin{gather}
    \partial_j(\partial_iHu_iu_j)=-\lambda H,\quad \text{on } \Omega\label{eq:pde}\\
    \partial_iHu_iu_jn_j\delta H=(\nabla H\cdot \mathbf{u})(\mathbf{u}\cdot \mathbf{n})\delta H=0,\quad \text{on }\partial\Omega,\quad\mathrm{and}\,\,
    \int_\Omega H^2dV=1.\label{eq:bcs}
\end{gather}
This system of equations defines an eigenvalue problem. Let $\partial\Omega=\partial\Omega_\mathrm{H}\cup\partial\Omega_\mathrm{F}$ with $\partial\Omega_\mathrm{H}\cap\partial\Omega_\mathrm{F}=\emptyset$, where $\partial\Omega_\mathrm{H}$ denotes the part of boundary where $H$ is prescribed. So the boundary conditions can be further specified as
\begin{equation}
    H(\mathbf{x})=0,\,\,\forall \mathbf{x}\in\partial\Omega_\mathrm{H},\quad (\nabla H\cdot \mathbf{u})(\mathbf{u}\cdot\mathbf{n})=0,\,\,\forall \mathbf{x}\in\partial\Omega_\mathrm{F}.
\end{equation}
We have assumed homogeneous boundary conditions for $\mathbf{x}\in\partial\Omega_\mathrm{H}$ without loss of generality, since we can add an arbitrary constant to the minimum solution and such an updated solution is still a minimum solution. Note that whenever $\partial \Omega_\text{H} \neq \emptyset$ the constant solution $H= 1/ \text{Vol}(\Omega)$ is no longer a solution to the minimiazation problem, since it cannot satisfy the homogeneous boundary condition.

\section{Weak form}
\label{sec:weak-form}
We select the trial function space $\mathcal{H}_\mathrm{trial}$ for solving the eigenvalue problem~\eqref{eq:pde}-\eqref{eq:bcs} as 
\begin{equation}
\label{eq:Htrial}
    \mathcal{H}_\mathrm{trial}=\{H\in H^1(\Omega),\,\, H(\mathbf{x})=0,\,\,\forall \mathbf{x}\in\partial\Omega_\mathrm{H}\}.
\end{equation}
We also introduce the test function space \textcolor{black}{that is the same as the trial function space}. To obtain the weak form of the PDE~\eqref{eq:pde}, we multiply both sides of the equation by $h\in\mathcal{H}_\mathrm{trial}$ and perform integration over the domain $\Omega$ to obtain
\begin{equation}
    \int_\Omega \partial_j(\partial_iHu_iu_j) hdV=-{\lambda} \int_\Omega Hh dV.
\end{equation}
For the left-hand side, we have
\begin{align}
    \int_\Omega \partial_j(\partial_iHu_iu_j) hdV
    & =\int_\Omega \partial_j(\partial_iHu_iu_jh)dV-\int_\Omega \partial_iHu_iu_j\partial_j hdV\nonumber\\
    & =\int_{\partial\Omega}\partial_iHu_iu_jh n_j dS-\int_\Omega \partial_iHu_iu_j\partial_j hdV\nonumber\\
    & =\int_{\partial\Omega}(\nabla H\cdot \mathbf{u})(\mathbf{u}\cdot \mathbf{n}) hdS-\int_\Omega (\nabla H\cdot\mathbf{u})(\nabla h\cdot\mathbf{u}) dV\nonumber\\
    & =-\int_\Omega (\nabla H\cdot \mathbf{u})(\nabla h\cdot \mathbf{u}) dV:=-a(H,h),\label{eq:left-weak}
\end{align}
where we have used the facts that $h(\mathbf{x})=0$ for $\mathbf{x}\in\Omega_\mathrm{H}$ and $(\nabla H\cdot \mathbf{u})(\mathbf{u}\cdot \mathbf{n})=0$ for $\mathbf{x}\in\Omega_\mathrm{F}$. So the weak form is obtained as follows
\begin{equation}
\label{eq:weak-form}
    a(H,h)=\lambda\langle H,h\rangle,
\end{equation}
where $\langle H,h\rangle = \int_\Omega H hdV$.

\section{Eigensolutions and minimum solution}
\label{sec:eig-to-min}
The eigenvalue problem~\eqref{eq:pde}-\eqref{eq:bcs} has a set of eigensolutions $\{(H_i,\lambda_i)\}$ that satisfies
\begin{equation}
    a(H_i,h)=\lambda_i\langle H_i,h\rangle.
\end{equation}
Since $a(\cdot,\cdot)$ is a symmetric bi-linear operator (namely, $a(H,h)=a(h,H)$), the following hold:
\begin{itemize}
\item The eigenvalue $\lambda_i$ is real and hence $H_i$ is also a real-valued function. To see that, we note that $a(\bar{H}_i,h)=\bar{\lambda}_i\langle \bar{H}_i,h\rangle$.
It follows that
\begin{equation}
    a(H_i,\bar{H}_i)=\lambda_i\langle H_i,\bar{H}_i\rangle,\quad a(\bar{H}_i,H_i)=\bar{\lambda}_i\langle \bar{H}_i,H_i\rangle
\end{equation}
Since $a(H_i,\bar{H}_i)=a(\bar{H}_i,H_i)$ and $\langle H_i,\bar{H}_i\rangle=\langle \bar{H}_i,{H}_i\rangle$, we have $\lambda_i=\bar{\lambda}_i$ and hence $\lambda_i$ is real.
\item The eigenvalues are all non-negative,  $\lambda_i\geq0$. This follows from the fact that $a(\cdot,\cdot)$ is a positive semi-definite  operator, since 

\begin{equation}
    a(H, H) = \int_\Omega |\nabla H \cdot \mathbf{u}|^2 dV \geq 0.
\end{equation}

As a result, we can arrange these eigenvalues as $0\leq \lambda_1\leq\lambda_2\leq\cdots$.
\item The eigenfunctions are orthogonal to each other, i.e.  $\langle H_i,H_j\rangle=0$ if $\lambda_i\neq\lambda_j$. Likewise, we have $a(H_i,H_j)=0$ if $\lambda_i\neq\lambda_j$.
\end{itemize}

We now introduce a Rayleigh quotient
\begin{equation}
    R(h)={a(h,h)},\quad\text{with}\,\, {\langle h,h\rangle}=1.
\end{equation}
If we let $h=\sum_i c_i H_i$, then ${\langle h,h\rangle}=1$ implies that $\sum_i c_i^2=1$. After the orthonormalization of $\{H_i\}$, we have
\begin{equation}
    R(h)=a\left(\sum_i c_i H_i,\sum_j c_j H_j\right)=\sum_i c_i^2a(H_i,H_i)=\sum_i \lambda_i c_i^2\geq \lambda_1.
\end{equation}
The minimum is achieved when $h=H_1$. Similarly, if we restrict $c_1=\cdots c_{i-1}=0$, we have
\begin{equation}
    R(h)\geq \lambda_i.
\end{equation}
In the case that $\Omega_\mathrm{H}=\emptyset$, the constant $H=1/\text{Vol}(\Omega)$ is a minimal solution, therefore we have $\lambda_1=0$. However, this trivial solution is not the first integral we seek. Thus, we should restrict $c_1=0$ and look for $H_2$. The eigenvalue $\lambda_2$ characterizes the minimal value of the objective functional.

\section{Finite-element implementation in FEniCS}
\label{sec:fenics}
We use FEniCS~\cite{alnaes2015fenics,logg2012automated} to solve the eigenvalue problem described in Sect.~\ref{sec:formulation}. FEniCS is an open source package for finite-element analysis. The main steps of using FEniCS to solve the eigenvalue problem are as follows:
\begin{itemize}
    \item Specify the domain $\Omega$ and create a mesh to discretize the domain. Users can use built-in functions of FEniCS to generate a mesh or load mesh files generated by other packages into FEniCS. We use a tetrahedron mesh throughout this study. 
    \item Specify test and trial function spaces shown in~\eqref{eq:Htrial}. In particular, elements along with boundary conditions need to be specified to define the spaces. Here
    we use Lagrange elements of interpolation order $r$ in our computations. We choose $r=2$ unless otherwise stated.
    \item Specify $\mathbf{A}$ and $\mathbf{B}$ in a generalized eigenvalue problem $\mathbf{A}\mathbf{v}=\lambda \mathbf{B}\mathbf{v}$. Here $\mathbf{A}$ and $\mathbf{B}$ are matrices from $a(H,h)$ and $\langle H,h\rangle$, as seen in the weak form~\eqref{eq:weak-form}. Since both $\mathbf{A}$ and $\mathbf{B}$ are symmetric, this problem is a generalized Hermitian eigenvalue problem.
    \item Call \texttt{SLEPcEigenSolver} of FEniCS to solve for the eigenvalue problem. \texttt{SLEPcEigenSolver} is a wrapper for the SLEPc eigenvalue solver~\cite{hernandez2005slepc}.
    It should be pointed out that we do not need to solve for all eigenvalues, only for a small subset of them that are closest to zero. A spectral transform (shift-and-invert) is used to enhance the convergence of computing these target eigenvalues~\cite{hernandez2005slepc}. A parameter called spectral shift needs to be specified in the transform. This parameter should be close to the target eigenvalues. We set this parameter to be a negative number of small norm.
\end{itemize}

\section{Benchmark studies: generalized axisymmetric flows}
\label{sec:benchmark}
In this section, we construct representative streamsurfaces for generalized axisymmetric flows of the form
\begin{equation}
\label{eq:gaxiflow-cyl}
    \dot{r} = u_r(r,z),\quad \dot{\theta}=u_\theta(r,z),\quad \dot{z}=u_z(r,z), \quad (r,\theta,z)\in\Omega.
\end{equation}
Such flows are \emph{generalized} axisymmetric because we also allow for a non-zero angular velocity component. In \emph{classic} axisymmetric flow, by contrast, we have $u_\theta=0$. We will consider a sphere, a cylinder, and a hollow cylinder for the domain $\Omega$ to demonstrate the use of our methodology. In contrast, the Fourier representation used in~\cite{stergios} would not be able to handle these geometries.

We can find an exact first integral for any generalized axisymmetric flow as follows. For a given generalized axisymmetric flow, we have the corresponding restricted axisymmetric flow with the same $u_r$ and $u_z$ but zero angular velocity. The Stokes stream function $\psi(r,z)$ for the restricted axisymmetric flow is then an exact first integral. To see this, we recall that
\begin{equation}
\label{eq:phi-to-u}
    u_z=\frac{1}{r}\frac{\partial\psi}{\partial r},\quad u_r=-\frac{1}{r}\frac{\partial\psi}{\partial z}.
\end{equation}
We then have
\begin{equation}
   \frac{d\psi}{dt}=\nabla\psi\cdot{u} = \frac{\partial\psi}{\partial z}\dot{z}+\frac{\partial\psi}{\partial r}\dot{r}=\frac{\partial\psi}{\partial z}u_z+\frac{\partial\psi}{\partial r}u_r\equiv0. 
\end{equation}
Thus, $\psi(r,z)$ is an exact first integral, independently of the angular velocity $u_\theta$.

We note that the generalized axisymmetric flow with $(u_z,u_r)$ induced by the Stokes stream function $\psi(r,z)$ satisfies the continuity equation automatically by construction:
\begin{equation}
    \nabla\cdot u = \frac{1}{r}\frac{\partial(ru_r)}{\partial r}+\frac{1}{r}\frac{\partial u_\theta}{\partial\theta}+\frac{\partial u_z}{\partial z}= \frac{1}{r}\frac{\partial(ru_r)}{\partial r}+\frac{\partial u_z}{\partial z}\equiv0.
\end{equation}
Since the existence of $\psi$ is guaranteed, we can solve for $\psi(r,z)$ analytically provided that we have analytical expressions for $u_r(r,z)$ and $u_z(r,z)$.

In our implementation in FEniCS, we consistently use a Cartesian coordinate system for computations. So we need transform the velocity field~\eqref{eq:gaxiflow-cyl} to Cartesian coordinates. We have
\begin{gather}
    u_x=d(r\cos\theta)/dt=\dot{r}\cos\theta-r\sin\theta\dot{\theta}=u_r\cos\theta-r\sin\theta u_\theta=u_rx/\sqrt{x^2+y^2}-yu_\theta,\\
    u_y =d(r\sin\theta)/dt=\dot{r}\sin\theta+r\cos\theta\dot{\theta}=u_r\sin\theta+r\cos\theta u_\theta=u_ry/\sqrt{x^2+y^2}+xu_\theta,\\
    u_z = u_z(r,z).
\end{gather}

As we do not impose any Dirichlet boundary conditions for the generalized flow, we have $\Omega_\mathrm{H}=\emptyset$. Following the discussion in Sect.~\ref{sec:eig-to-min}, we seek the eigenvector $H_2$ that corresponds to the second smallest eigenvalue $\lambda_2$. Since the existence of a nontrivial first integral is guaranteed for the generalized axisymmetric flow, we expect that the eigenvalue $\lambda_2$ is numerically close to zero. In particular, $\lambda_2\to0$ as the resolution of the mesh increases.

The first integral $H_2$ will generically not be equal to $\psi$ because $c_1H_2+c_2$ is also a stationary solution to the optimization problem for all constants $c_1$ and $c_2$. However, we expect that there exists constants $c_1$ and $c_2$ such that
\begin{equation}
\label{eq:fit-H}
    \psi\approx c_1H_2+c_2=:\hat{H}_2
\end{equation}
For the purpose of validation, we will use the least squares method to fit these two coefficients with $\psi$ and $H_2$ evaluated at a collection of grid points. We expect the coefficient of determination for the linear regression to be close to one, i.e., $R^2\approx 1$.

\begin{figure*}[!ht]
\centering
\includegraphics[width=0.3\textwidth]{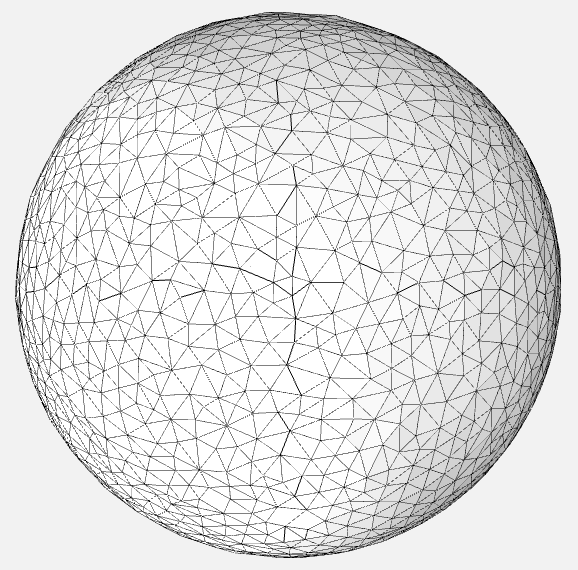}
\includegraphics[width=0.3\textwidth]{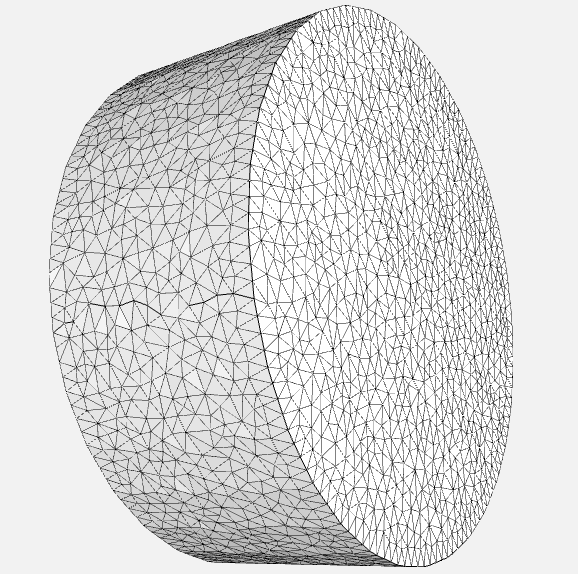}
\includegraphics[width=0.3\textwidth]{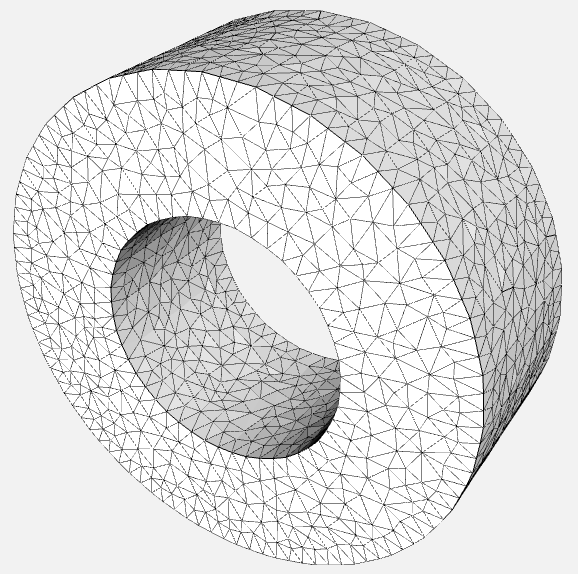}
\caption{Meshes used in the computation of the benchmark generalized axisymmetric flows in the domain of a sphere (left), a cylinder (middle) and a hollow cylinder (right).}
\label{fig:mesh-axis}
\end{figure*}

\begin{figure*}[!ht]
\centering
\includegraphics[width=0.32\textwidth]{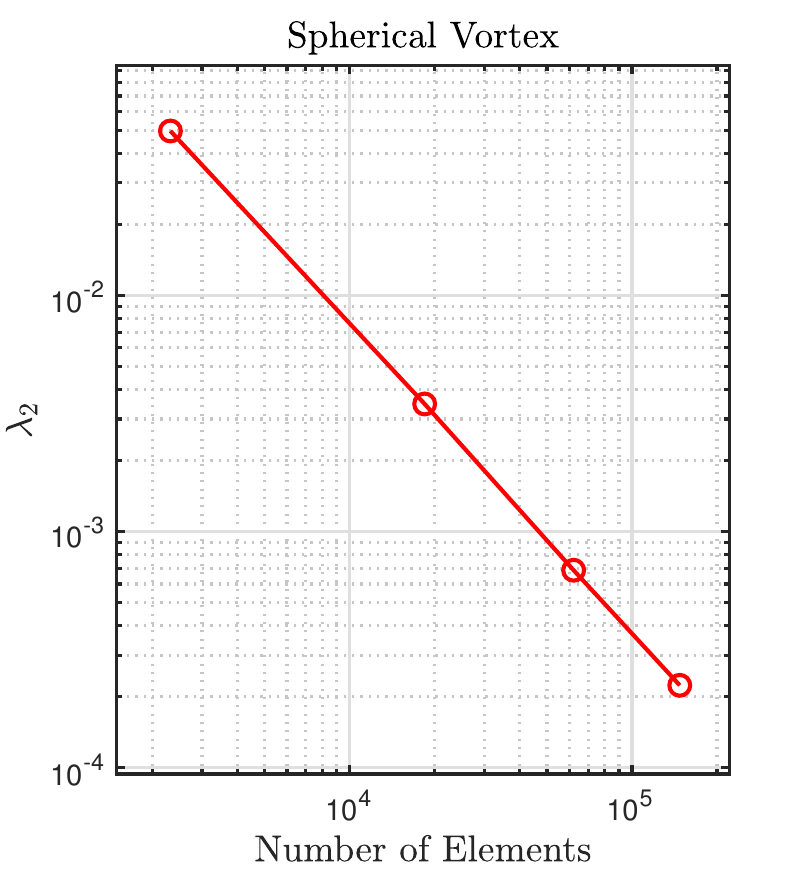}
\includegraphics[width=0.32\textwidth]{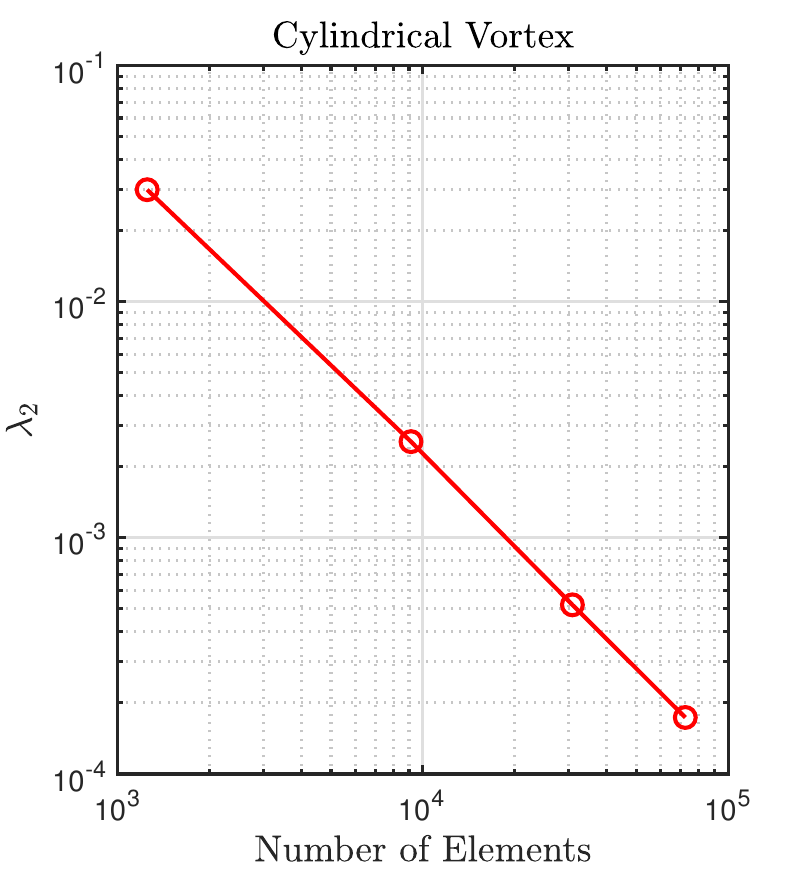}
\includegraphics[width=0.32\textwidth]{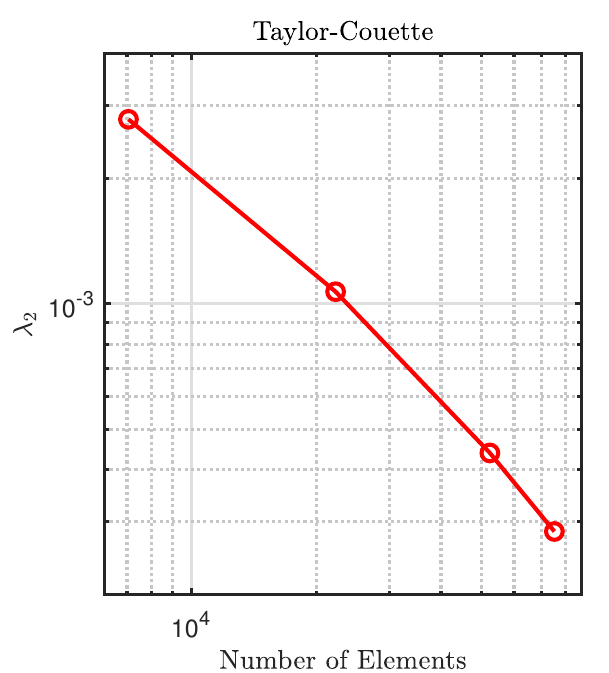}
\caption{Second smallest eigenvalues as functions of the number of elements used in the  discretization of the generalized axisymmetric flows: spherical vortex (left), cylindrical vortex (middle) and Taylor-Couette flow (right).}
\label{fig:lamd2-axis}
\end{figure*}

\subsection{Spherical vortex}
Let us now consider the domain $\Omega=\{(x,y,z):x^2+y^2+z^2\leq1\}$ and
\begin{equation}
    \dot{r} = rz,\quad \dot{z} = 1-2r^2-z^2,\quad \dot{\theta} = {2c}/({r^2+\epsilon}),
\end{equation}
where $c$ is an arbitrary constant. The flow above is a superposition of the well-known Hill's spherical vortex with a line vortex on the $z$ axis, which induces a swirl velocity $\dot{\theta}$~\cite{haller1998reduction}. We have added $\epsilon$ to avoid singularity of the swirl velocity on the $z$ axis. In Cartesian coordinates, we have
\begin{equation}
    u_x = xz-\frac{2cy}{x^2 + y^2+\epsilon},\quad
    u_y = yz+\frac{2cx}{x^2 + y^2+\epsilon},\quad
    u_z = 1-2(x^2 + y^2)-z^2.
\end{equation}
The Stokes stream function for this generalized axisymmetric flow is given by 
\begin{equation}
\label{eq:psi-sph}
\psi(r,z) = 0.5 r^2(1-z^2-r^2).
\end{equation}

\begin{figure*}[!ht]
\centering
\includegraphics[width=0.45\textwidth]{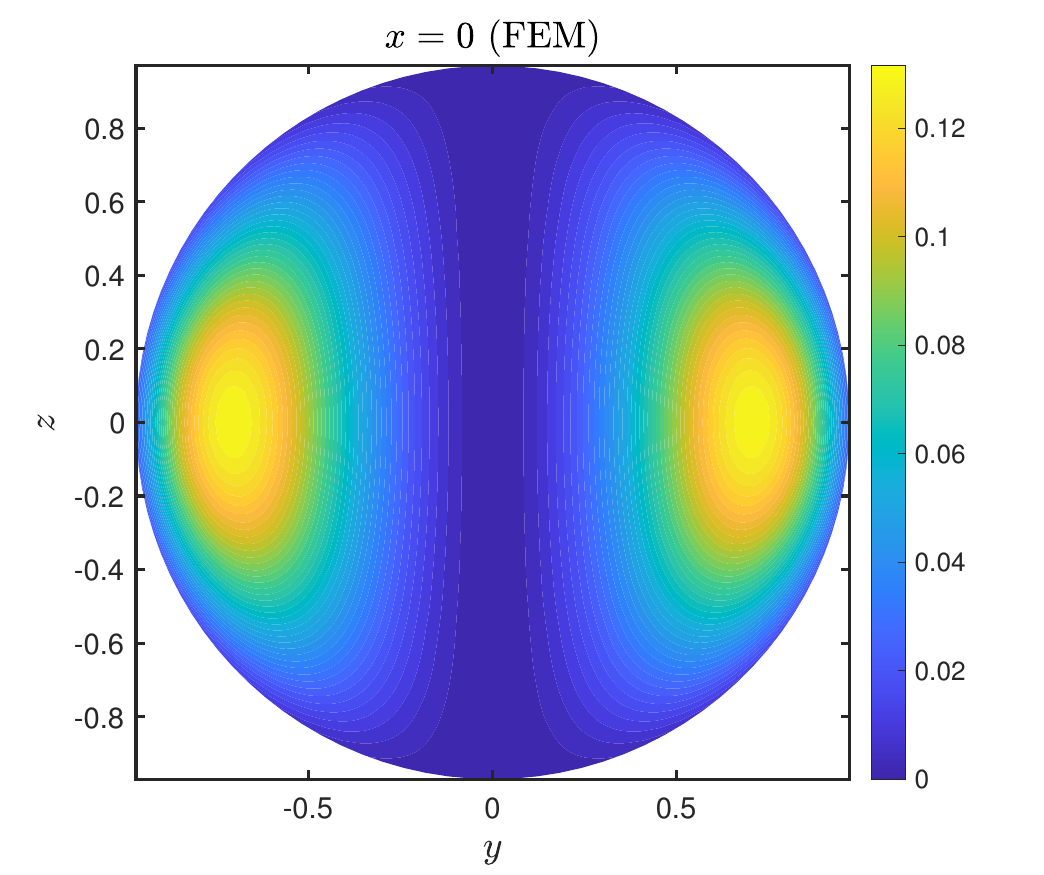}
\includegraphics[width=0.45\textwidth]{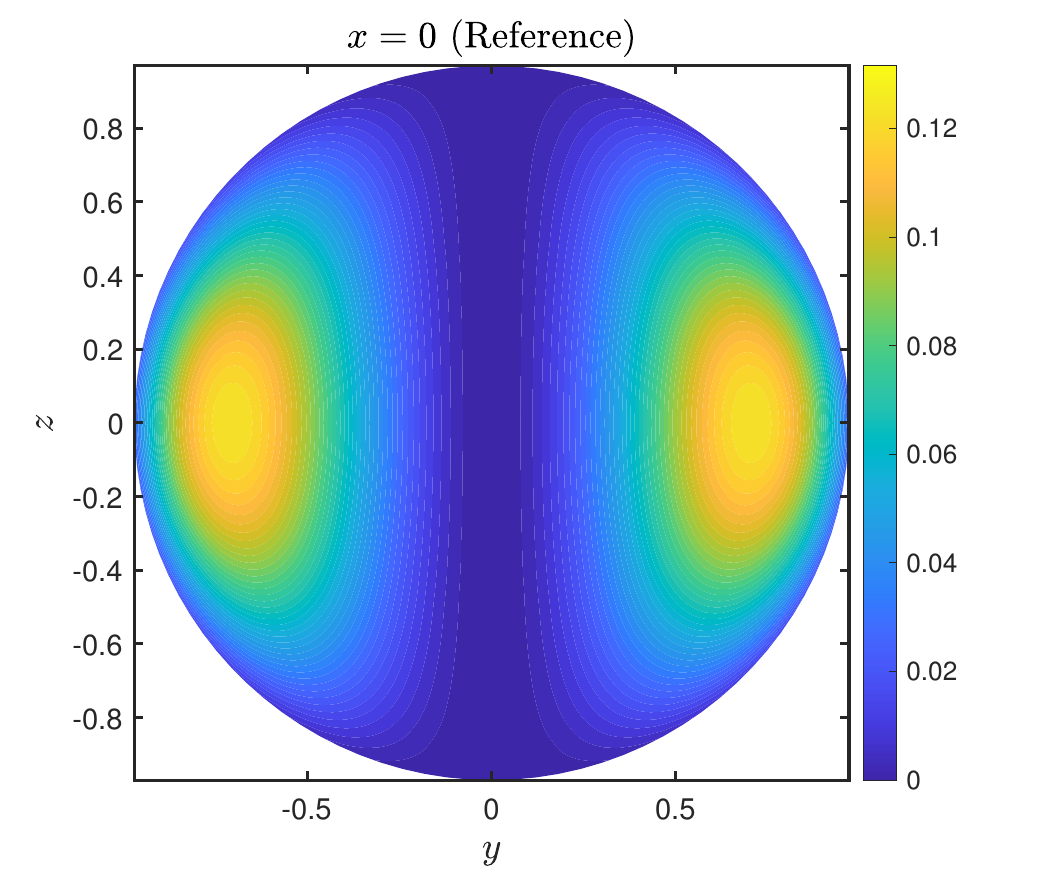}\\
\includegraphics[width=0.45\textwidth]{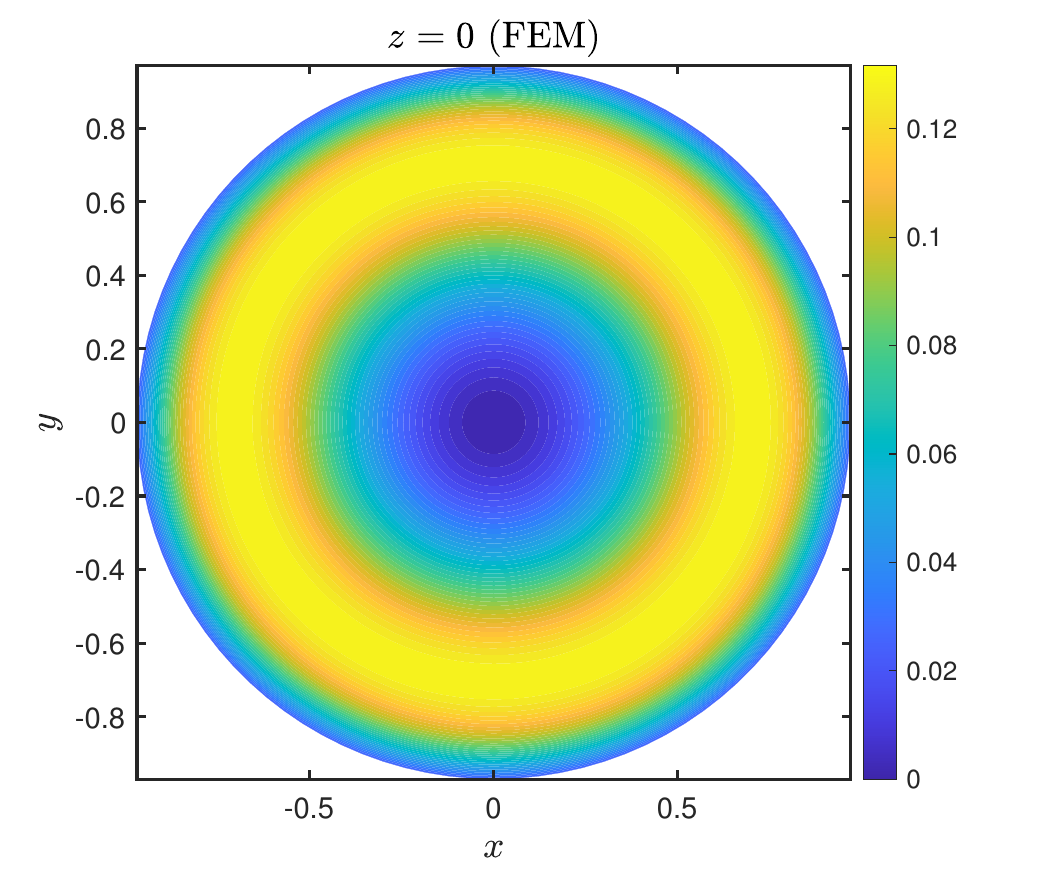}
\includegraphics[width=0.45\textwidth]{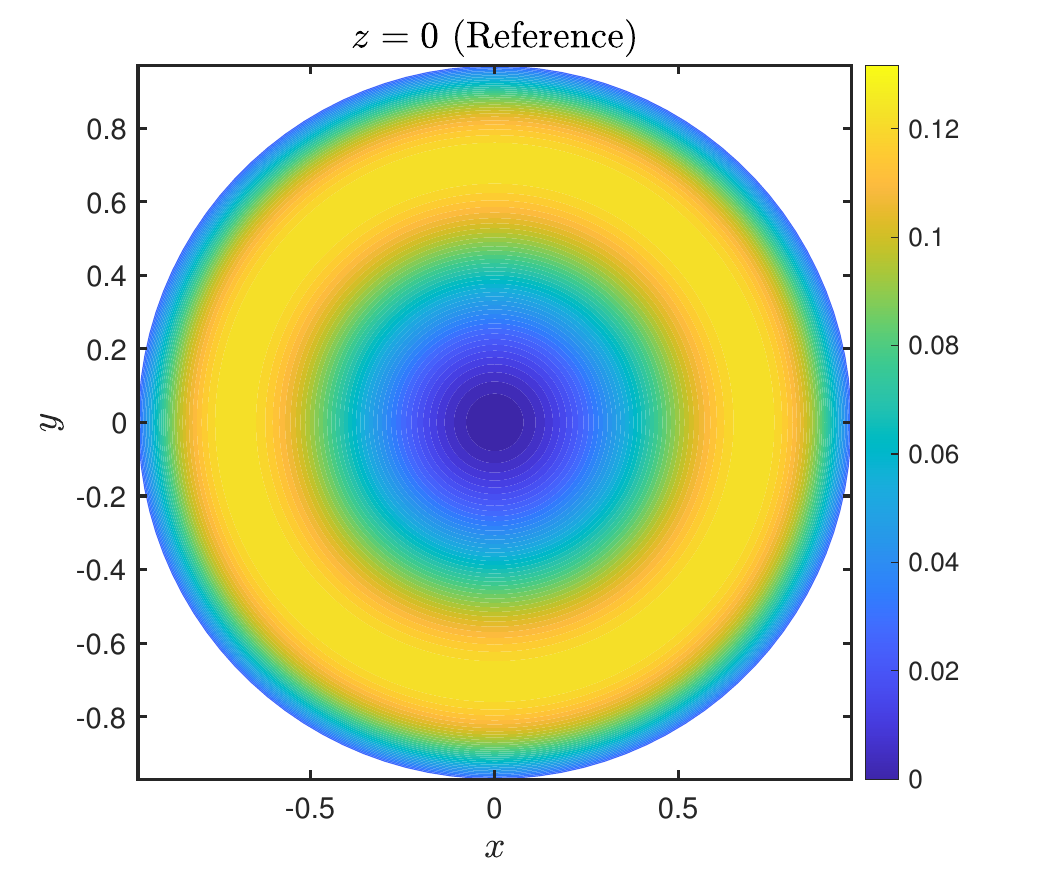}
\caption{Contour plots of $\hat{H}_2$ for the spherical vortex obtained from finite-element methods with 62,105 elements (left panels) and $\psi$ in~\eqref{eq:psi-sph} (right panels), at cross section $x=0$ (upper panels) and $z=0$ (lower panels). Here, the solutions from the finite-element computation are denoted by FEM, and that of analytical expressions are denoted by Reference.}
\label{fig:sph-2d}
\end{figure*}

In the following computations, we take $c=0.1$ and $\epsilon=0.1$. As motioned earlier, we use a tetrahedron mesh to discretize the sphere, as shown in the left panel of Fig.~\ref{fig:mesh-axis}. We use quadratic Lagrange elements to interpolate the unknown function $H$. As predicted, we obtain $\lambda_1=0$ with a constant eigenvector. As seen in the left panel of Fig.~\ref{fig:lamd2-axis}, $\lambda_2$ indeed converges to zero when the number of elements is increased, indicating that the flow admits a nontrivial first integral.

As an illustration of the obtained nontrivial first integral, we plot $\hat{H}_2$ at the cross sections $x=0$ and $z=0$ in Fig.~\ref{fig:sph-2d}. Here we present the results for $\hat{H}_2$ defined in~\eqref{eq:fit-H} instead of $H_2$ to compare against the Stokes stream function~\eqref{eq:psi-sph}. Specifically, we first obtain $H_2$ discretized with 62,105 elements (see the third circle in the left panel of Fig.~\ref{fig:lamd2-axis}), and then fit a linear function following~\eqref{eq:fit-H}. Indeed, the linear relationship holds because the linear regression returns $R^2=0.9976$. As seen in Fig.~\ref{fig:sph-2d}, our numerical results match the reference solution given by~\eqref{eq:psi-sph} well.

\begin{figure*}[!ht]
\centering
\includegraphics[width=0.45\textwidth]{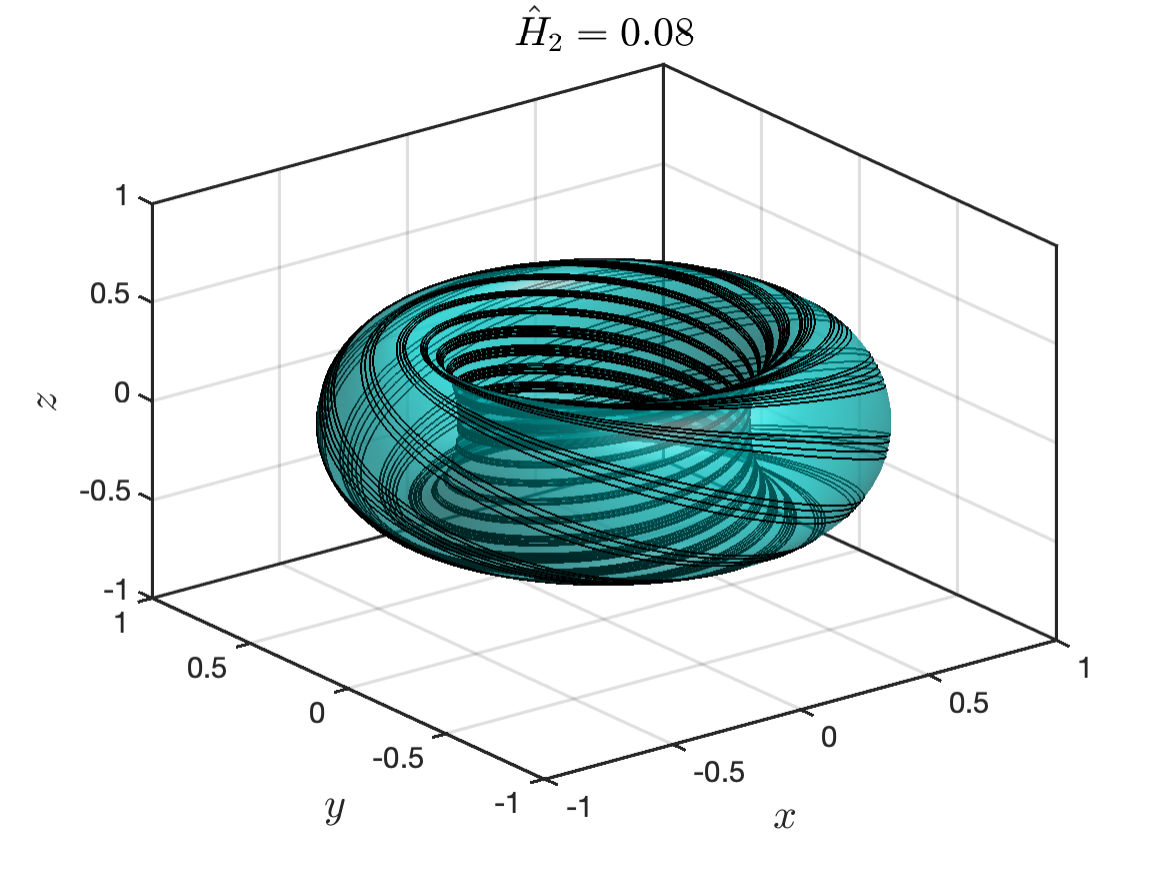}
\includegraphics[width=0.45\textwidth]{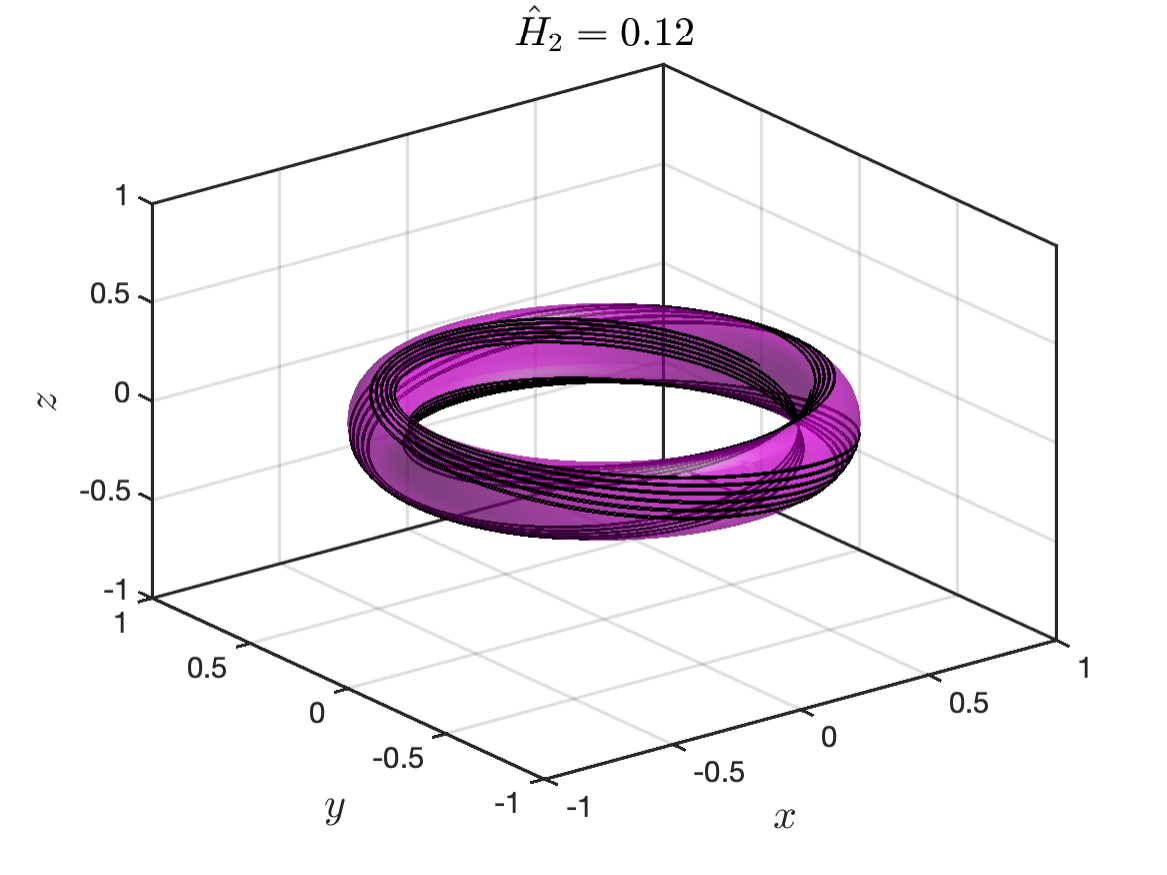}
\caption{Contour plots of isosurfaces for $\hat{H}_2$ of the spherical vortex obtained from finite-element methods with 62105 elements. Here we have $\hat{H}_2=0.08$ (left panels) and $\hat{H}_2=0.12$ (right panels). The black lines are streamlines from forward simulations with initial points on the isosurfaces.}
\label{fig:sph-3d}
\end{figure*}

We infer from Fig.~\ref{fig:sph-2d} that the flow has a family of vortex rings. To illustrate this, we plot the isosurfaces for $\hat{H}_2=0.08$ and $\hat{H}_2=0.12$ in Fig.~\ref{fig:sph-3d}, from which we see torus-shaped isosurfaces. Given these surfaces are streamsurfaces, they should be invariant under the flow. To validate the invariance of these isosurfaces, we launch streamlines of the flow. In particular, we take a point on each of these isosurface as the initial condition and integrate the flow forward in time. The generated trajectories indeed stay on the isosurfaces, which again validates our results.

\subsection{Cylindrical vortex}
Next, we consider the cylindrical domain $\Omega=\{(x,y,z):x^2+y^2\leq1, -0.4\leq z\leq0.4\}$ and the flow
\begin{equation}
\label{eq:vel-cylin}
    \dot{r} = 4rz,\quad \dot{z} = 1-2r^2-4z^2,\quad \dot{\theta} = \omega,
\end{equation}
where $\omega$ denotes a rigid body angular velocity. This flow is a superposition of a cylindrical vortex with a rigid body rotation. In Cartesian coordinates, we have
\begin{equation}
u_x = 4xz-y\omega,\quad
u_y = 4yz+x\omega,\quad
u_z = 1-2(x^2+y^2)-4z^2.
\end{equation}
The Stokes stream function for this generalized axisymmetric flow is 
\begin{equation}
\label{eq:psi-cylin}
\psi(r,z) = 0.5r^2\left(1-{r^2}-4{z^2}\right).
\end{equation}
One can easily check that the function above indeed induces the $\dot{r}$ and $\dot{z}$ in~\eqref{eq:vel-cylin} (cf.~\eqref{eq:phi-to-u}).

In the following computations, we take $\omega=1$. We use a tetrahedron mesh to discretize the cylinder, as shown in the middle panel of Fig.~\ref{fig:mesh-axis}. We use quadratic Lagrange elements to interpolate the unknown function $H$. We again obtain $\lambda_1=0$ with a constant eigenvector. As seen in the middle panel of Fig.~\ref{fig:lamd2-axis}, $\lambda_2$ converges to zero when the number of elements is increased, indicating that the flow indeed admits a nontrivial first integral.

\begin{figure*}[!ht]
\centering
\includegraphics[width=0.45\textwidth]{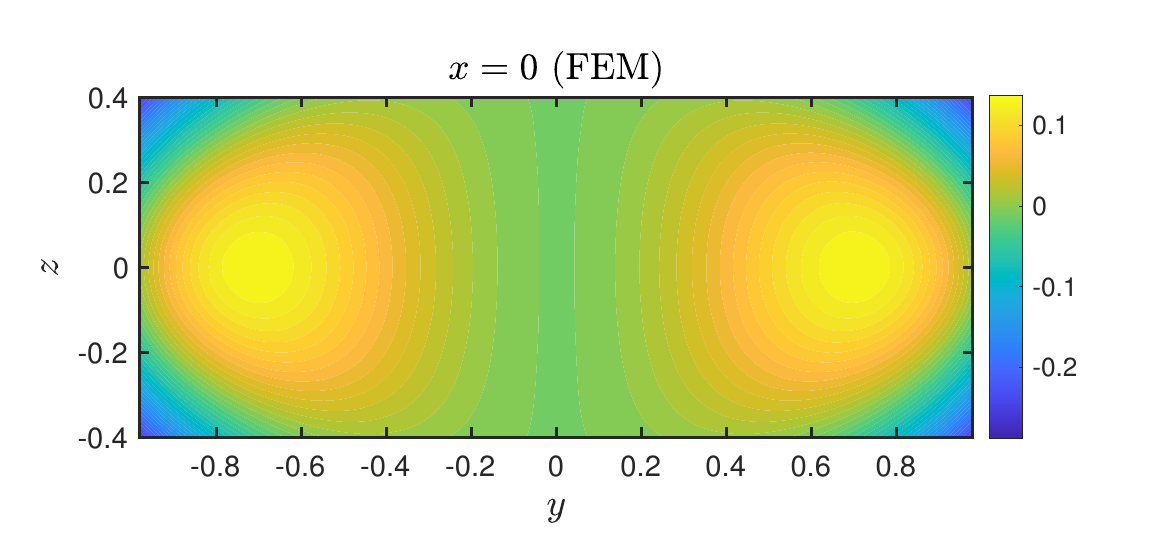}
\includegraphics[width=0.45\textwidth]{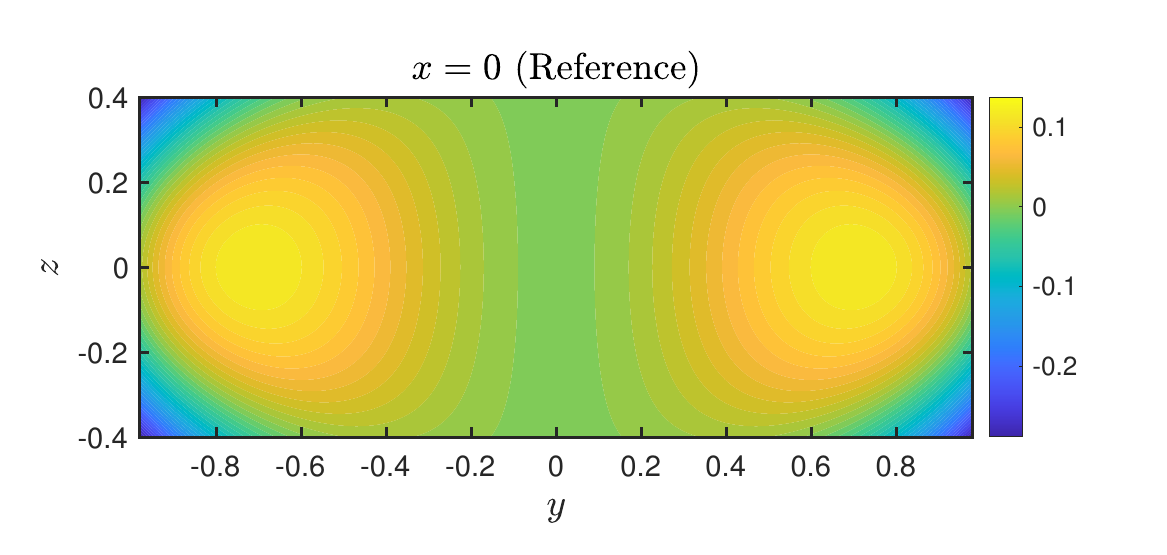}\\
\includegraphics[width=0.45\textwidth]{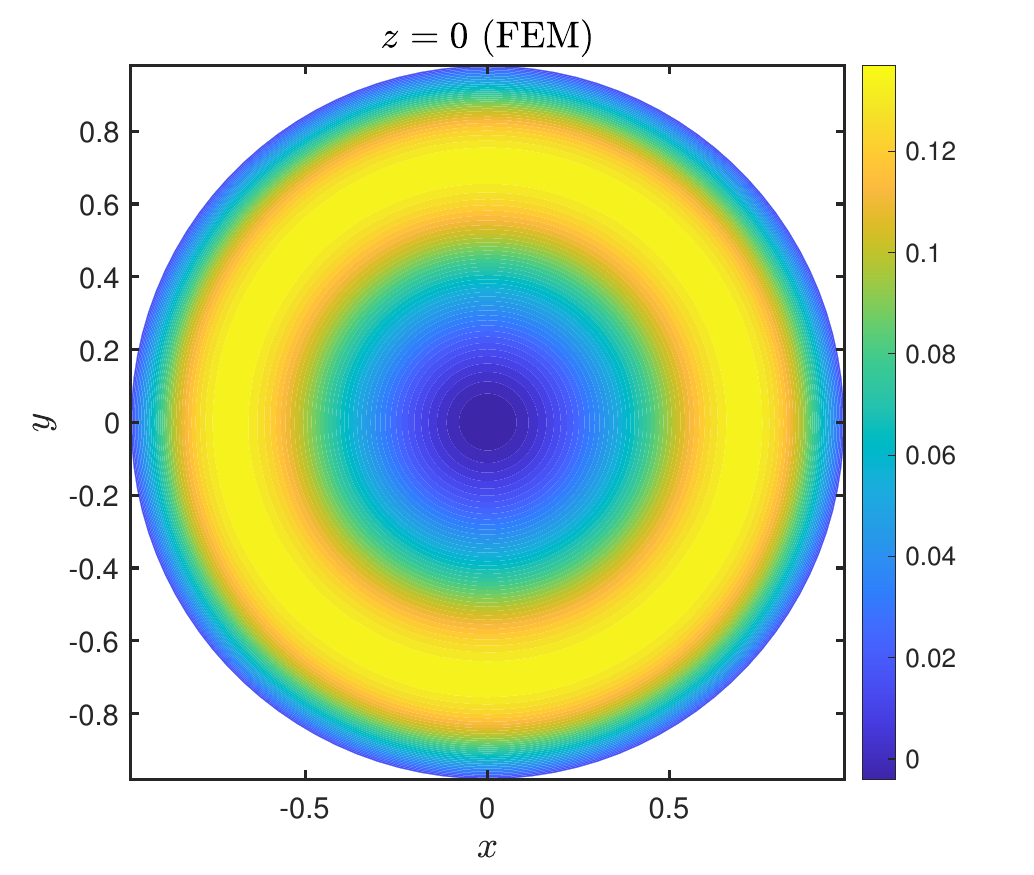}
\includegraphics[width=0.45\textwidth]{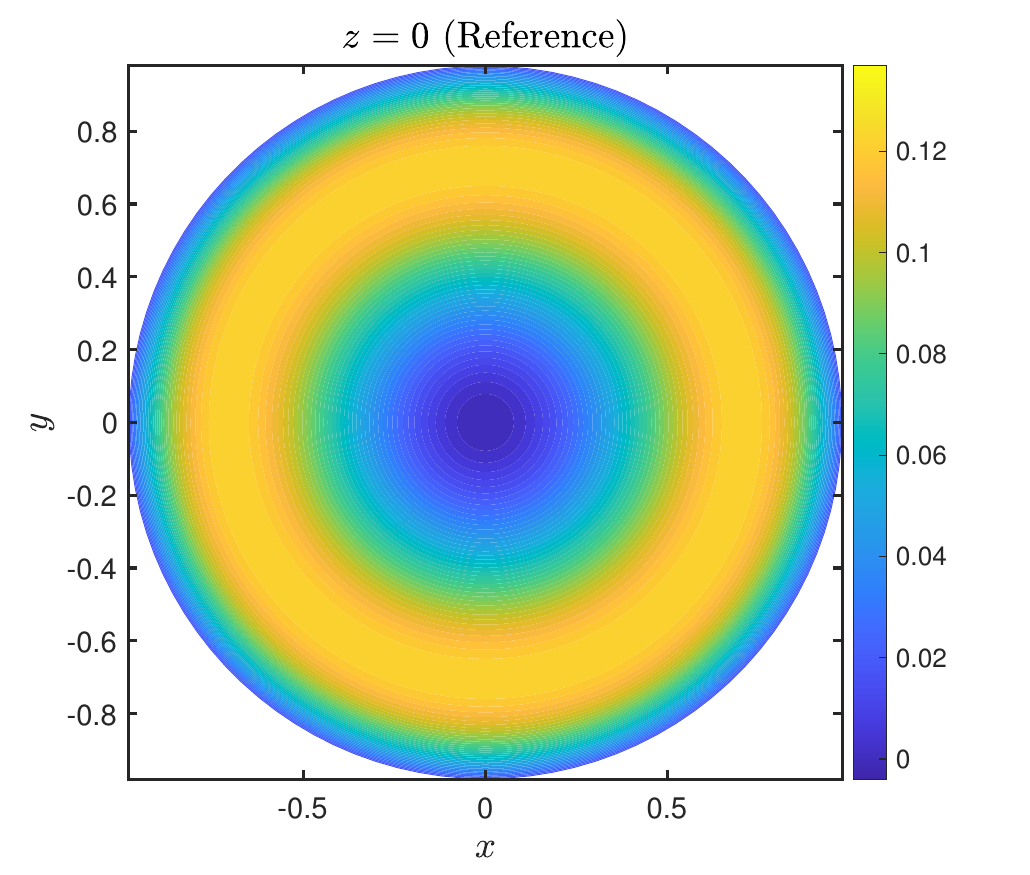}
\caption{Contour plots of $\hat{H}_2$ for the cylindrical vortex obtained from finite-element methods with 30,888 elements (left panels) and $\psi$ in~\eqref{eq:psi-cylin} (right panels), at cross section $x=0$ (upper panels) and $z=0$ (lower panels). Here, the solutions from finite-element computation are denoted by FEM, and that of analytical expressions are denoted by Reference.}
\label{fig:cylin-2d}
\end{figure*}

As an illustration of the nontrivial first integral obtained in this fashion, we plot $\hat{H}_2$ at the cross sections $x=0$ and $z=0$ in Fig.~\ref{fig:cylin-2d}. Similarly, we obtain $\hat{H}_2$ from $H_2$ via a linear fit shown in~\eqref{eq:fit-H}. Here $H_2$ is discretized with 30,888 elements. The linear fitting returns $R^2=0.9920$. As seen in Fig.~\ref{fig:cylin-2d}, our numerical results closely match the reference solution given by~\eqref{eq:psi-cylin}.

\begin{figure*}[!ht]
\centering
\includegraphics[width=0.45\textwidth]{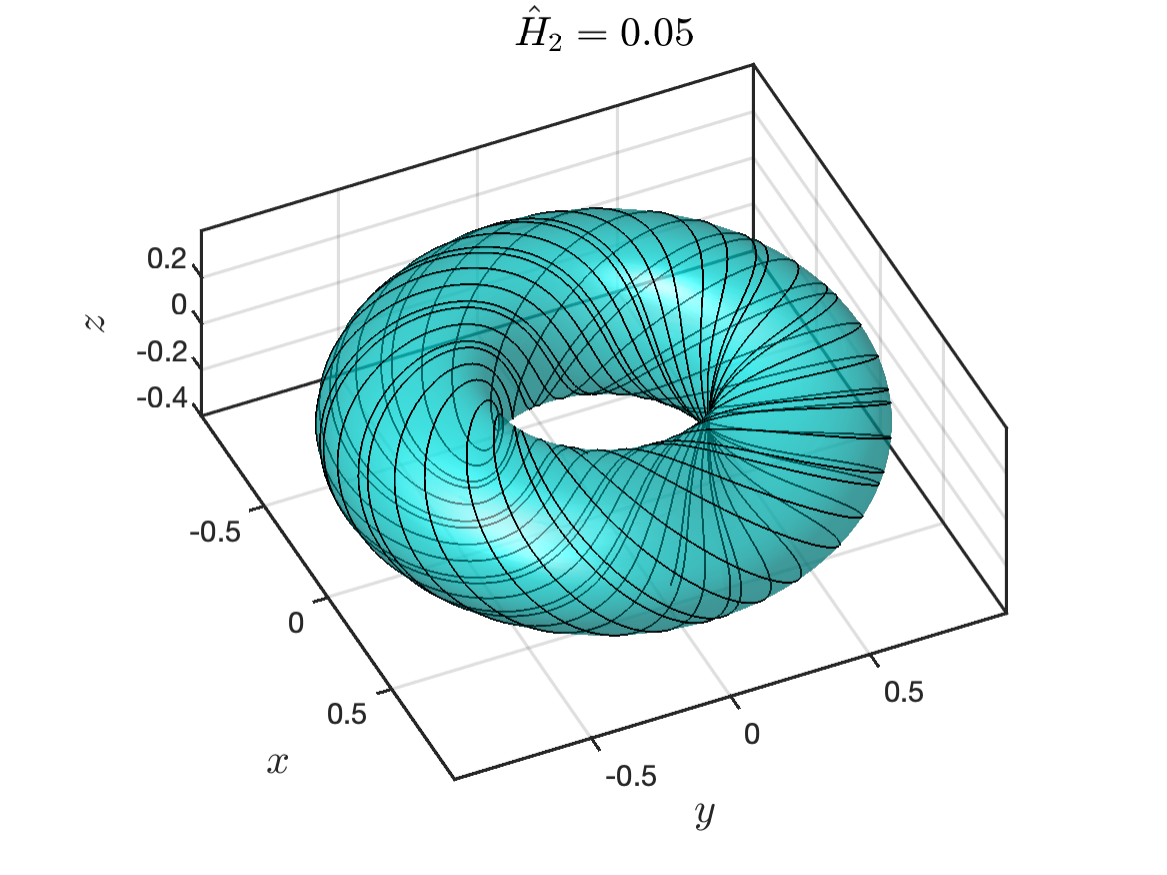}
\includegraphics[width=0.45\textwidth]{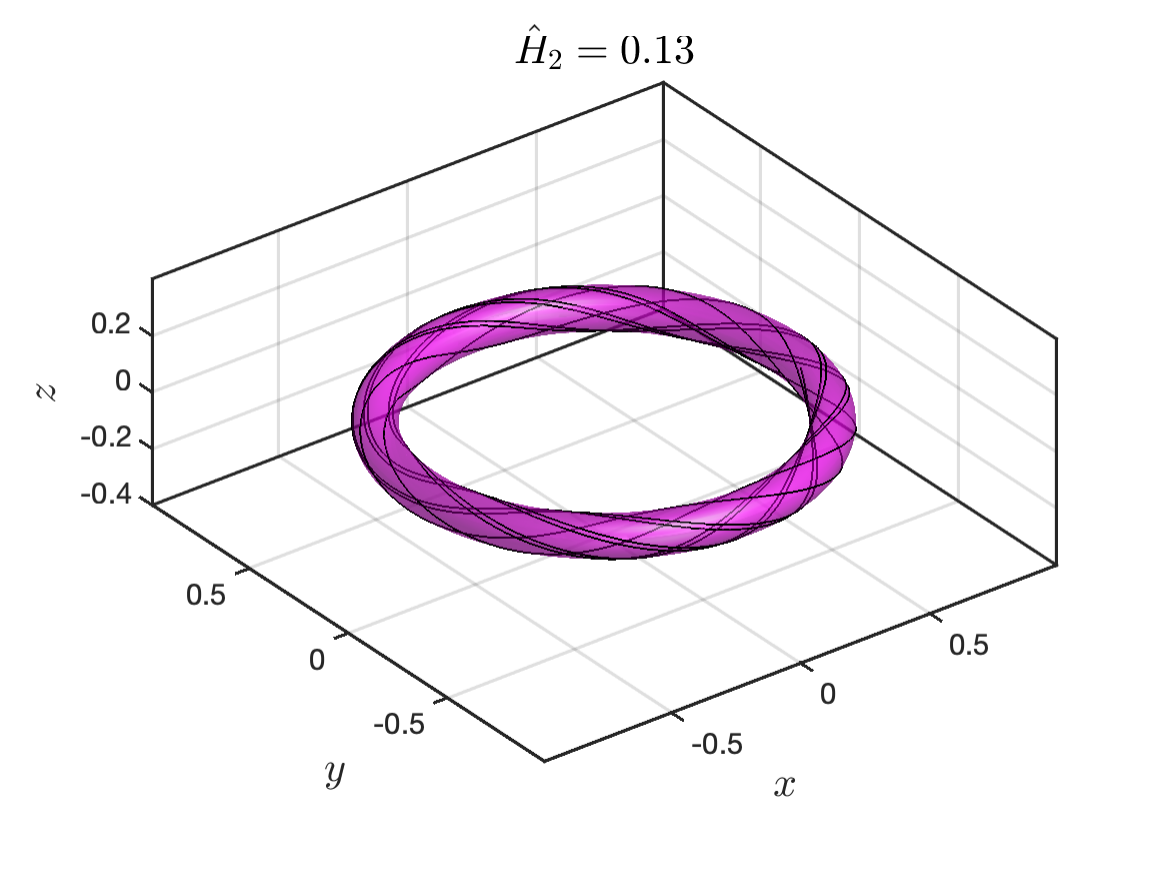}
\caption{Contour plots of isosurfaces for $\hat{H}_2$ of the cylindrical vortex obtained from finite-element methods with 30,888 elements. Here we have $\hat{H}_2=0.05$ (left panels) and $\hat{H}_2=0.13$ (right panels). The black lines are streamlines from forward simulations with initial points on the isosurfaces.}
\label{fig:cylin-3d}
\end{figure*}

We see from Fig.~\ref{fig:cylin-2d} that the flow has a family of vortex rings, but now these vortex rings are constrained into the cylinder instead of a sphere. We plot the isosurfaces for $\hat{H}_2=0.05$ and $\hat{H}_2=0.13$ in Fig.~\ref{fig:cylin-3d}, from which we see torus-shaped isosurfaces. We again launch streamlines of the flow to validate the invariance of these isosurfaces. We take a point on each of these isosurface as the initial condition and integrate them forward in time. The generated trajectories indeed stay close to the isosurfaces, as seen in Fig.~\ref{fig:cylin-3d}. This again serves as a validation of the results from our finite-element calculations.

\subsection{Taylor-Couette flow}
We now consider a Taylor-Couette flow of a viscous fluid between two rotating cylinders. Linear stability theory successfully explains many of the flow transitions in this standard geometry~\cite{gollub1975}. For low Reynolds numbers, the flow is fully laminar and has a closed form analytic expression. 

Let us consider the domain $\Omega=\{(x,y,z): 1\leq x^2+y^2\leq 4,\,\, 0\leq z\leq \pi\}$, which describes the volume between two concentric cylinders with height $\pi$ and radii $r_\mathrm{in}=1$, $r_\mathrm{out} = 2$, respectively. We consider the case of a stationary outer wall and a steadily rotating inner wall. 
The steady solutions and their stability are determined by the Reynolds number defined as $\text{Re} = {\Omega_\mathrm{in} r_\mathrm{in} \delta}/{\nu}$,
where the radial velocity of the inner wall is $\Omega_\mathrm{in}$, the radius of the inner cylinder is $r_\mathrm{in}$, the distance between the concentric cylinders is $\delta = r_\mathrm{out}- r_\mathrm{in}$ and the kinematic viscosity is $\nu$.
For low Reynolds numbers, the steady flow that develops is steady and purely azimuthal. Using the distance, $r$, from the center line of the cylinders, the angle $\theta$ and the vertical coordinate $z$, the velocity field in cylindrical coordinates reads as $\mathbf{u}(r, \theta, z) = \left(u_r, u_\theta, u_z \right)(r, \theta, z)$.
The base flow, which is stable for low Reynolds-numbers, is called Couette flow \cite{landauFluidMechanics1987} and has the form
\begin{equation}
    \label{eq:taylorCouetteBase}
    u_\theta(r) = -\Omega_\mathrm{in}\frac{ r_\mathrm{in}^2}{r_\mathrm{out}^2-r_\mathrm{in}^2}r + \Omega_\mathrm{in}\frac{r_\mathrm{out}^2 r_\mathrm{in}^2}{r_\mathrm{out}^2-r_\mathrm{in}^2}\frac{1}{r}, \quad u_r = u_z = 0.
\end{equation}

For larger Reynolds numbers, the Couette flow loses its stability and the newly obtained stable flow exhibits the well-known Taylor vortices. This flow now has non-trivial radial and axial velocities but it is still axisymmetric, that is, we have $u_r(r,z),\,u_\theta(r,z)$ and $u_z(r, z)$, as in all the examples shown above. 

This generalized axisymmetric flow is already more complicated than the previous two, as there are no analytical solutions to the Stokes stream function. As a result,  we do not have analytical expressions for the velocity field in this case. 

\begin{figure*}[!ht]
\centering
\includegraphics[width=0.32\textwidth]{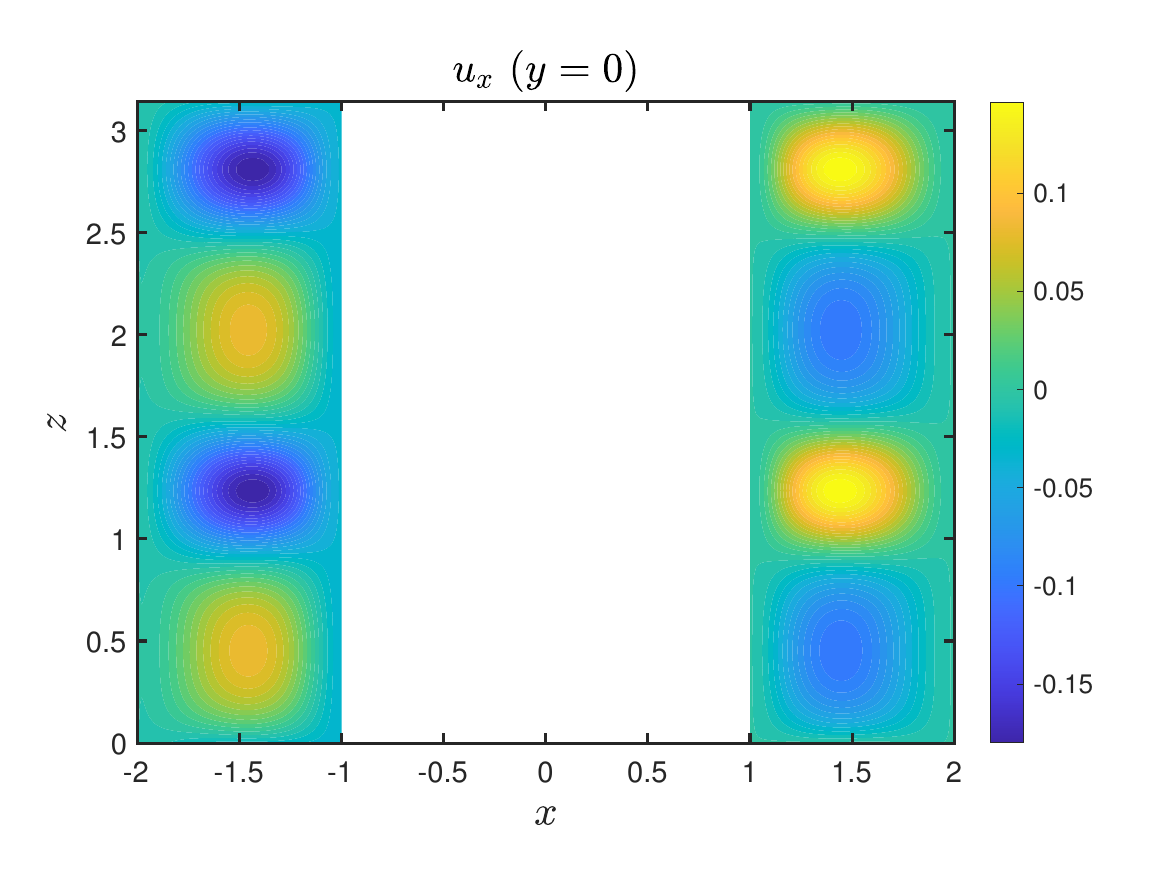}
\includegraphics[width=0.32\textwidth]{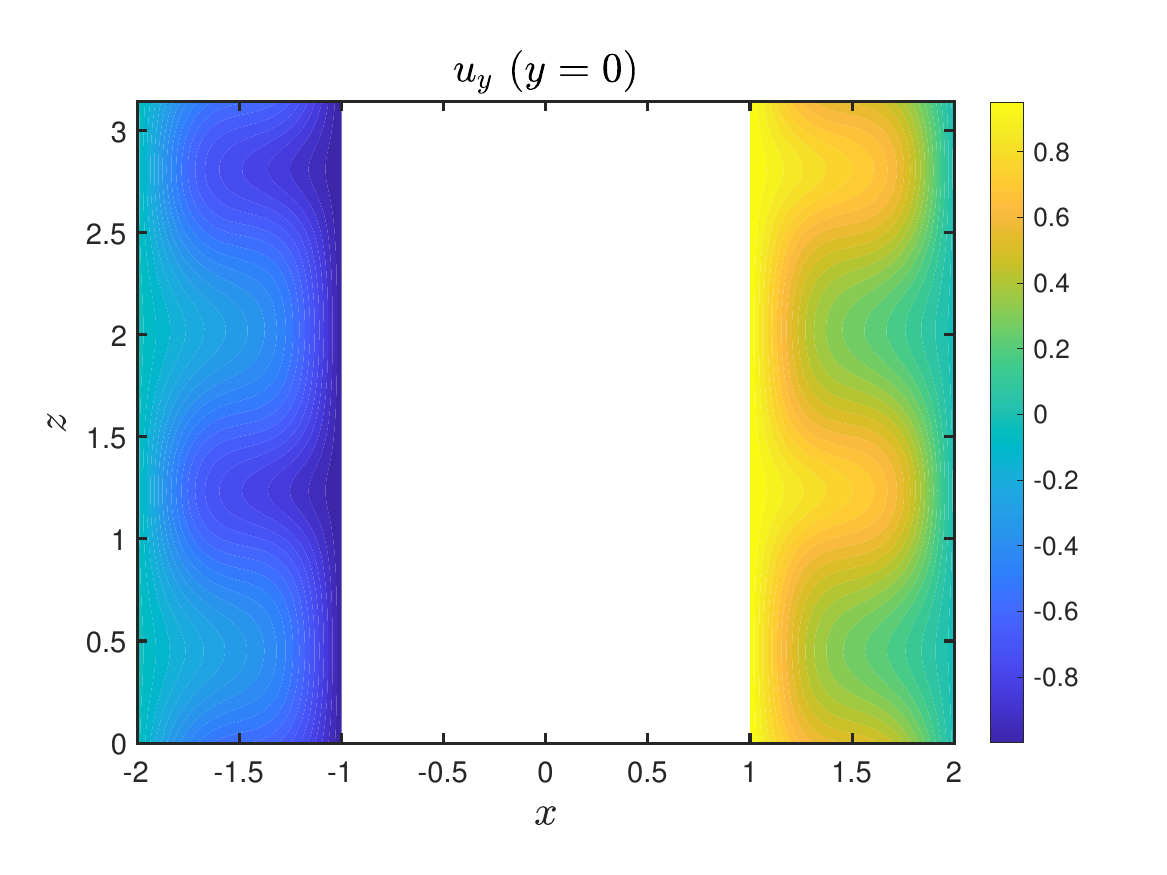}
\includegraphics[width=0.32\textwidth]{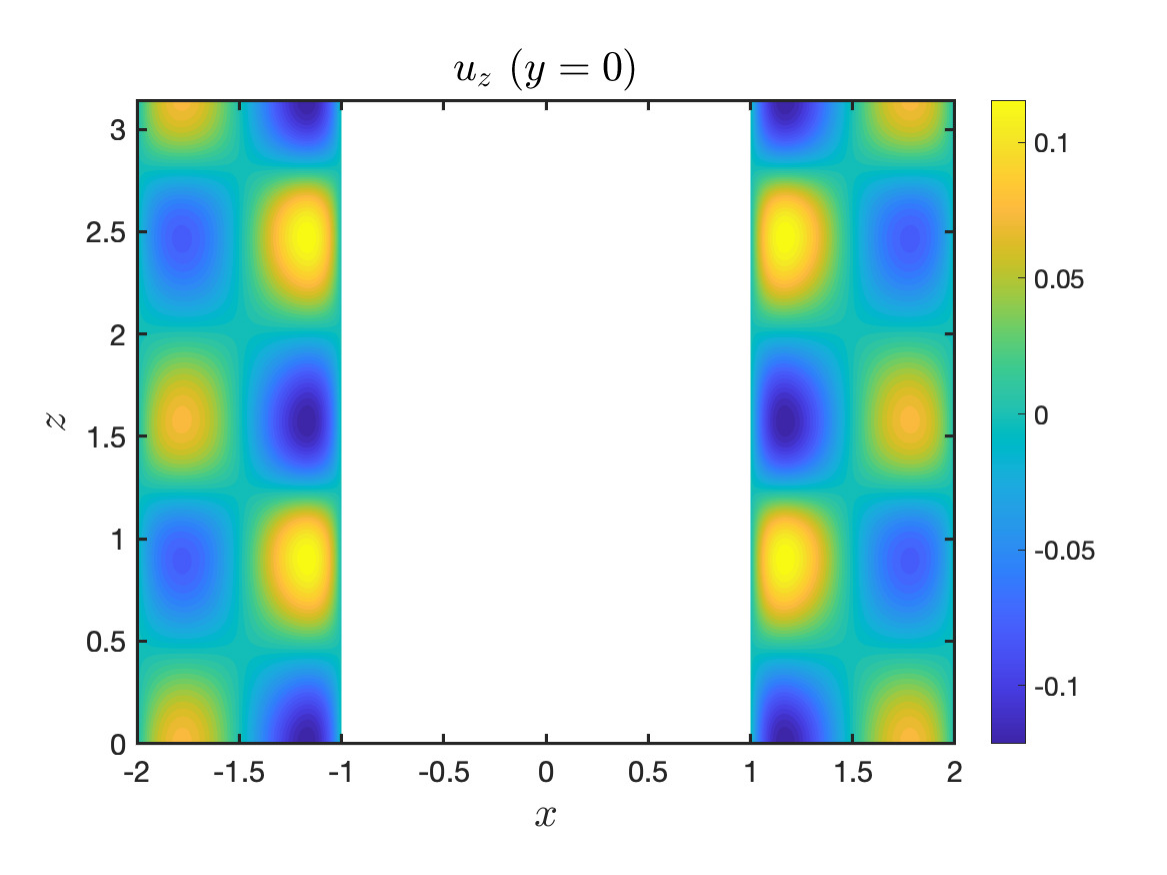}
\caption{Contour plots of the velocity field of Taylor-Couette flow at the cross section $y=0$.}
\label{fig:TR-uxyz}
\end{figure*}

We compute this steady flow field with periodic boundary conditions for the axial direction, as it is often done in the literature \cite{Liang2017}. This allows for a pseudo-spectral discretization via a Fourier decomposition in the $z$-direction and a Chebyshev decomposition in the $r$ direction. We use the open-source package Dedalus \cite{dedalus} to solve the discretized initial value problem at Re = 100. Contours of the three components of the steady flow that develops can be seen in Fig. \ref{fig:TR-uxyz}.

For the calculation of our approximate first integral, we restrict the domain to $z\in[0.5,2]$ because of the periodic pattern along the $z$ direction, as seen in Fig.~\ref{fig:TR-uxyz}. We use a tetrahedron mesh to discretize the hollow cylinder, as shown in the right panel of Fig.~\ref{fig:mesh-axis}. We again use quadratic Lagrange elements to interpolate the unknown function $H$. From the right panel of Fig.~\ref{fig:lamd2-axis}, we observe the monotonic decay of $\lambda_2$ with increasing number of elements. Such a decay indicates that the numerical solutions converge to the first integral we seek.

\begin{figure*}[!ht]
\centering
\includegraphics[width=0.45\textwidth]{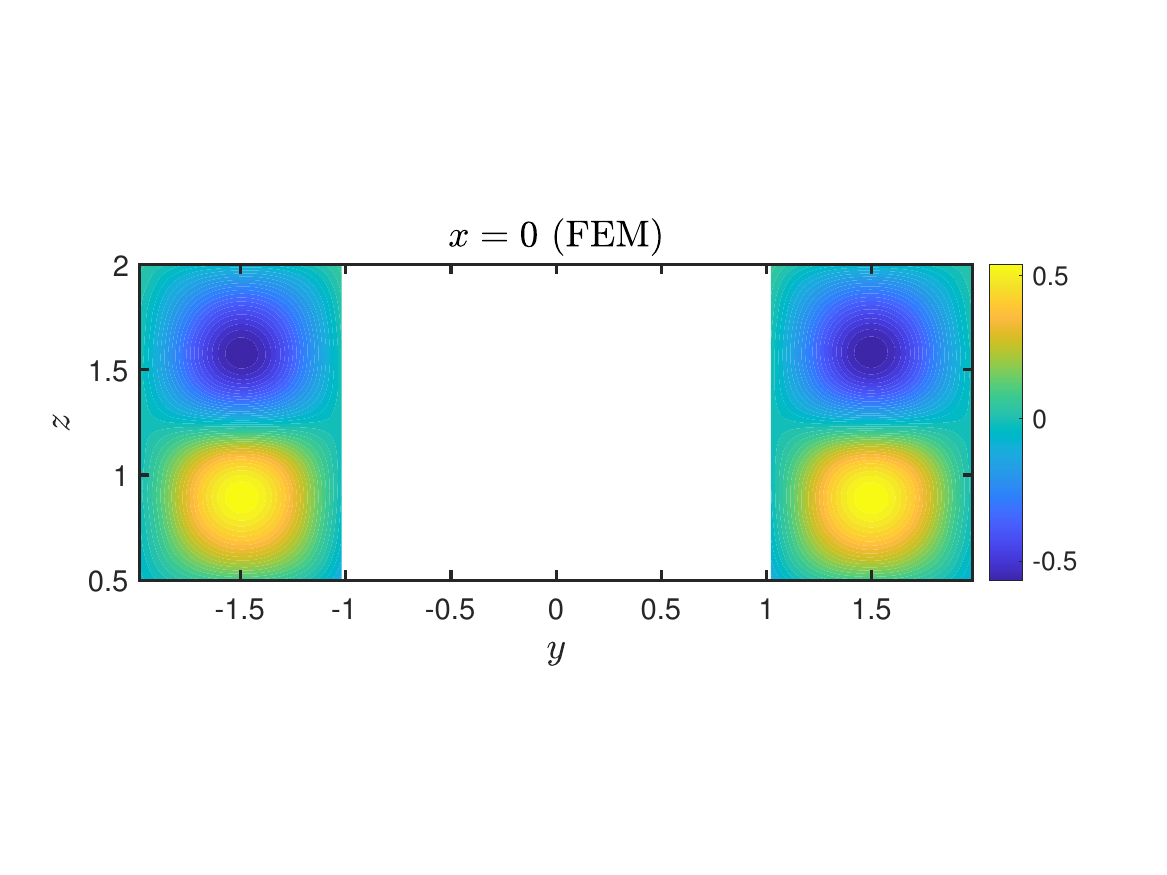}
\includegraphics[width=0.45\textwidth]{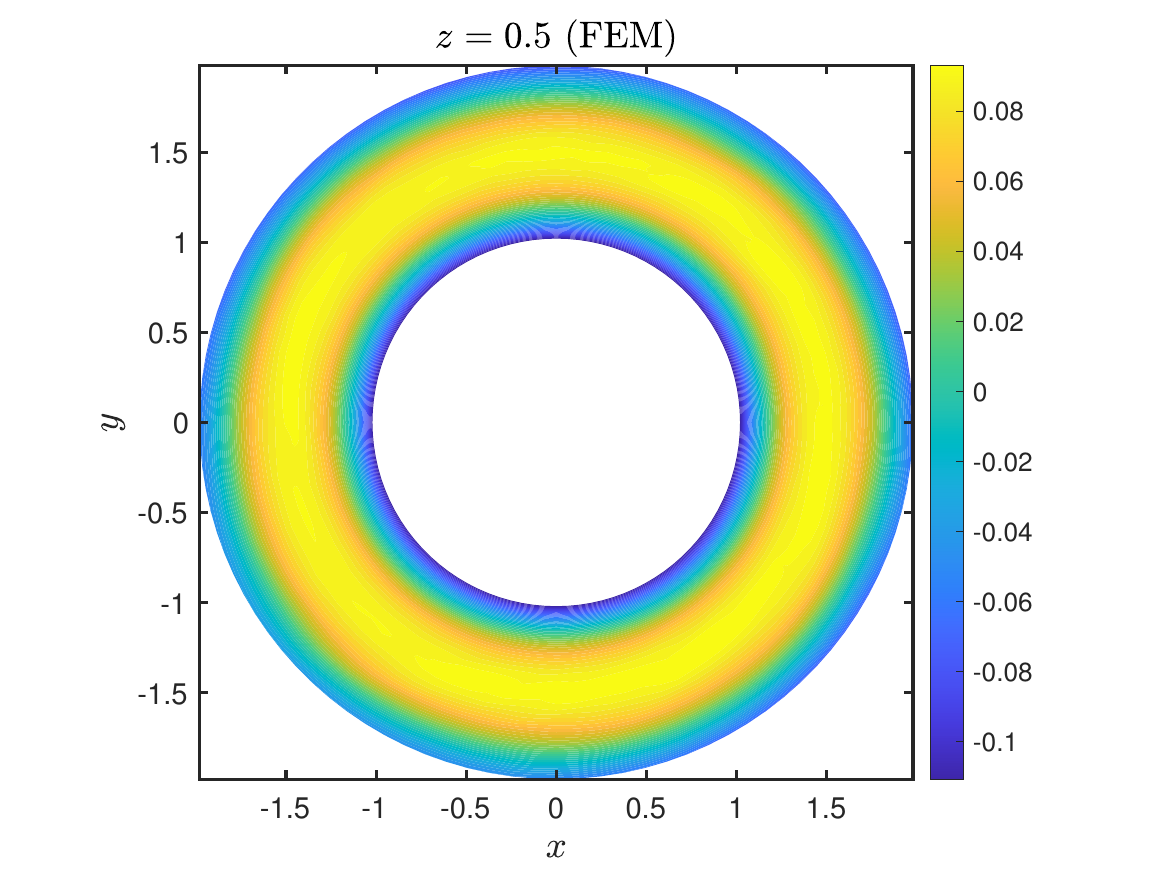}
\caption{Contour plots of ${H}_2$ of Taylor-Couette flow obtained from finite-element methods with 52,495 elements at cross section $x=0$ (left panel) and $z=0.5$ (right panel).}
\label{fig:TC-2d}
\end{figure*}

As we do not have analytical expressions for the Stokes stream function here, we simply plot $H_2$ instead of its linear transformation $\hat{H}_2$. The contour plots of the cross sections of $H_2$ obtained with 52,495 elements are shown in Fig.~\ref{fig:TC-2d}.

\begin{figure*}[!ht]
\centering
\includegraphics[width=0.45\textwidth]{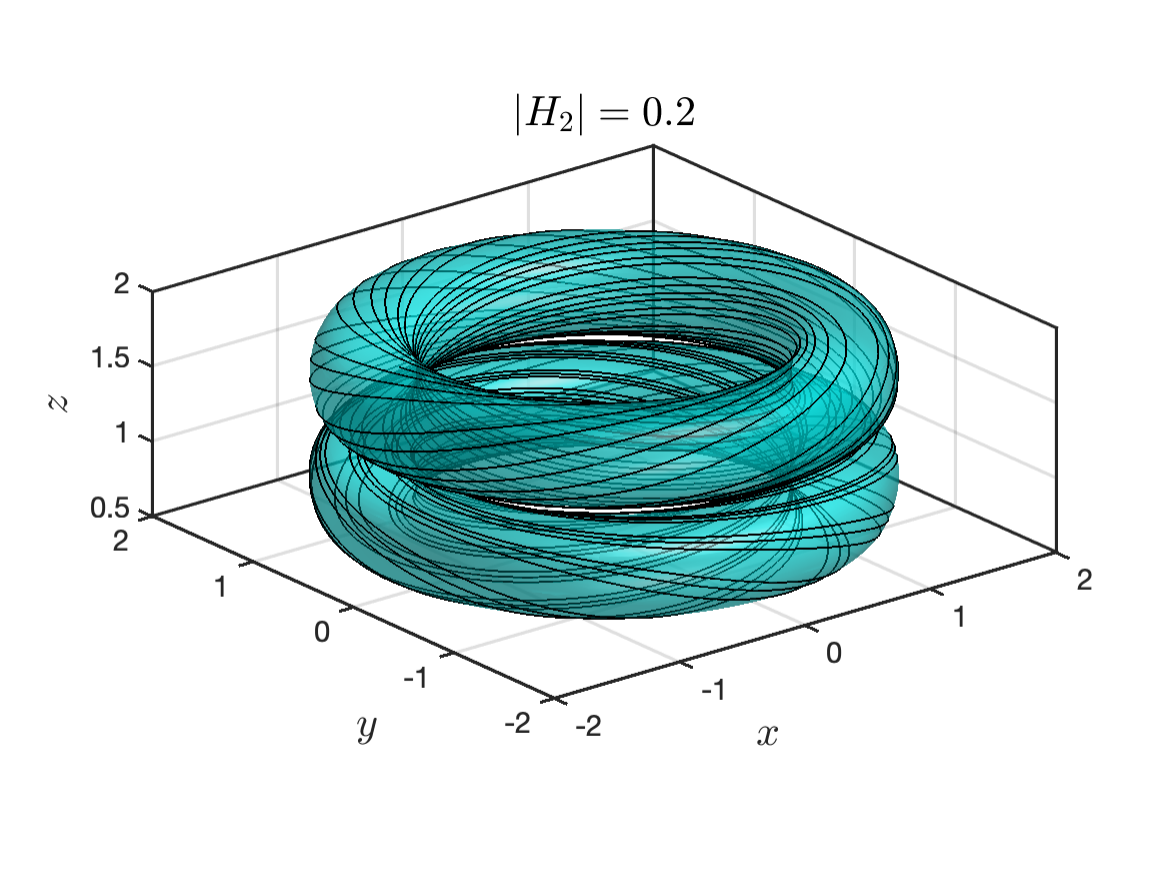}
\includegraphics[width=0.45\textwidth]{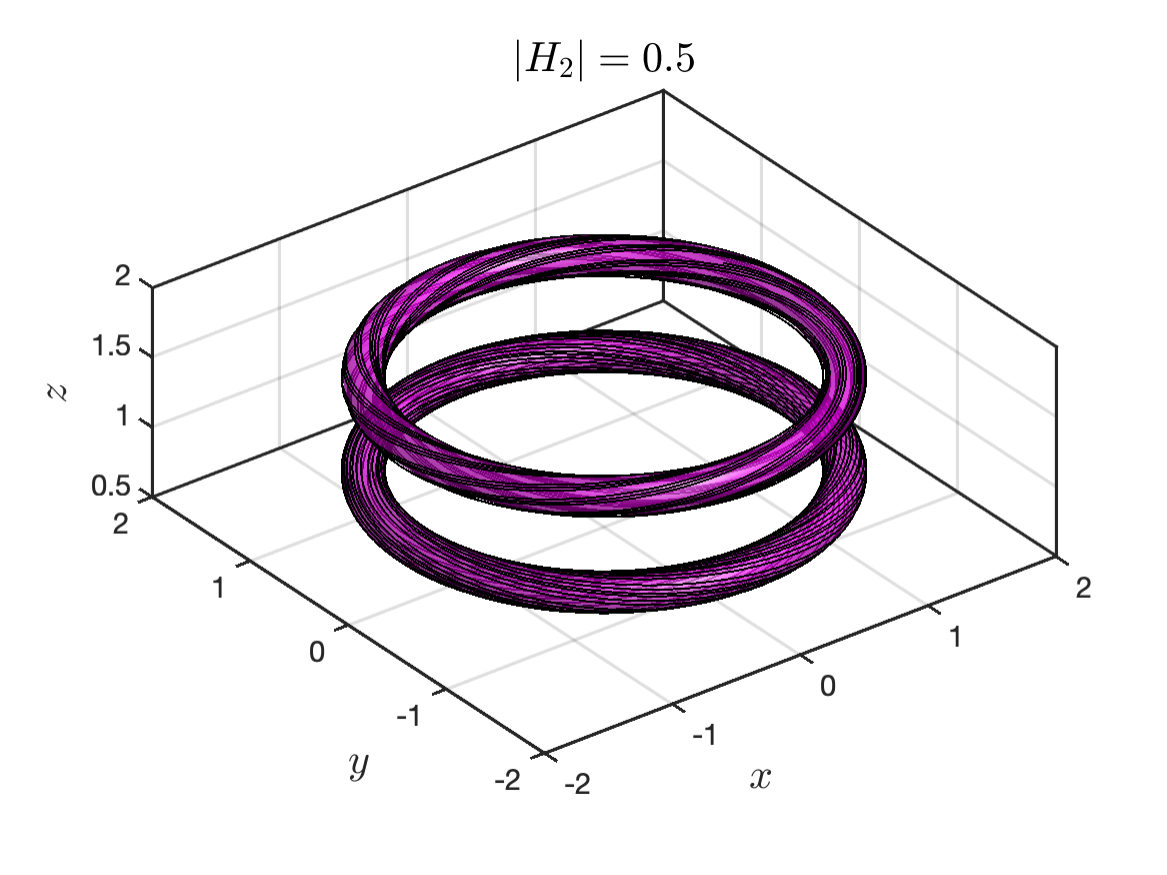}
\caption{Contour plots of isosurfaces for $|{H}_2|$ of the Taylor-Couette flow obtained from finite-element methods with 52495 elements. Here we have $|{H}_2|=0.2$ (left panel) and $|{H}_2|=0.5$ (right panel). The black lines are streamlines from forward simulations with initial points on the isosurfaces.}
\label{fig:TC-3d}
\end{figure*}

We infer from Fig.~\ref{fig:TC-2d} that there are two families of vortex rings in the restricted domain, one with $H_2>0$ and one with $H_2<0$. This is consistent with the observation of Taylor-vortices in experimental \cite{gollub1975} and numerical studies \cite{barenghi1991}. We plot the isosurfaces for $|H_2|=0.2$ and $|H_2|=0.5$. For each case, we indeed have two vortex rings, as seen in Fig.~\ref{fig:TC-3d}. We launch a few streamlines with initial conditions on these surfaces. These streamlines stay close to the corresponding isosurfaces, as illustrated in Fig.~\ref{fig:TC-3d}.

\section{Periodic flows}
\label{sec:periodicflows}
In this section, we consider periodic flows in the domain $\Omega=[0,2\pi]\times[0,2\pi]\times[0,2\pi]$. Specifically, we will consider both the ABC (Arnold-Beltrami-Childress)~\cite{dombre1986chaotic} flow and the  Euler flow~\cite{antuono2020tri}. The velocity field of these flows is periodic in all three directions. We note that periodic flows have been treated in~\cite{stergios}, where the approximate first integral is represented by Fourier series. In the previous section, we have demonstrated the power of our finite-element computations for flows in spherical and cylindrical domains which cannot be treated via the Fourier representation. Here we illustrate that the finite-element approach can also be applied to periodic flows. Thus our finite-element implementation provides a unified treatment for both periodic and aperiodic flows.~\textcolor{black}{In addition, thanks to the sparsity of finite-element methods, our finite-element implementation outperforms the Fourier series schemes in~\cite{stergios}, as we illustrate in Appendix (see Sect.~\ref{sec:comp-fea-fourier}).}

For the periodic flows above, the trial function space (see~\eqref{eq:Htrial}) is
\begin{equation}
\label{eq:Htrial-po}
    \mathcal{H}_\mathrm{trial}=\{H\in H^1(\Omega),\,\, H|_{x=0}=H|_{x=2\pi},\,H|_{y=0}=H|_{y=2\pi},\,H|_{z=0}=H|_{z=2\pi}\},
\end{equation}
where $H|_{x_i=a}$ denotes the evaluation of $H$ on the plane $x_i=a$. It follows that~\eqref{eq:left-weak} still holds because both $H,h$ and $\mathbf{u}$ are periodic and 
\begin{equation}
     \mathbf{n}|_{x=0}=-\mathbf{n}|_{x=2\pi},\quad \mathbf{n}|_{y=0}=-\mathbf{n}|_{y=2\pi},\quad \mathbf{n}|_{z=0}=-\mathbf{n}|_{z=2\pi}.
\end{equation}
Indeed, the integral over the boundary of $\Omega$ vanishes (see~\eqref{eq:left-weak}) because of the opposite orientation of the normal vectors on opposite faces of the cube. Therefore, the weak form~\eqref{eq:weak-form} still holds and the discussions in Sect.~\ref{sec:eig-to-min} are still true.

Here we use \texttt{BoxMesh} in FEniCS to generate a mesh for $\Omega$. Given the number of cells $(N_x,N_y,N_z)$ in each direction, the total number of tetrahedrons is $6N_xN_yN_z$ and the total number of vertices is $(N_x+1)(N_y+1)(N_z+1)$. In the following computations, we simply set $N_x=N_y=N_z=N$ for the cubic domain. Since $\Omega_\mathrm{H}=\emptyset$, we again have $\lambda_1=0$ with constant eignvector, so we look for $(\lambda_2,H_2)$.

\begin{figure*}[!ht]
\centering
\includegraphics[width=0.45\textwidth]{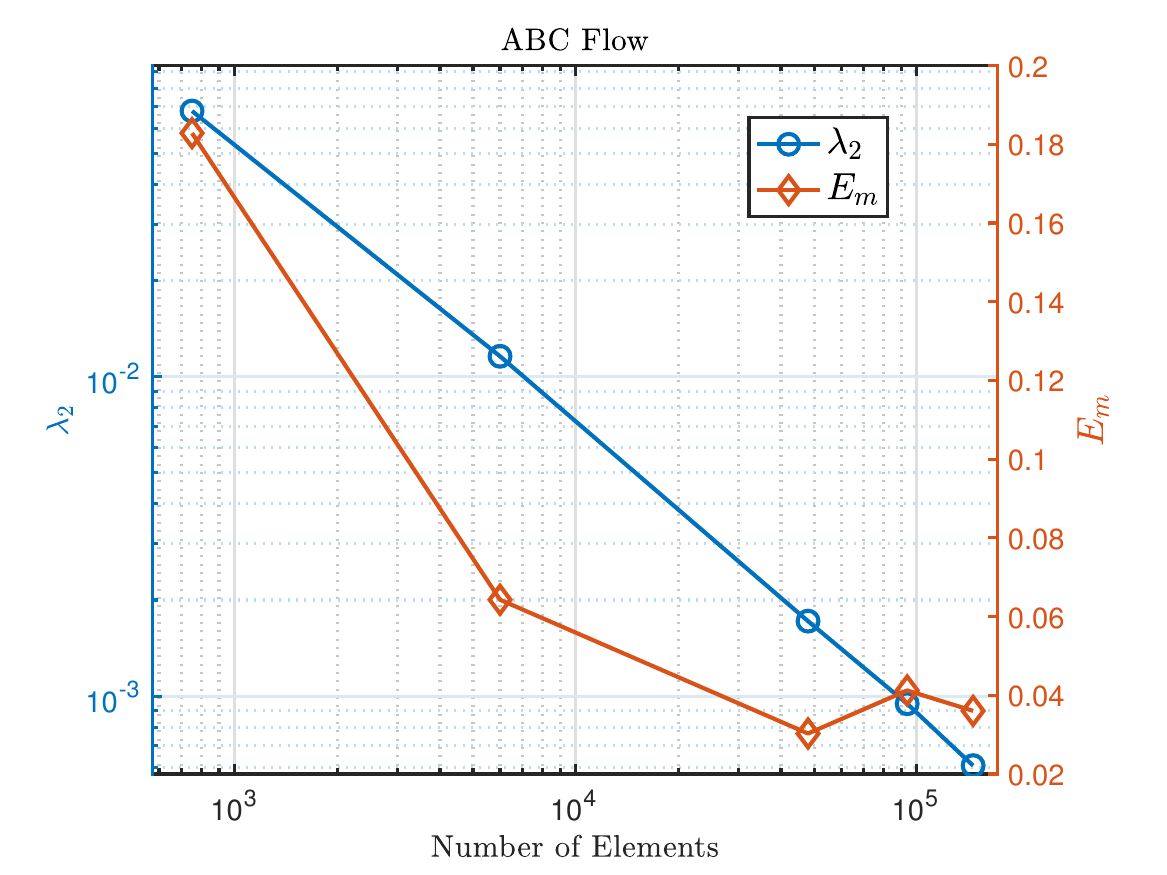}
\includegraphics[width=0.45\textwidth]{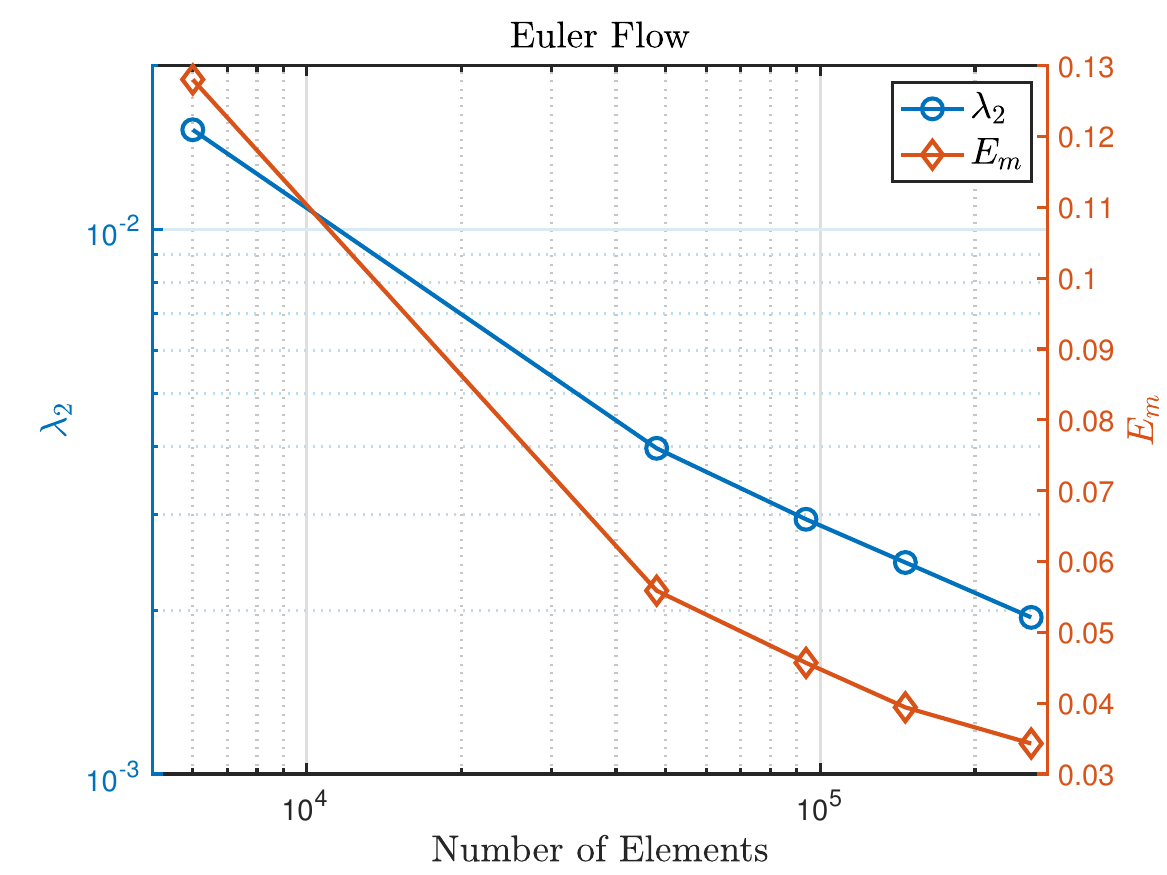}
\caption{Second smallest eigenvalues and mean invariance errors as a function of the number of elements to discretize the periodic flows: ABC flow (left) and Euler flow (right).}
\label{fig:lamd2-em-po}
\end{figure*}

\subsection{ABC flow}
Consider the classic ABC flow
\begin{equation}
    u_x = A\sin z+C\cos y,\quad
    u_y = B\sin x+A\cos z,\quad
    u_z = C\sin y+B\cos x.
\end{equation}
We choose $A=\sqrt{3}$, $B=\sqrt{2}$ and $C=1$, for which the ABC flow is known to be non-integrable, i. e. there is no nontrivial exact first integral for this flow in that case~\cite{dombre1986chaotic,stergios}.

\begin{figure*}[!ht]
\centering
\includegraphics[width=0.32\textwidth]{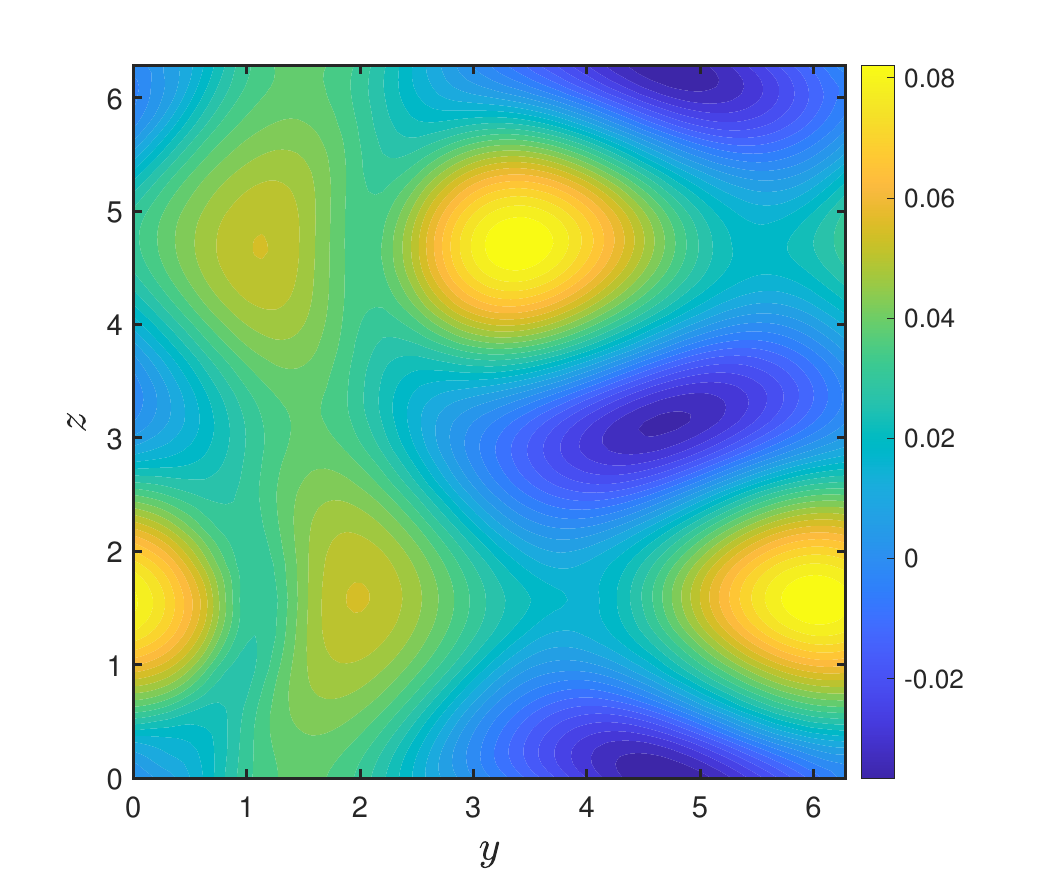}
\includegraphics[width=0.32\textwidth]{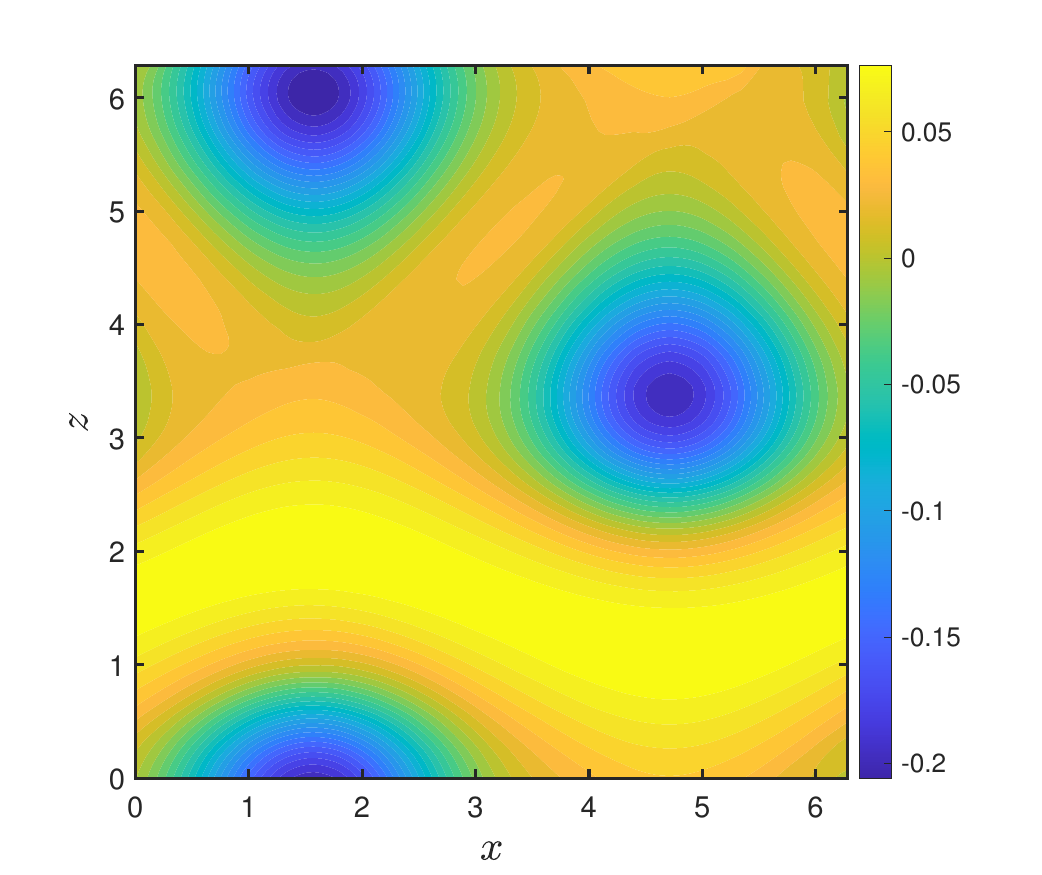}
\includegraphics[width=0.32\textwidth]{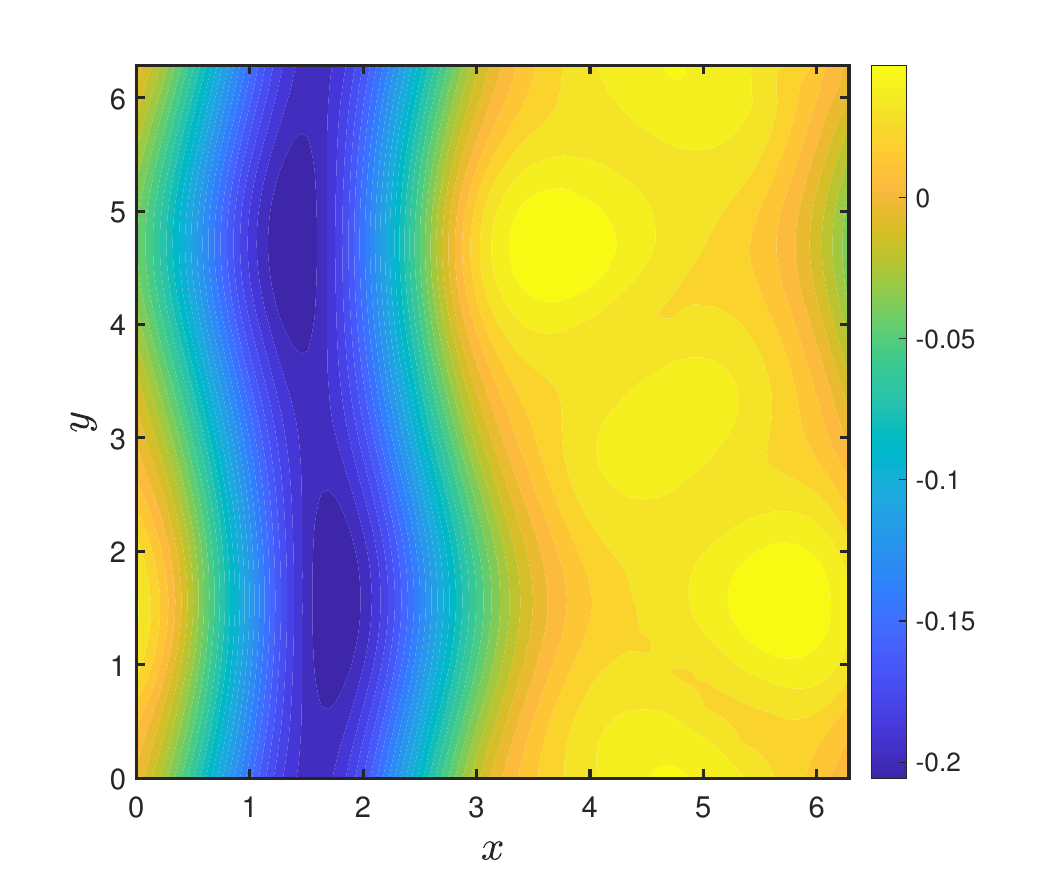}\\
\includegraphics[width=0.32\textwidth]{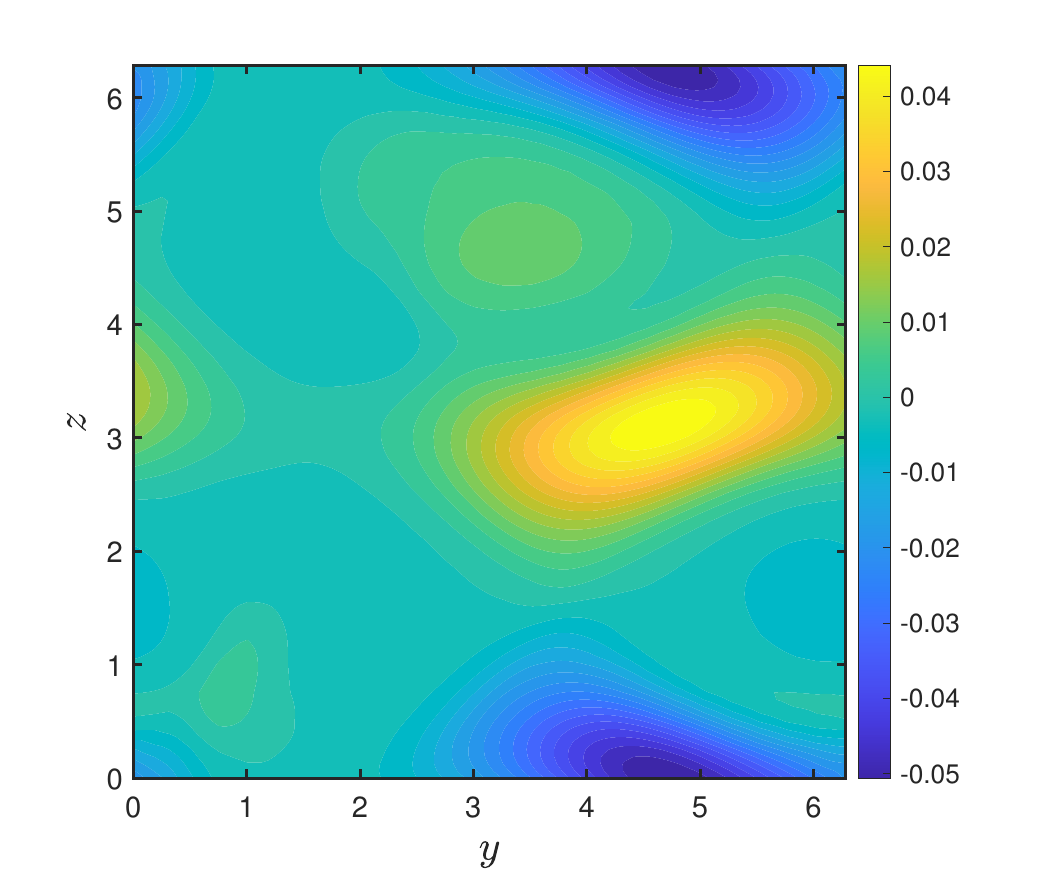}
\includegraphics[width=0.32\textwidth]{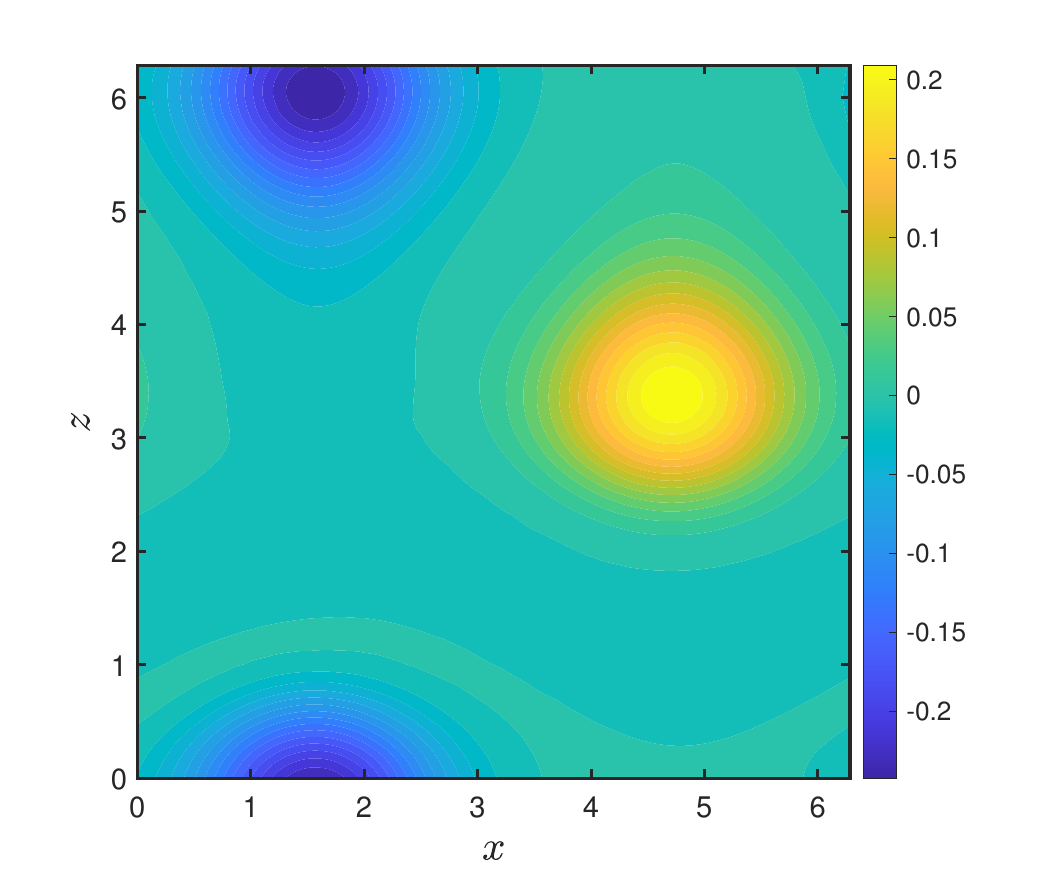}
\includegraphics[width=0.32\textwidth]{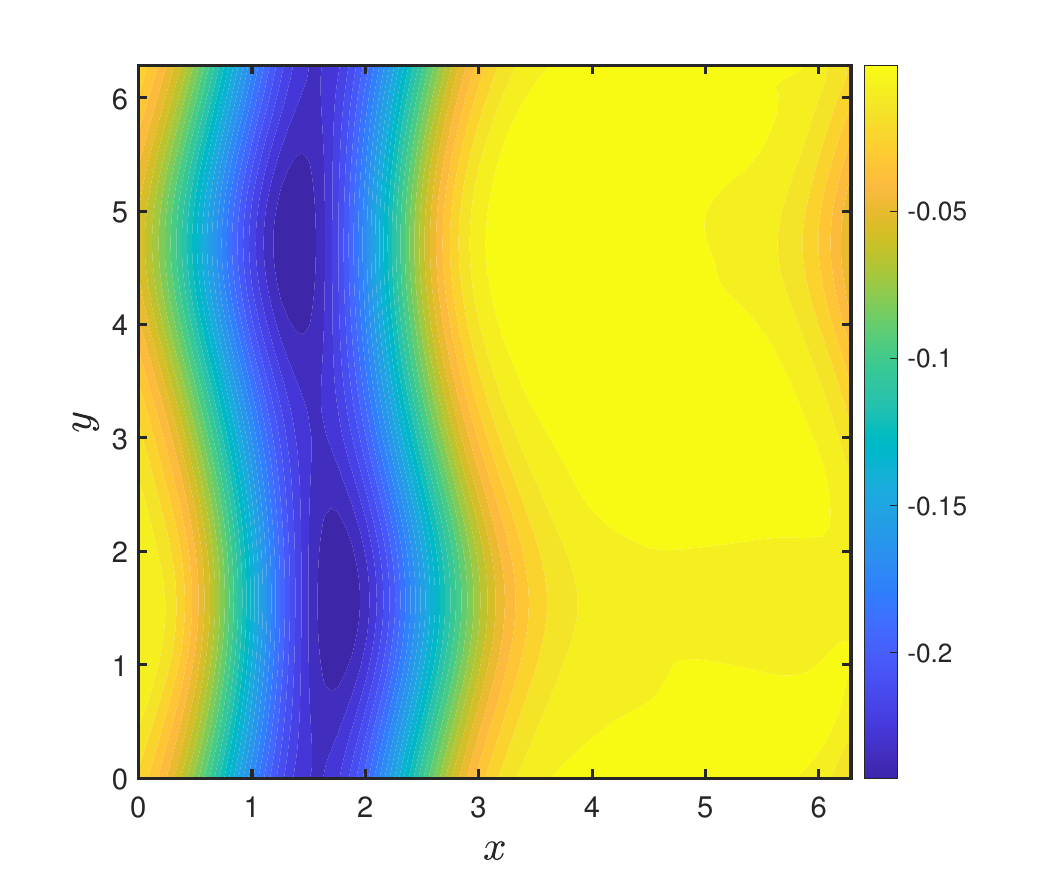}
\caption{Contour plots of $H_2$ of the ABC flow at cross sections $x=0$ (left column), $y=0$ (middle column) and $z=0$ (right column). Here the upper and lower rows correspond to the results with $N=20$ (48,000 elements) and $N=25$ (93,750 elements).}
\label{fig:abc-2d}
\end{figure*}

We take $N=\{5,10,20,25,29\}$ and perform the computations with refined meshes. In the left panel of Fig.~\ref{fig:lamd2-em-po}, we observe the monotonic decay of $\lambda_2$ as a power-law with respect to the numbers of elements. While we have observed the same decay in the previous results for the generalized axisymmetric flows, the ABC flow does not admit an exact first integral. To gain a better understanding of the behavior of $\lambda_2$, we plot the contours of $H_2$ for $N=20$ and $N=25$ (see the third and fourth points in the left panel of Fig.~\ref{fig:lamd2-em-po}) at cross sections $x=0$, $y=0$ and $z=0$ in Fig.~\ref{fig:abc-2d}. By comparing subplots in the upper and lower panels, we find that both primary and secondary vortical regions are captured with $N=20$. In contrast, when we increase $N$ to 25, the secondary vortex structures disappear and the variation of $H_2$ is aggregated around the primary vortex regions. In other words, $H_2$ barely changes outside the primary vortical regions (see the lower panels in Fig.~\ref{fig:abc-2d}). So we have $\nabla H\approx0$ outside the primary vortical regions while $H$ converges to a first integral inside the primary vortical regions. This explains the monotonic decay of $\lambda_2$ as $N$ increases.

As a brief summary, if we only want to extract primary vortex structures, a higher fidelity discretization is helpful. On the other hand, if we also want to extract secondary vortex structures where $\nabla H$ is of small magnitude, we should instead use a relative invariance measure $\sqrt{\int_\Omega \frac{(\nabla H\cdot\mathbf{u})^2}{|\nabla H|^2|\mathbf{u}|^2} dV}$.
As a variant of the above measure, we consider the mean invariance of the entire solution as an error measure~\cite{stergios} by defining
\begin{equation}
    \label{eq:Em}
    E_m = \frac{1}{m}\sum_{i=1}^{m} \frac{|\nabla H_i\cdot \mathbf{u}_i|}{|\nabla H_i|\cdot|\mathbf{u}_i|},
\end{equation}
where the summation takes place over all grid points. Here and in the example below, we take 101 grid points in each direction so that $m=101^3$. We also plot $E_m$ as a function of the number of elements in the left panel of Fig.~\ref{fig:lamd2-em-po}, from which we see that $E_m$ for $N=20$ is the smallest among all the five cases. Therefore, one should use $E_m$ as an error measure to choose the proper discretization in order to obtain both primary and secondary vortex structures.

Motivated by \eqref{eq:Em}, we introduce a filter to efficiently extract approximate streamsurfaces in vortical regions. In particular, we extract some level surfaces of $H$ to represent the approximate streamsurfaces. To identify whether a level surface of $H$ is an approximate streamsurface in vortical regions, we introduce the surface-averaged invariance error~\cite{stergios}
\begin{equation}
\label{eq:EA}
    E_A =  \frac{1}{p}\sum_{i=1}^p \frac{|\nabla H_i\cdot \mathbf{u}_i|}{|\nabla H_i|\cdot|\mathbf{u}_i|},
\end{equation}
where $p$ is the number of points on the surface of the level set. These points are determined by surface meshing algorithms embedded in commonly used routines, e.g., \texttt{isosurface} in \textsc{matlab} and \textsc{python}.

\begin{figure*}[!ht]
\centering
\includegraphics[width=0.32\textwidth]{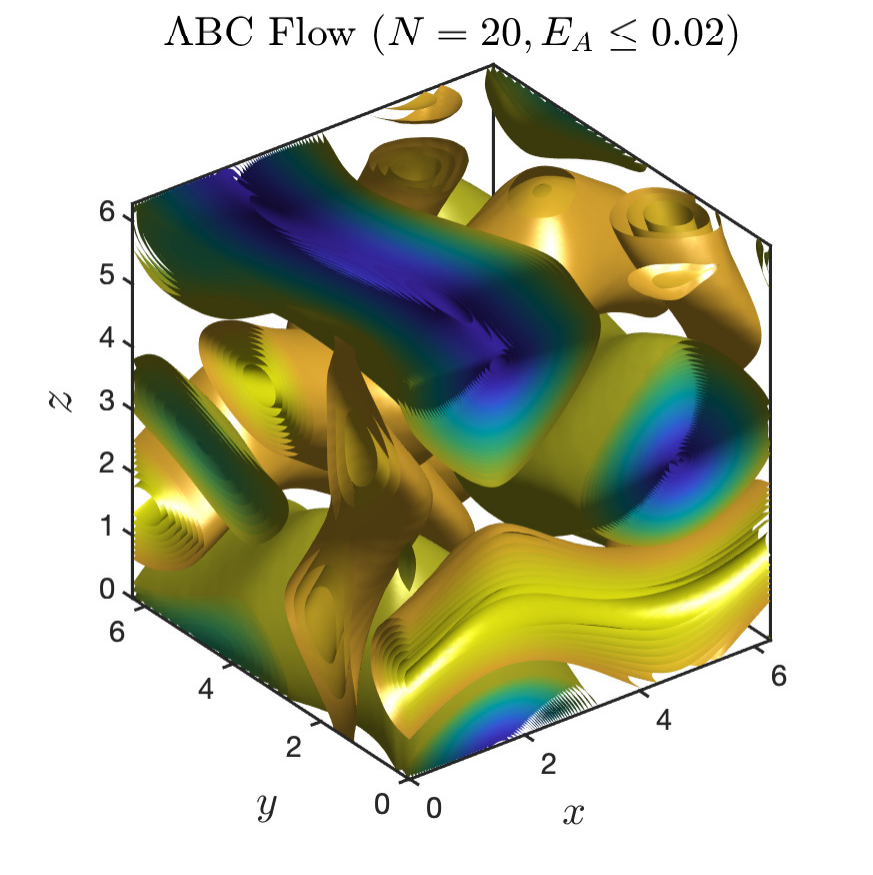}
\includegraphics[width=0.32\textwidth]{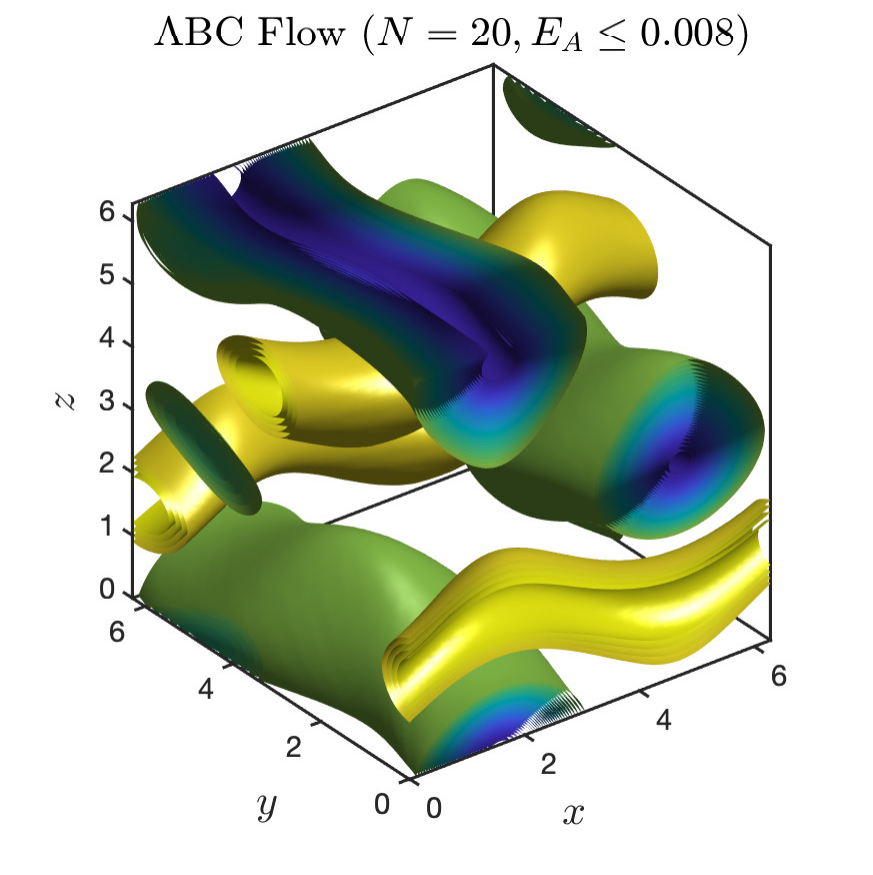}
\includegraphics[width=0.32\textwidth]{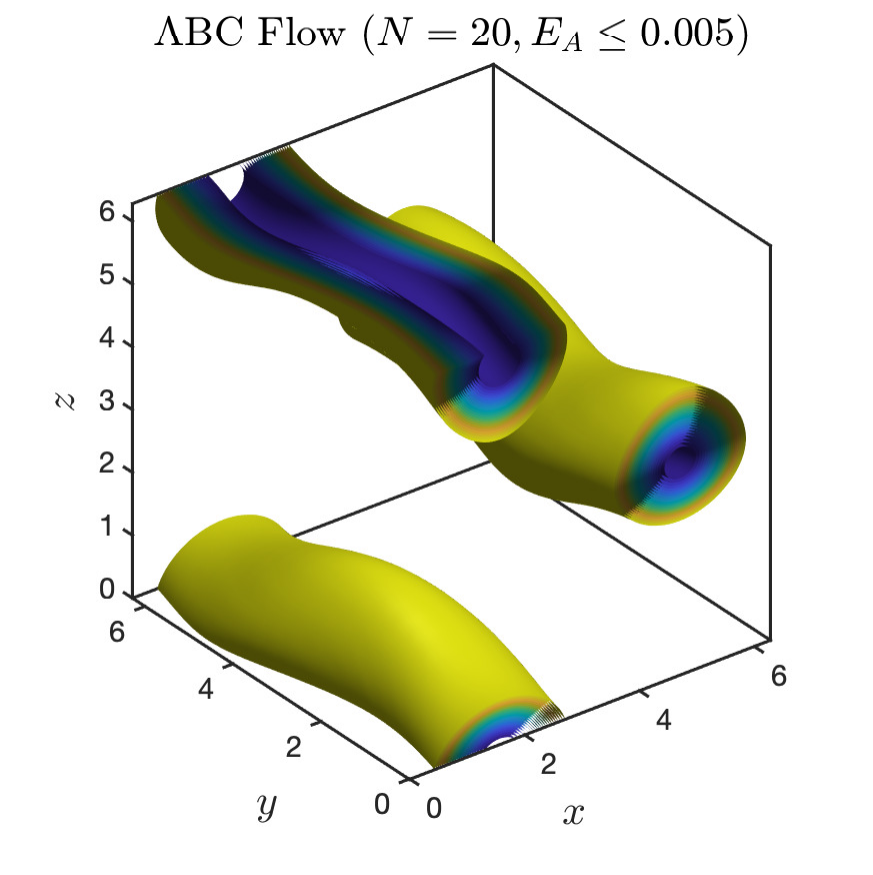}\\
\includegraphics[width=0.32\textwidth]{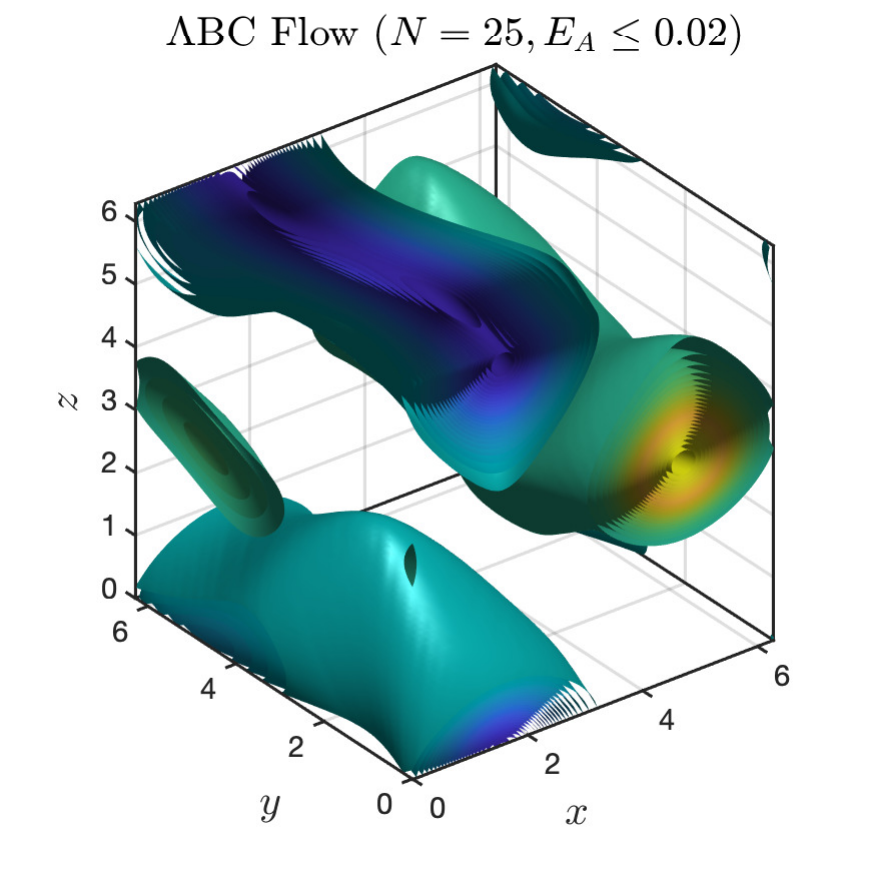}
\includegraphics[width=0.32\textwidth]{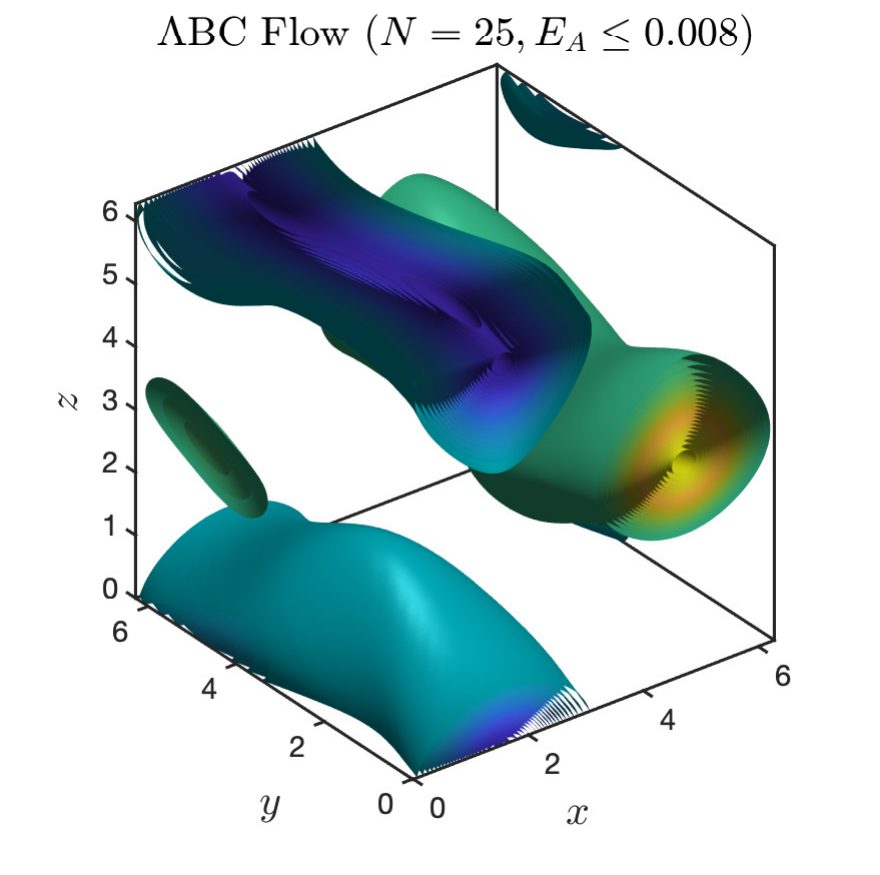}
\includegraphics[width=0.32\textwidth]{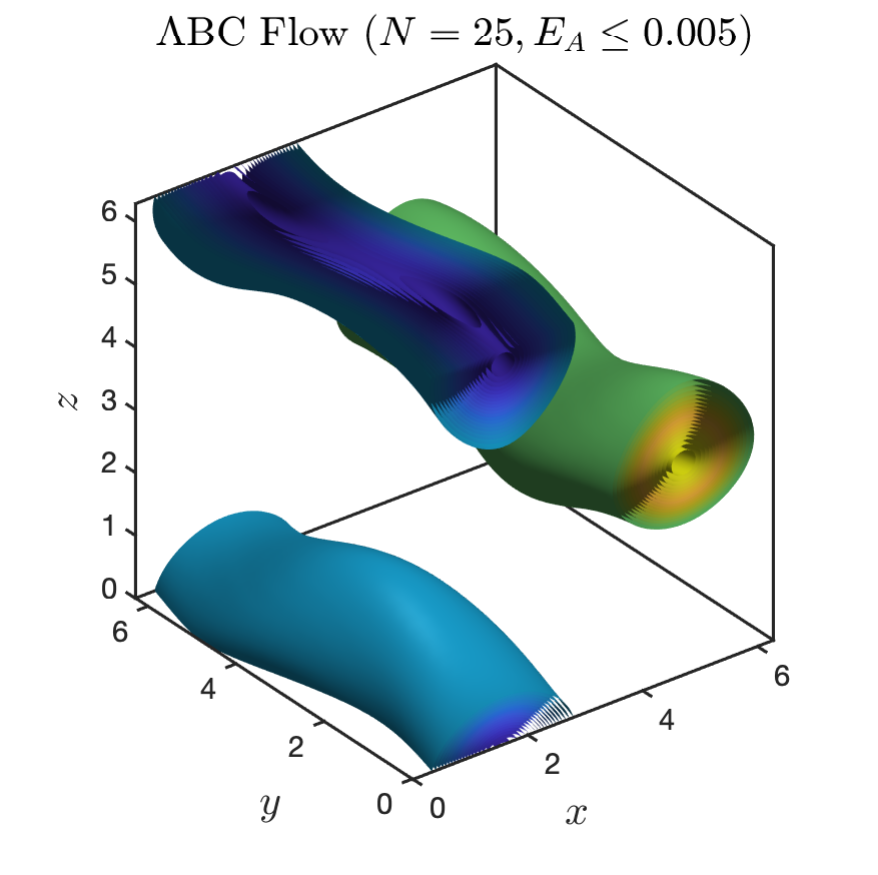}
\caption{Isosurfaces of $H_2$ for the ABC flow with various filter thresholds: $E_A\leq0.02$ (left column), $E_A\leq0.008$ (middle column), and $E_A\leq0.005$ (right column). Here the upper and lower rows correspond to the results with $N=20$ (48,000 elements) and $N=25$ (93,750 elements).}
\label{fig:abc-3d}
\end{figure*}

The isosurfaces of $H_2$ with various thresholds for $E_A$ and different discretizations are shown in Fig.~\ref{fig:abc-3d}. By comparing the upper panels and corresponding lower panels (especially the first two columns), we see that the results for $N=20$ extract both primary and secondary vortical regions while that for $N=25$ only extract the primary vortical regions. This observation is consistent with the one we made from Fig.~\ref{fig:abc-2d}. From the upper panels, we also see that secondary vortical regions are filtered out when we decrease the threshold for $E_A$. This indicates that one can use a lower threshold to extract primary vortical regions that are robust with respect to the change of mesh fidelities.

\begin{figure*}[!ht]
\centering
\includegraphics[width=0.45\textwidth]{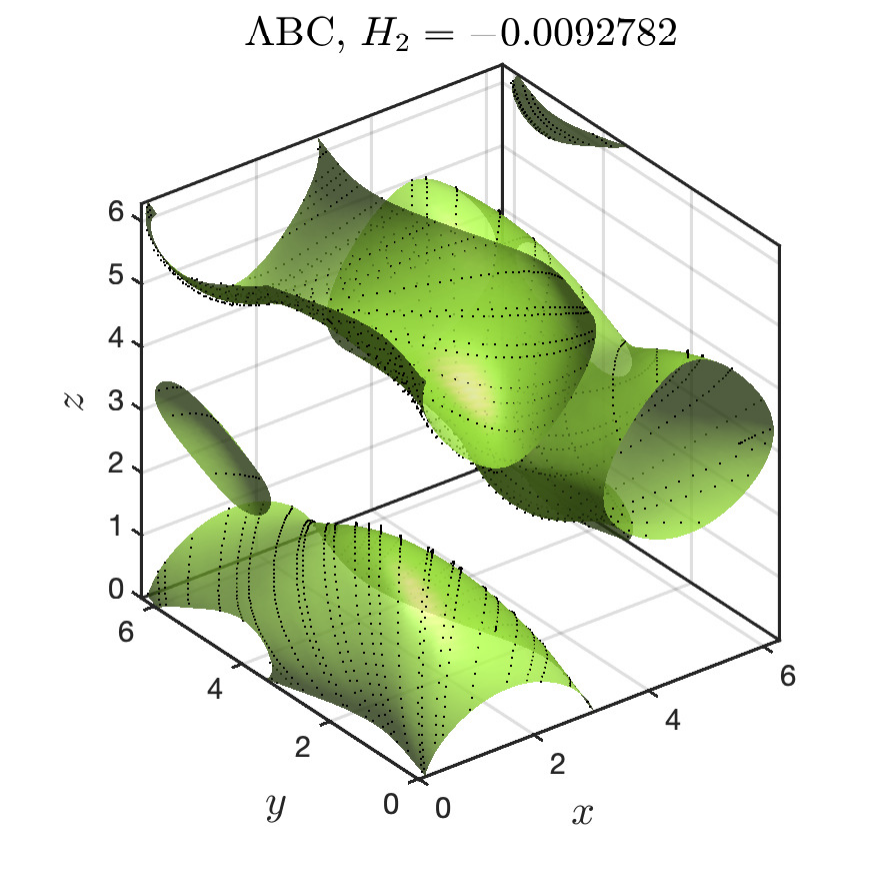}
\includegraphics[width=0.45\textwidth]{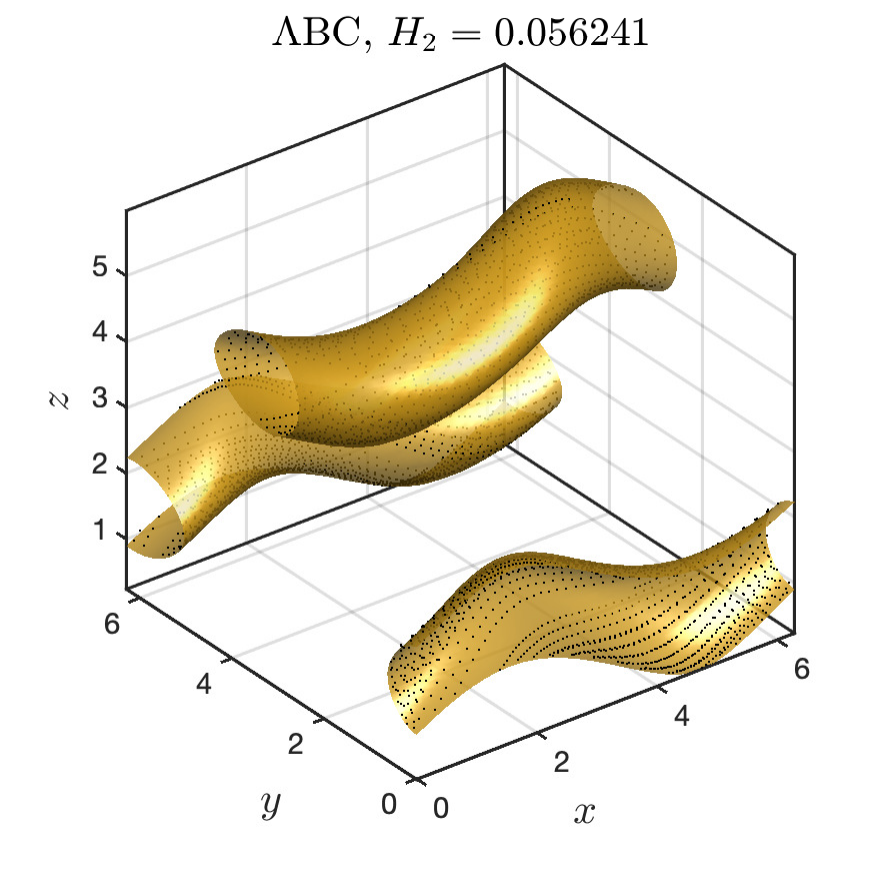}
\caption{Contour plots of isosurfaces for ${H}_2$ of the ABC flow obtained from finite-element methods with $N=20$ (48,000 elements). The isosurfaces in the left and right panels correspond to the outermost layer of the primary and secondary vortical regions of the upper-middle panel of Fig.~\ref{fig:abc-3d}. Here the black dots denote simulated trajectories.}
\label{fig:abc-3d-streamline}
\end{figure*}

We conclude this example by validating some of the approximate streamsurfaces we have obtained. In Fig.~\ref{fig:abc-3d}, we see that there are primary and secondary vortex regions. We take the outermost layers of these two regions to perform the validation. As seen in Fig.~\ref{fig:abc-3d-streamline}, the streamlines obtained from forward simulation stay close to the approximate streamsurfaces. The little patches in the left panel of Fig.~\ref{fig:abc-3d-streamline} are results of the periodic boundary conditions.

\subsection{Euler flow}
Consider the Euler flow
\begin{gather}
    u_x = \tfrac{4\sqrt{2}}{3\sqrt{3}}\left(\sin\left(x-{\tfrac{5\pi}{6}} \right)\cos\left(y-\tfrac{\pi}{6}\right)\sin\left(z\right)-\cos\left(z-{\tfrac{5\pi}{6}} \right)\sin\left(x-\tfrac{\pi}{6}\right)\sin\left(y\right)\right),\nonumber\\
    u_y = \tfrac{4\sqrt{2}}{3\sqrt{3}}\left(\sin\left(y-{\tfrac{5\pi}{6}} \right)\cos\left(z-\tfrac{\pi}{6}\right)\sin\left(x\right)-\cos\left(x-{\tfrac{5\pi}{6}} \right)\sin\left(y-\tfrac{\pi}{6}\right)\sin\left(z\right)\right),\nonumber\\
    u_z = \tfrac{4\sqrt{2}}{3\sqrt{3}}\left(\sin\left(z-{\tfrac{5\pi}{6}} \right)\cos\left(x-\tfrac{\pi}{6}\right)\sin\left(y\right)-\cos\left(y-{\tfrac{5\pi}{6}} \right)\sin\left(z-\tfrac{\pi}{6}\right)\sin\left(x\right)\right),
\end{gather}
which is also non-integrable~\cite{antuono2020tri,stergios}. We take $N=\{10,20,25,29,35\}$ and perform the computations with refined meshes. Within each cell, we use quadratic Lagrange polynomials to approximate $H$. In the right panel of Fig.~\ref{fig:lamd2-em-po}, we observe the monotonic decay of $\lambda_2$ as well as the mean invariance error $E_m$ with increasing numbers of elements. This decay indicates that more accurate results are obtained with increasing $N$.

\begin{figure*}[!ht]
\centering
\includegraphics[width=0.45\textwidth]{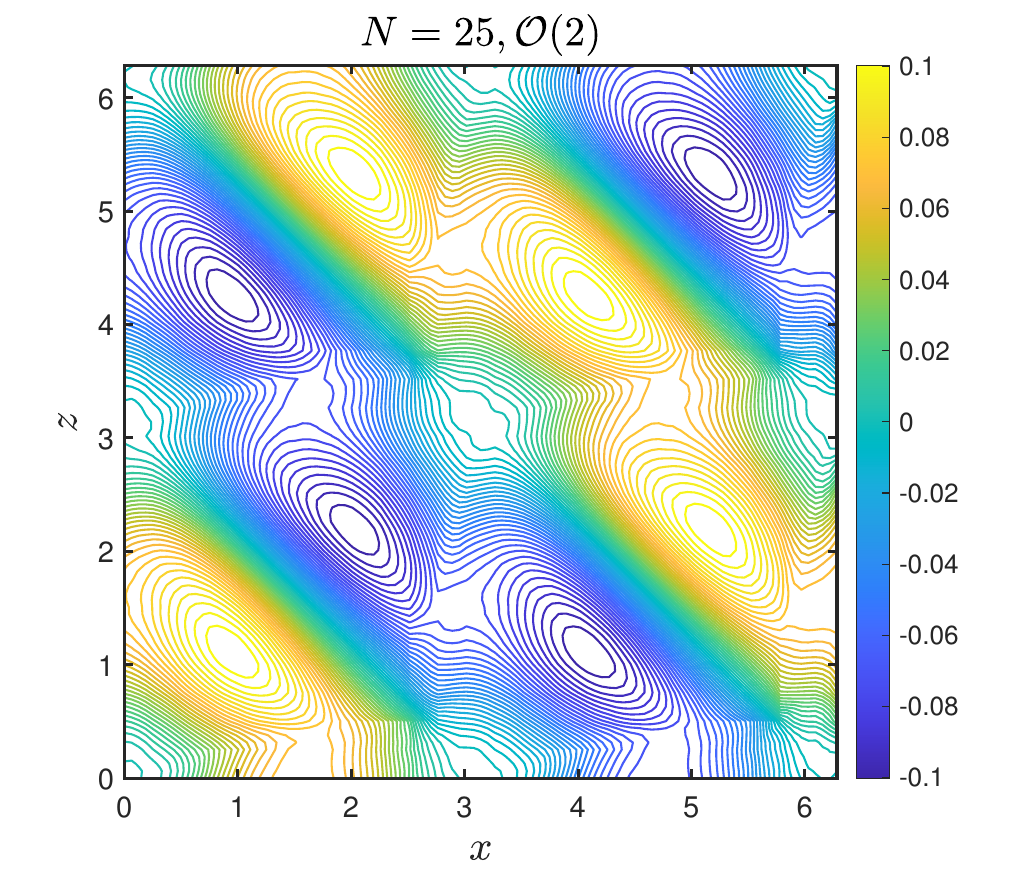}
\includegraphics[width=0.45\textwidth]{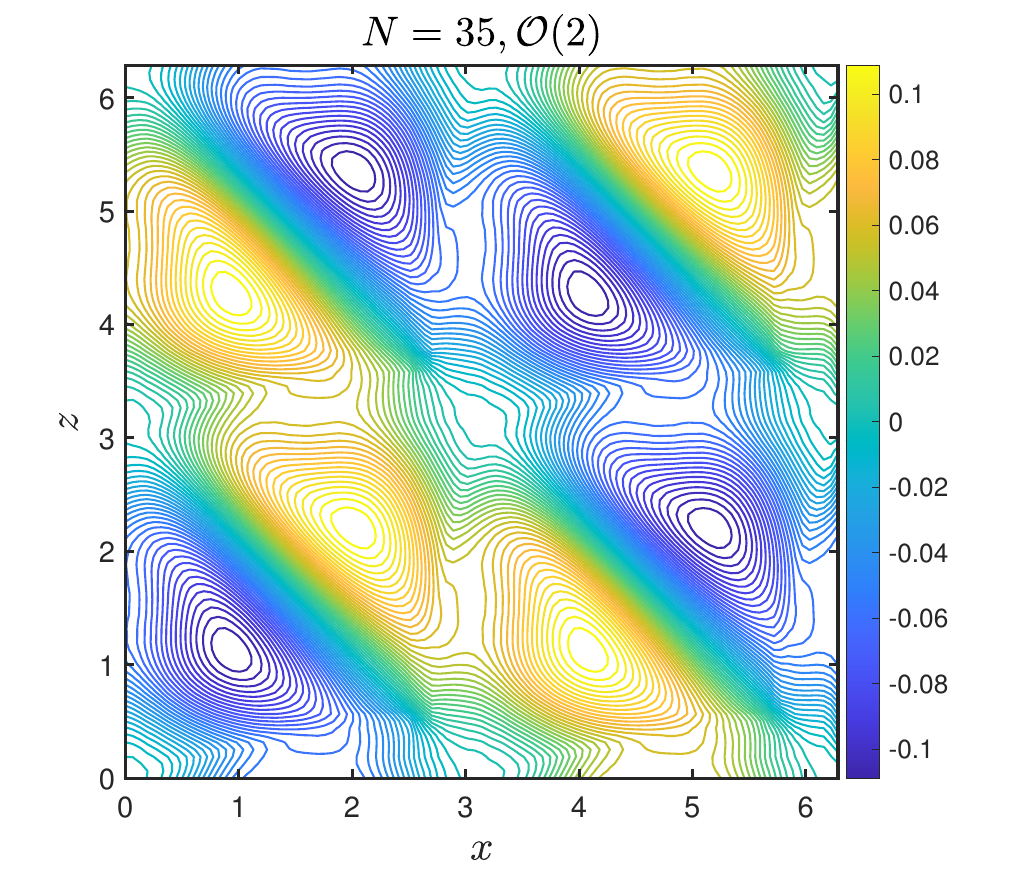}\\
\includegraphics[width=0.45\textwidth]{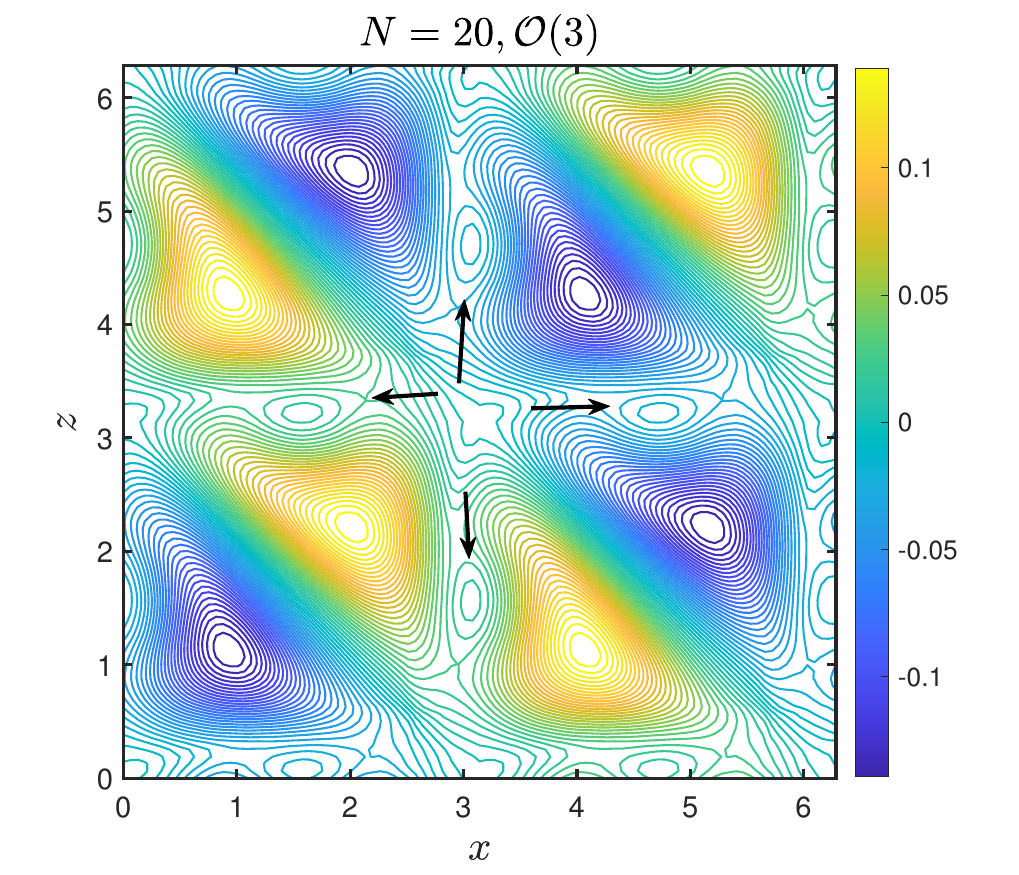}
\includegraphics[width=0.45\textwidth]{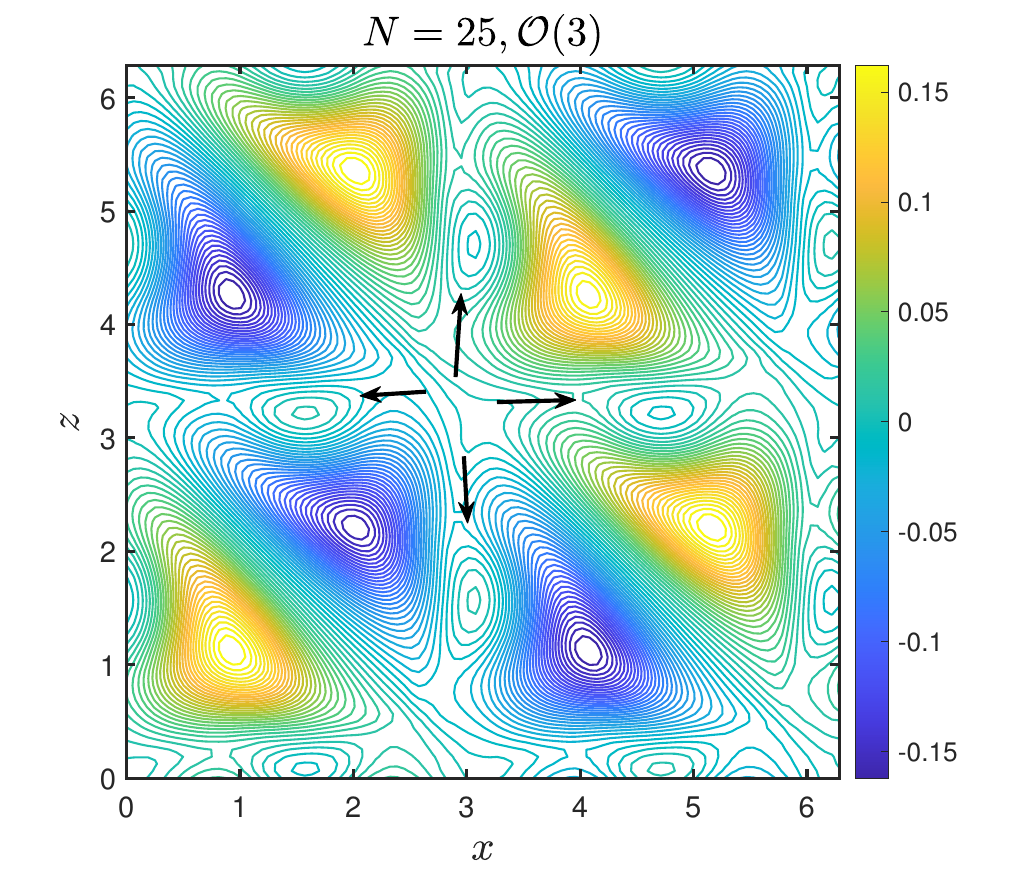}
\caption{Contour plots of $H_2$ of Euler flow at cross section $y=0$. Here $H_2$ is obtained with various discretizations. Here $\mathcal{O}(2)$ and $\mathcal{O}(3)$ represent quadratic and cubic interpolations for $H$ within each cell respectively. The number of degrees of freedom (DOFs) for these four cases are 125,297 (upper-left), 343,417 (upper-right), 216,357 (lower-left), and 422,322 (lower-right).}
\label{fig:euler-2d}
\end{figure*}

The contour plots of $H_2$ obtained with $N=25$ and $N=35$ (quadratic interpolation) at the cross section $y=0$ are presented in the upper panels of Fig.~\ref{fig:euler-2d}. We infer from these two plots that there are 8 primary vortical regions. No secondary vortical regions are observed in these two panels, which explains the monotonic decay of $E_m$ in the right panel of Fig.~\ref{fig:lamd2-em-po}.

\begin{figure*}[!ht]
\centering
\includegraphics[width=0.32\textwidth]{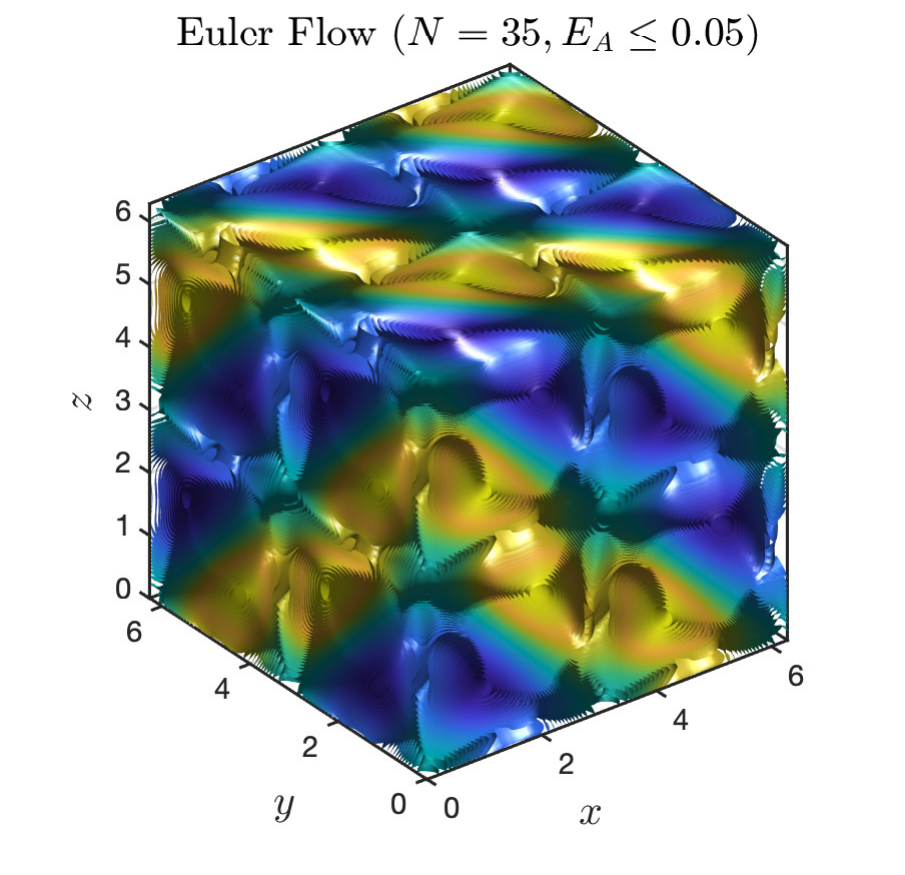}
\includegraphics[width=0.32\textwidth]{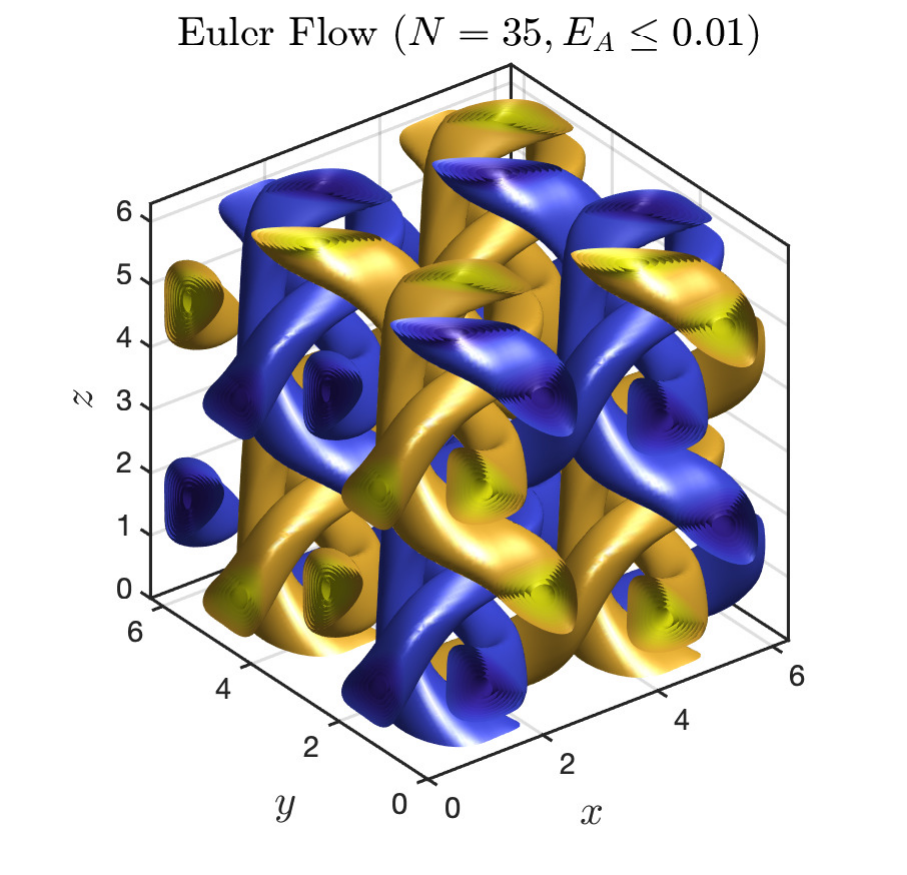}
\includegraphics[width=0.32\textwidth]{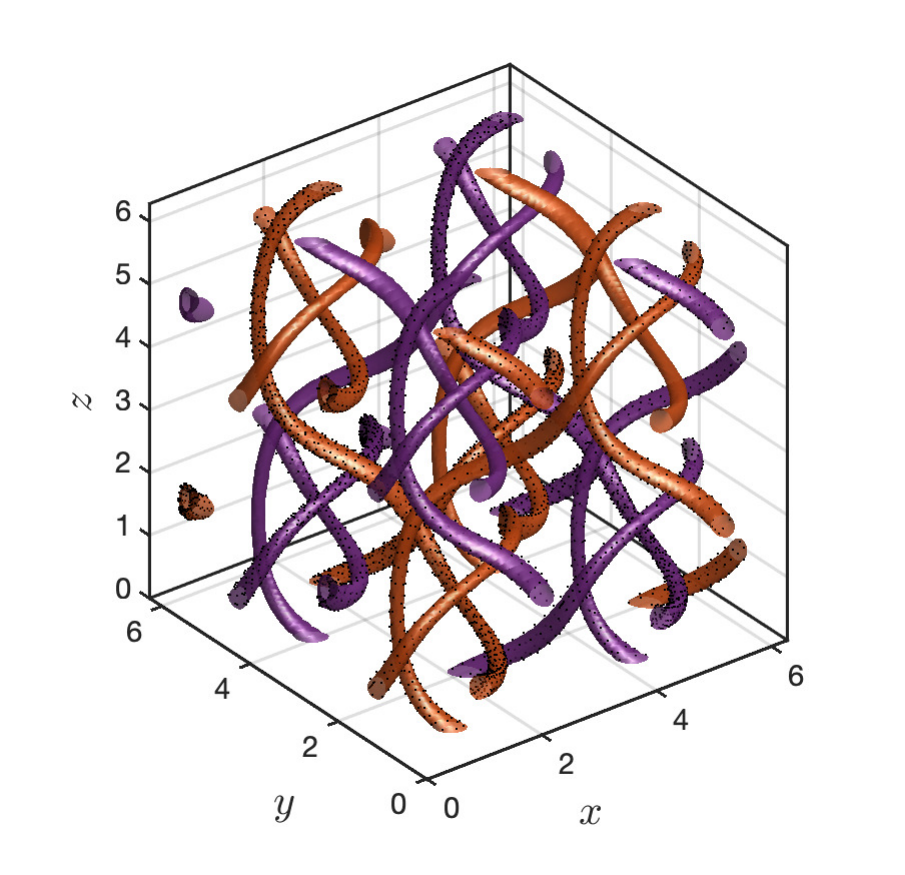}
\caption{Isosurfaces of $H_2$ for the Euler flow with various filter thresholds: $E_A\leq0.05$ (left panel) and $E_A\leq0.01$ (middle panel). The right panel gives the innermost layers of the 8 vortex tubes in the middle panel, along with some simulated trajectories (black dots).}
\label{fig:euler-3d}
\end{figure*}

To extract the approximate streamsurfaces of the eight vortex structures, we again apply the $E_A$-based (see~\eqref{eq:EA}) filter. The isosurfaces of $H_2$ obtained with $N=35$ (quadratic interpolation) under various thresholds for the filter are shown in Fig.~\ref{fig:euler-3d}. For $E_A\leq0.05$, these isosurfaces densely fill the cube. In contrast, for $E_A\leq0.01$, we clearly see the eight vortex tubes from the filtered isosurfaces in the middle panel of Fig.~\ref{fig:euler-3d}. These tubes are entangled with each other, as seen in the right panel of the figure. In the right panel, we also present the results from forward simulations with initial conditions on the selected approximate streamsurfaces. The trajectories obtained from forward simulation stay close to the approximate streamsurfaces, which illustrates the power of our method.

Based on a reference solution obtained from the Poincar\'e map~\cite{stergios}, we know that the system also has some delicate vortical regions between the primary vortical regions. Our method is able to extract even these vortex structures by increasing the interpolation order to cubic. Indeed, as shown in the lower panels of Fig.~\ref{fig:euler-2d}, delicate vortical regions are revealed and some of these small scale vortex structures are pointed out by the black arrows. These structures become more clear when we increase $N$ from 20 to 25. Note that the numbers of DOFs for the upper-right panel is more than that of the lower-left panel. This indicates that we may use higher-order interpolations to better extract delicate vortical regions.

\section{Rayleigh-Bénard convection}
\label{sec:rbc}
\begin{figure*}[!ht]
\centering
\includegraphics[width=0.4\textwidth]{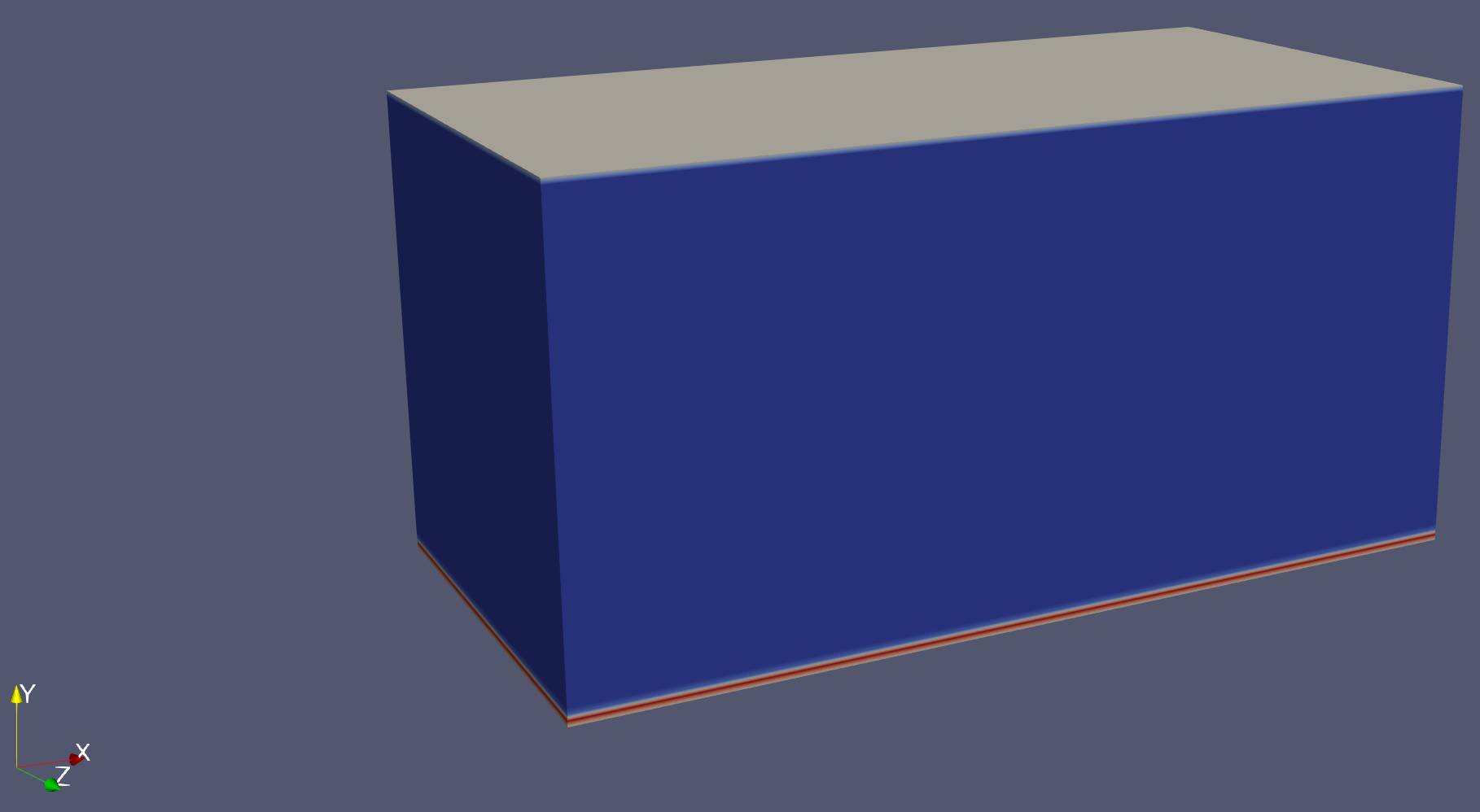}
\caption{A schematic plot of RBC.}
\label{fig:RBC}
\end{figure*}

In this section, we consider Rayleigh-Bénard convection (RBC) in the domain $\Omega=[0,l_x]\times[0,l_y]\times[0,l_z]$. This domain is constrained by a hot plate at the bottom ($y=0$) and a cold plate at the top ($y=l_y$), as illustrated in Fig.~\ref{fig:RBC}. In particular, the temperatures at the hot and cold plates are 274.15K and 273.15K, respectively. This temperature difference provides the driving force for the convection. In addition, periodic boundary conditions are imposed along the $x$ and $z$ directions. This flow is fully controlled by two dimensionless parameters. The first is the Prandtl number $Pr$, which describes the fluid properties as the ratio of the viscosity and the thermal diffusivity. The other parameter is the Rayleigh number $Ra$ characterizing the strength of the thermal driving. Here we fix $Pr=0.71$ (air at room temperature) but vary $Ra$ to extract approximate streamsurfaces for RBC with various dynamical behaviors.

We use the computational library OpenLB~\cite{krause2021openlb} to simulate the RBC. OpenLB is an open-source package that provides a flexible framework for lattice Boltzmann simulations. Let the resolution of the model be $N$, the number of grids of the discrete model is $(N+3)\times(2N+3)\times(N+3)$. More details about the simulations can be found in the \texttt{rayleighBenard3d} example of OpenLB.

Given there are two walls where $\mathbf{u}=\mathbf{0}$, we impose $H=0$ on the two walls. Accordingly, the trial function space (cf.~\eqref{eq:Htrial}) is updated as
\begin{equation}
\label{eq:Htrial-dpbc}
    \mathcal{H}_\mathrm{trial}=\{H\in H^1(\Omega),\,\, H|_{x=0}=H|_{x=l_x},\,H|_{y=0}=H|_{y=l_y}=0,\,H|_{z=0}=H|_{z=l_z}\}.
\end{equation}
One can easily see that~\eqref{eq:left-weak} still holds with this trial function space. Consequently, the weak form~\eqref{eq:left-weak} still holds and the discussions in Sect.~\ref{sec:eig-to-min} are still true.

We again use \texttt{BoxMesh} in FEniCS to generate a mesh for $\Omega$. Given the number of cells $(N_x,N_y,N_z)$ in each direction, the total number of tetrahedrons is $6N_xN_yN_z$ and the total number of vertices is $(N_x+1)(N_y+1)(N_z+1)$. Since $\Omega_\mathrm{H}\neq\emptyset$, we have $\lambda_1\neq0$ and look for $(\lambda_1,H_1)$.

\subsection{Quasi-two-dimensional flow}
\label{sec:quasi-2d}
\begin{figure*}[!ht]
\centering
\includegraphics[width=0.32\textwidth]{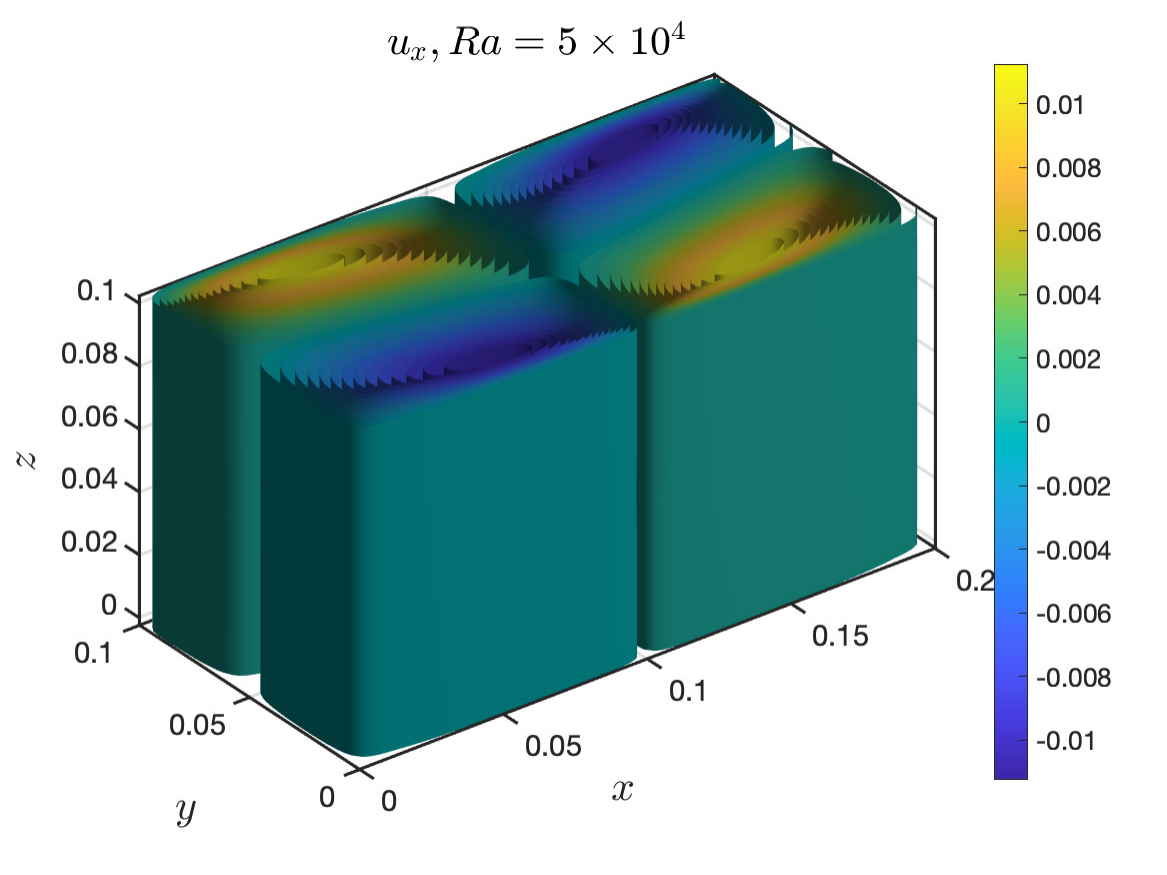}
\includegraphics[width=0.32\textwidth]{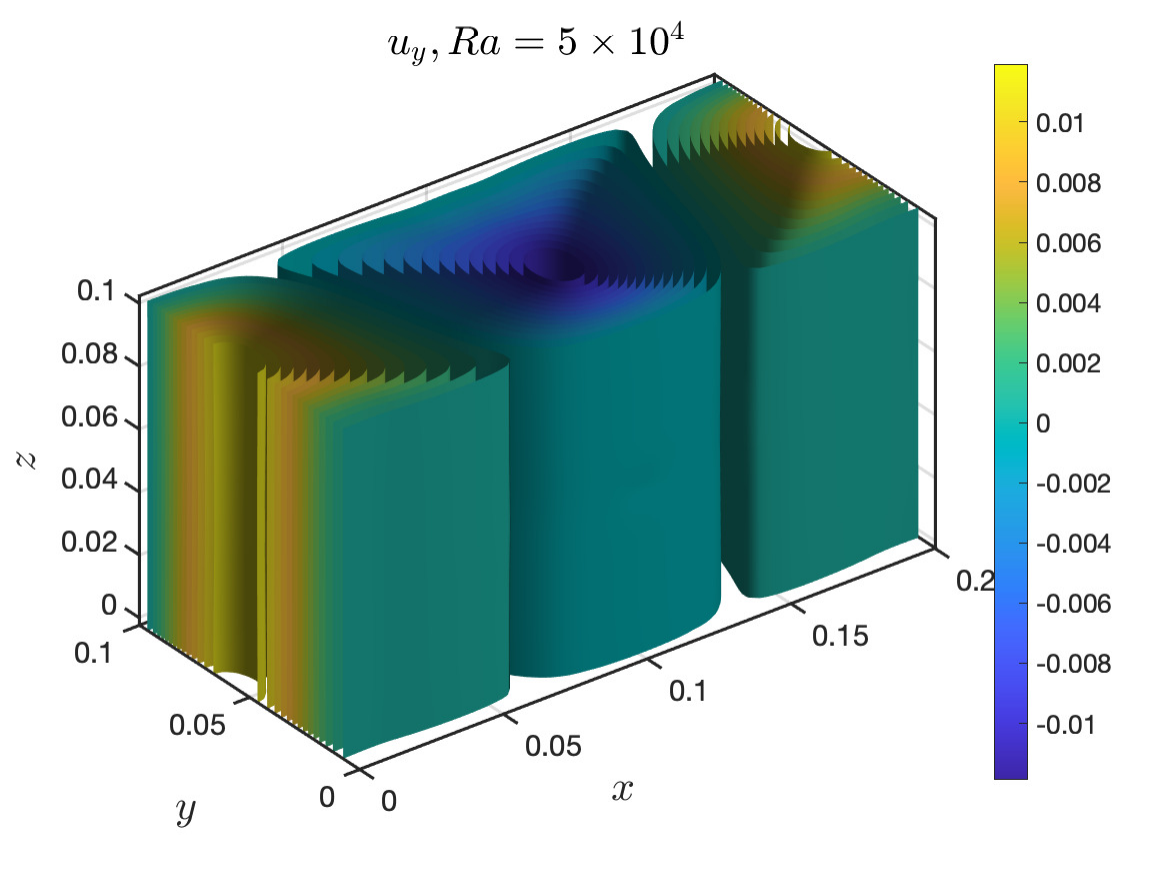}
\includegraphics[width=0.32\textwidth]{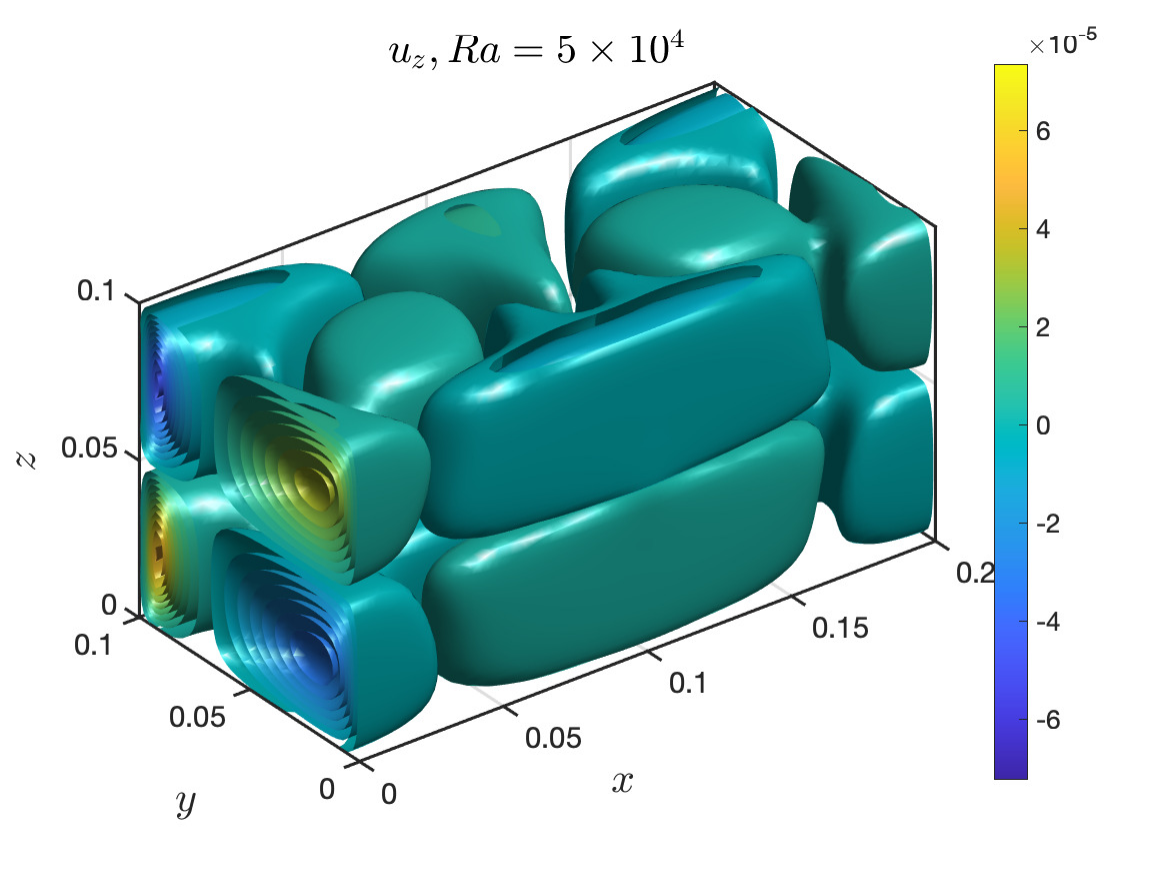}
\caption{Isosurfaces of the velocity components of the steady state of the RBC flow with $l_x=0.2$, $l_y=l_z=0.1$ and $Ra=5\times10^4$. The left, middle and right panels give the results for $u_x$, $u_y$ and $u_z$, respectively.}
\label{fig:RBC-5e4-vel}
\end{figure*}

Let $l_x=0.2$, $l_y=l_z=0.1$ and $Ra=5\times10^4$, in which case the flow converges to a steady velocity field with two large-scale rolls. This motion is quasi-two-dimensional as $|u_z|$ is much smaller than $|u_x|$ and $|u_y|$, and $(u_x,u_y)$ barely change along the $z$ direction, as seen in Fig.~\ref{fig:RBC-5e4-vel}. Therefore, we expect that the flow is close to integrable and  our approach is able to extract the approximate first integral.

\begin{figure*}[!ht]
\centering
\includegraphics[width=0.45\textwidth]{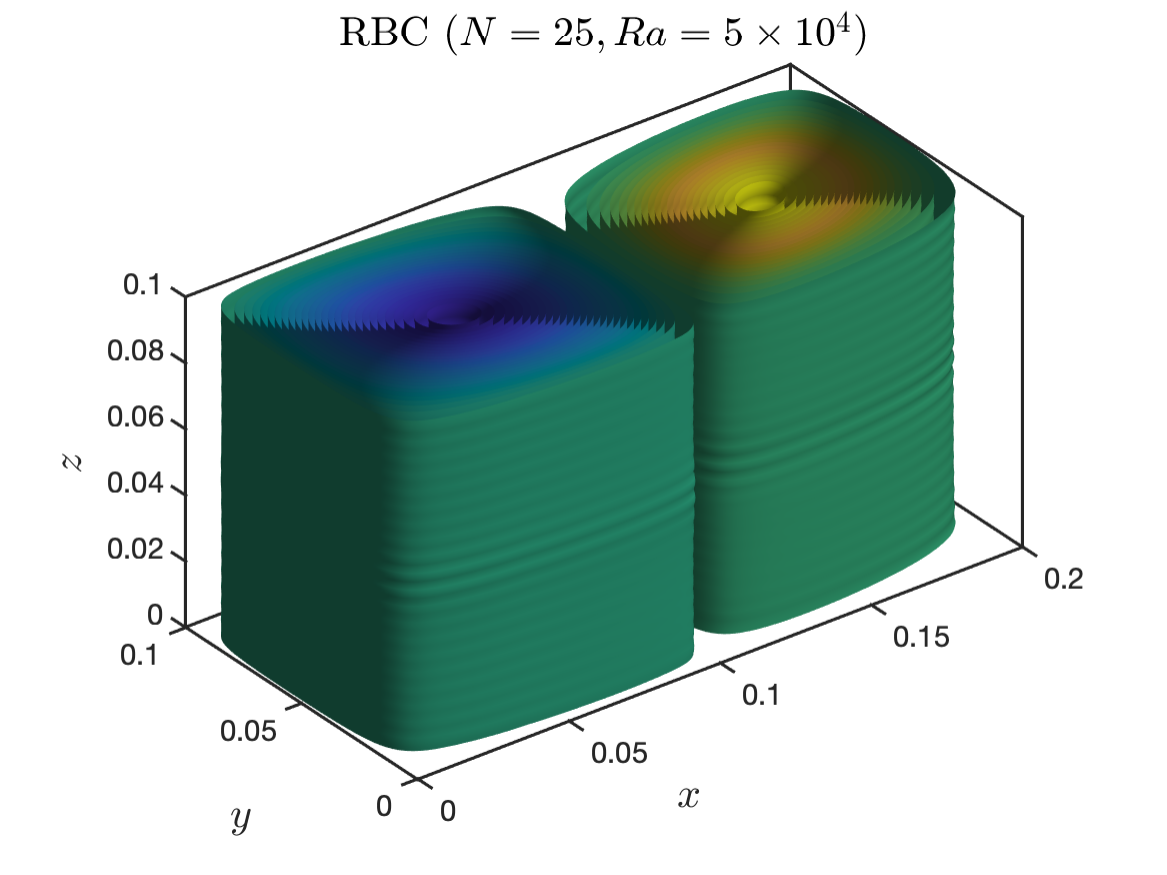}
\includegraphics[width=0.45\textwidth]{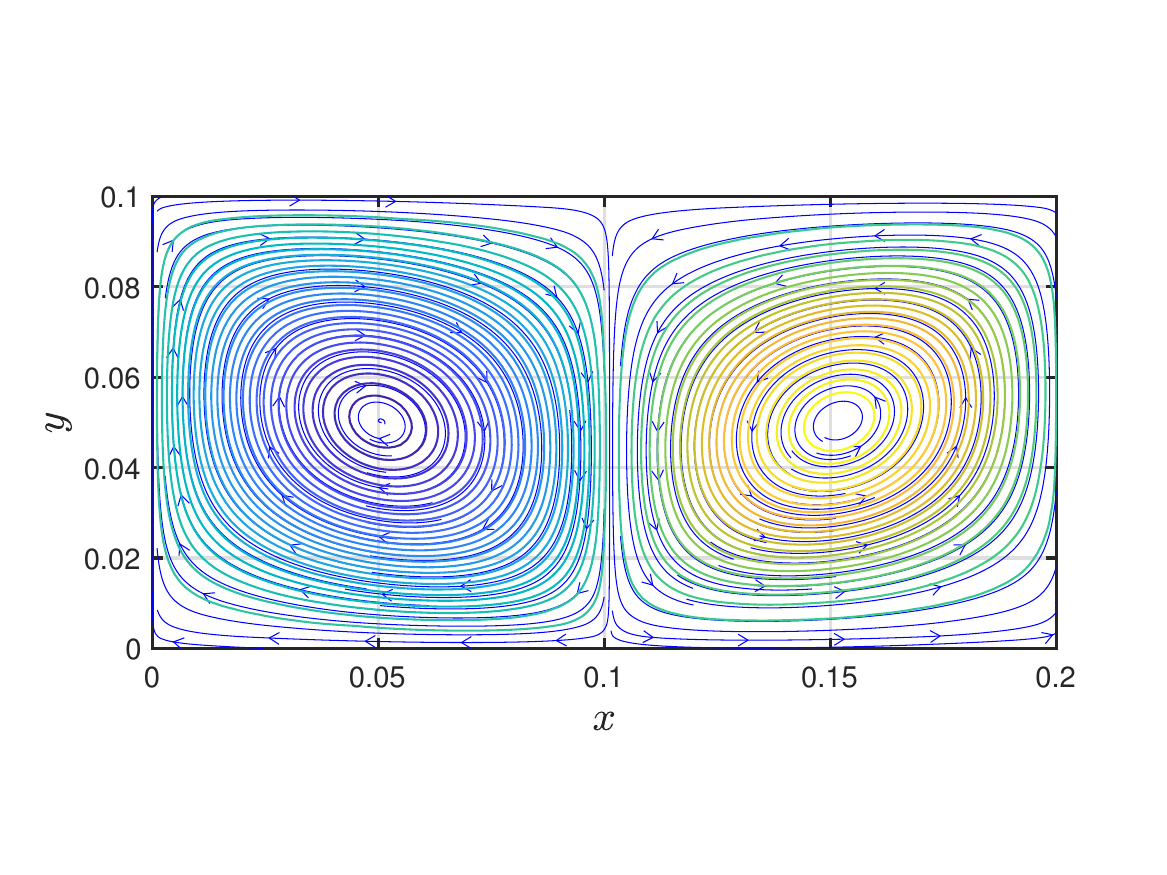}
\caption{(left) Isosurfaces of ${H}_1$ for the RBC at $Ra=5\times10^5$ obtained from finite-element method with $N=25$. (right) contour plot of $H_1$ at cross section $z=0$, along with streamlines of $(u_x,u_y)$ at the cross section.}
\label{fig:RBC-5e4-streamline}
\end{figure*}

With $N_x=50$ and $N_y=N_z=25$ and quadratic interpolation, we obtain $\lambda_1=5.9\times10^{-7}$ along with a mean invariance error of $E_m=0.02$. The isosurfaces of the corresponding $H_1$ are presented in Fig.~\ref{fig:RBC-5e4-streamline}. Indeed, the two primary rolls are revealed from the isosurfaces of $H_1$ and these isosurfaces barely change along the $z$ direction. To validate these results, we present the contour plot of $H_1$ at cross section $z=0$ along with the streamlines of the velocity field $(u_x,u_y)$ at the cross section in the right panel, from which we see that the streamlines match well with the contour plot.

\subsection{Unsteady three-dimensional flow}

\begin{figure*}[!ht]
\centering
\includegraphics[width=0.32\textwidth]{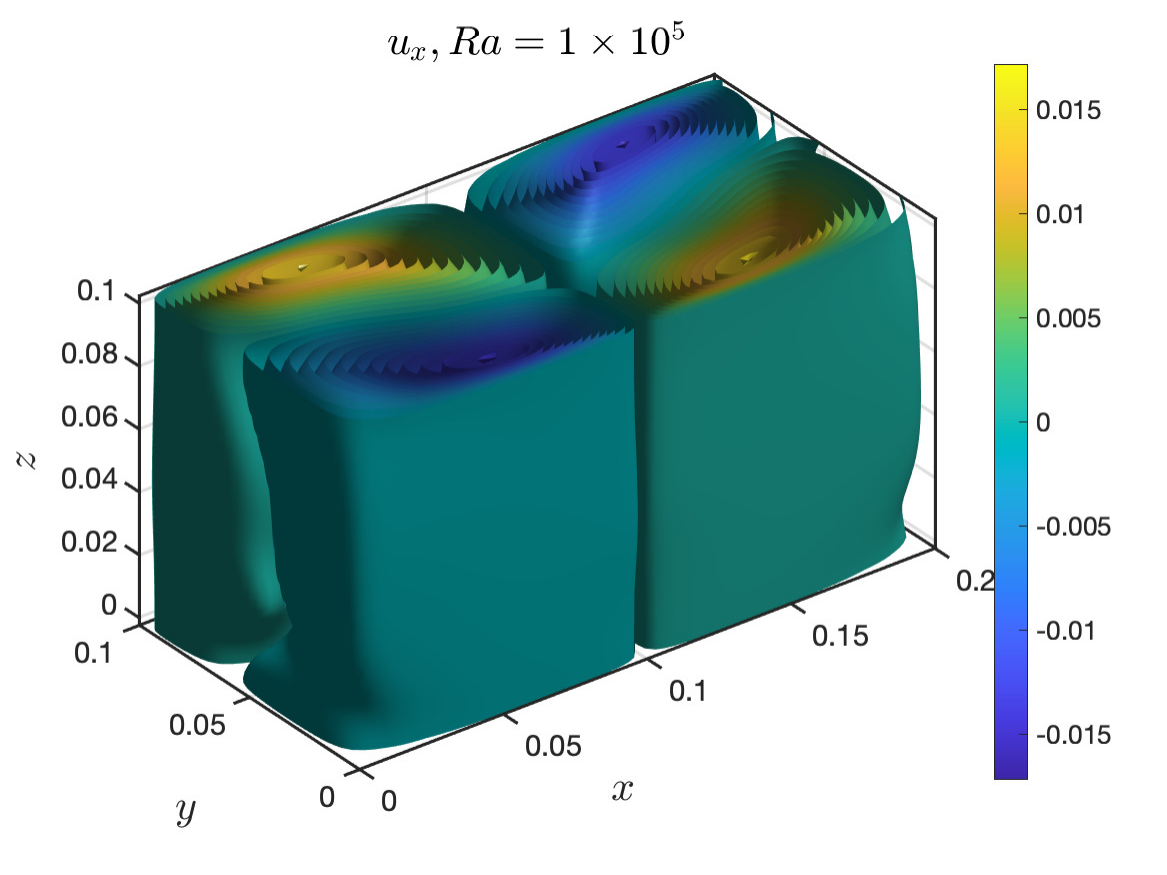}
\includegraphics[width=0.32\textwidth]{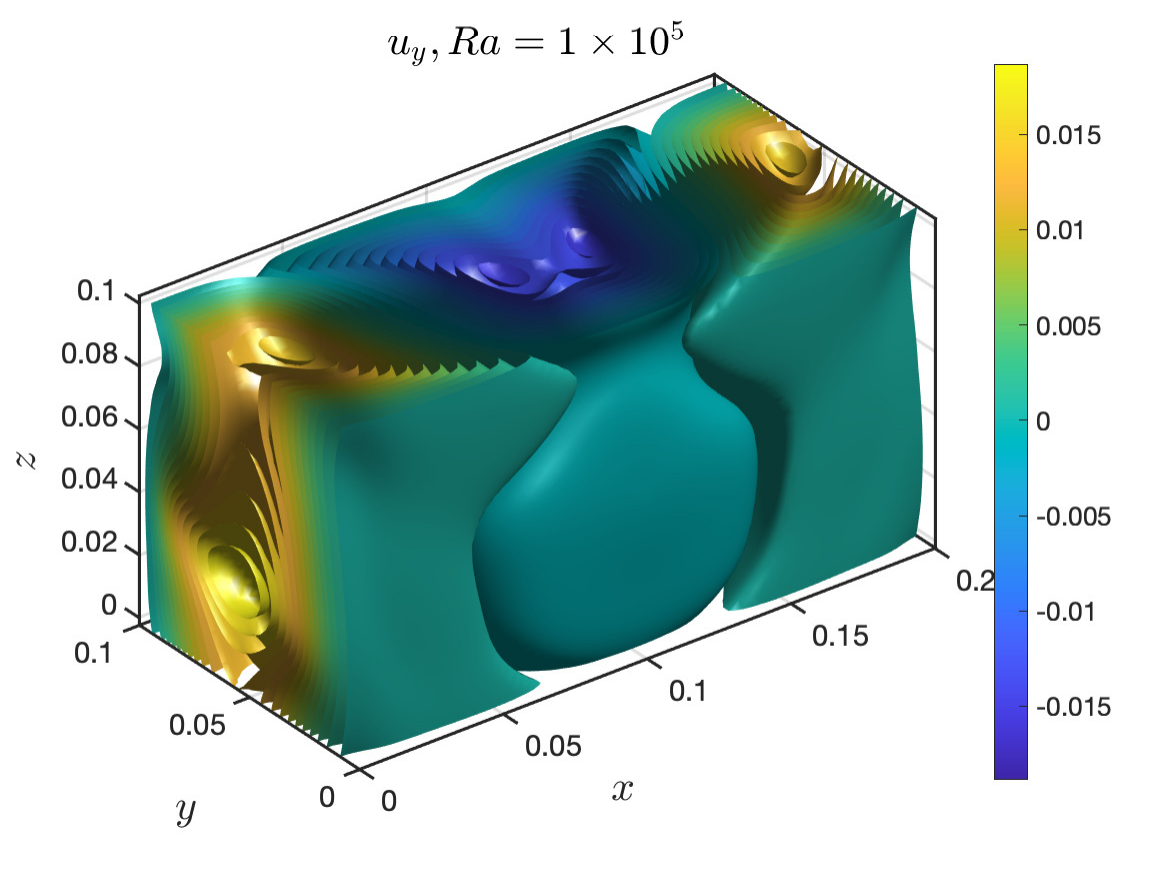}
\includegraphics[width=0.32\textwidth]{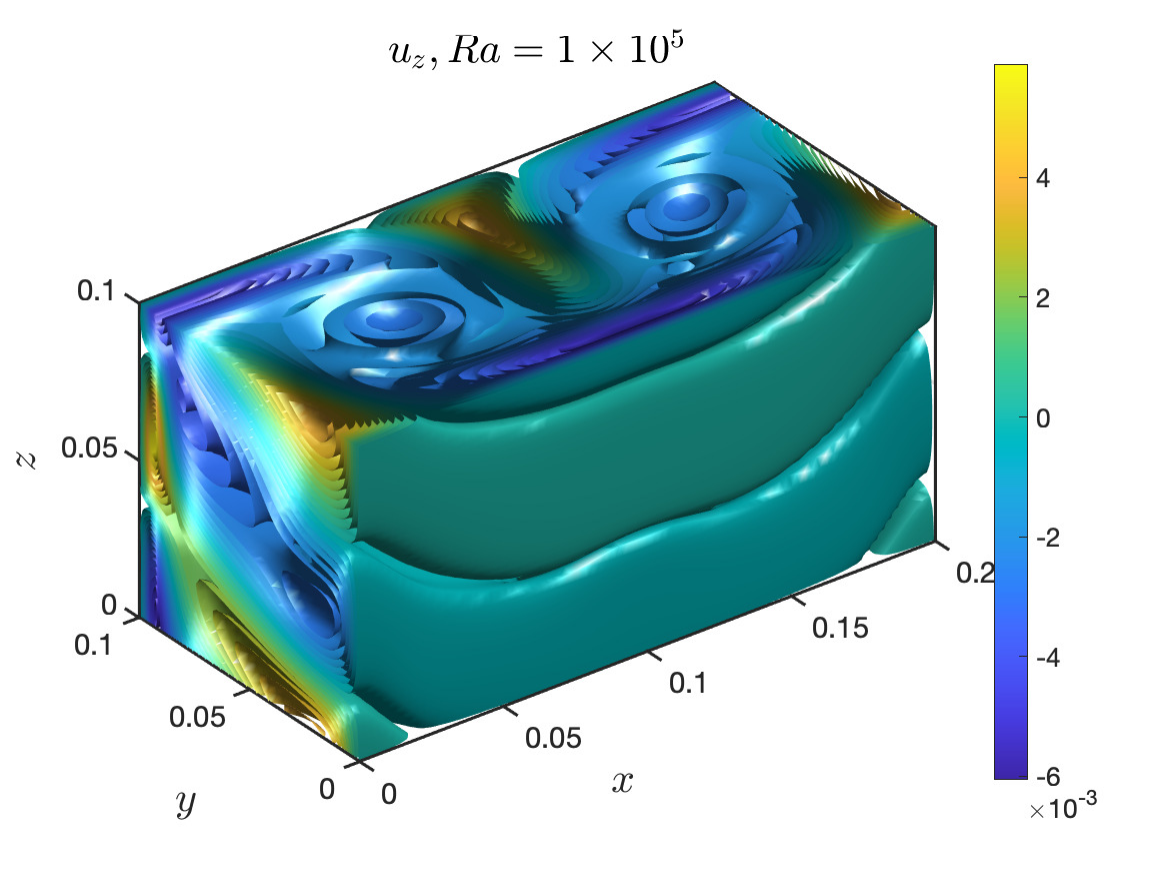}
\caption{Isosurfaces of velocity components of a snapshot ($t=614.519$) of the RBC flow with $l_x=0.2$, $l_y=l_z=0.1$ and $Ra=1\times10^5$. The left, middle and right panels show $u_x$, $u_y$ and $u_z$, respectively.}
\label{fig:RBC-1e5}
\end{figure*}

Next we still take $l_x=0.2$, $l_y=l_z=0.1$ but increase the Rayleigh number to $Ra=1\times10^5$. In this case the flow converges to a limit cycle, and hence the velocity field is unsteady but periodic. We take a snapshot of the velocity field and perform the computation of the approximate first integrals. The flow of this snapshot is three-dimensional, as suggested by the plots of isosurfaces of its three velocity components shown in Fig.~\ref{fig:RBC-1e5}. 

\begin{figure*}[!ht]
\centering
\includegraphics[width=0.45\textwidth]{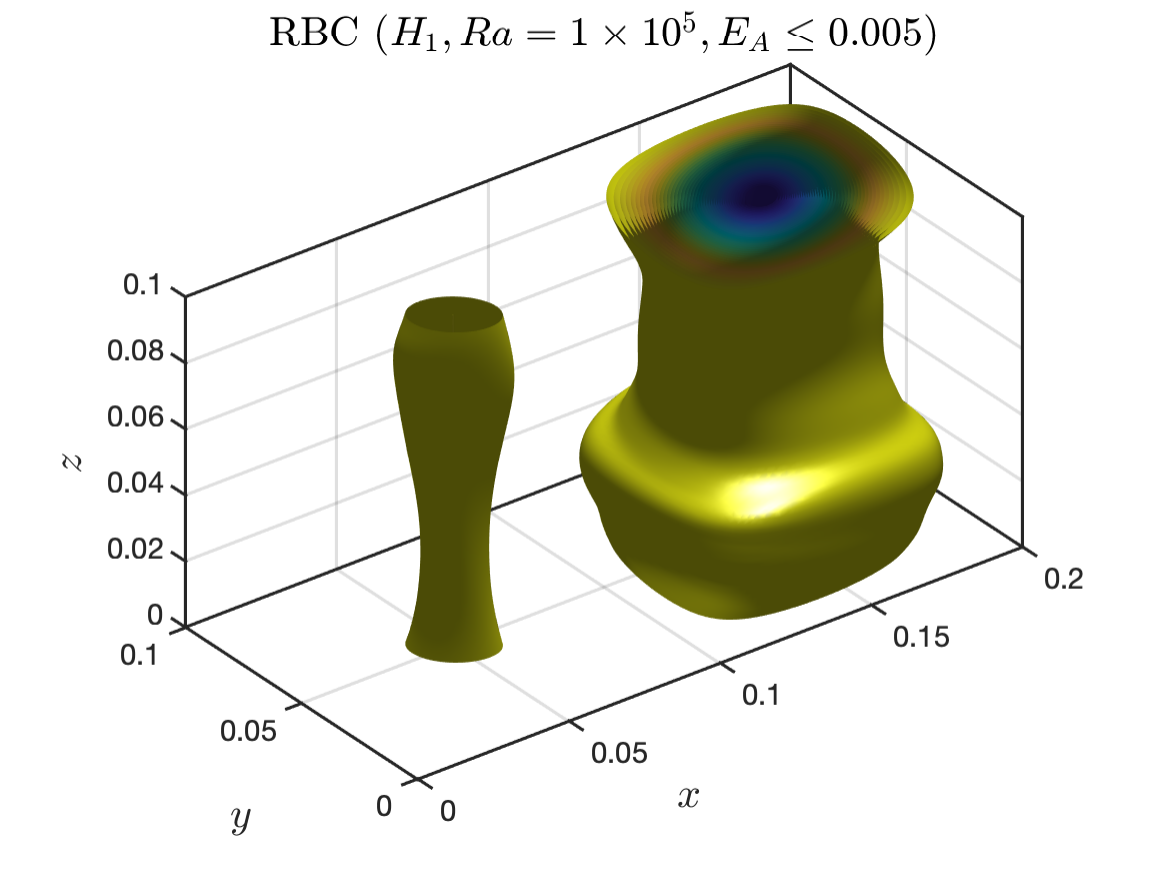}
\includegraphics[width=0.45\textwidth]{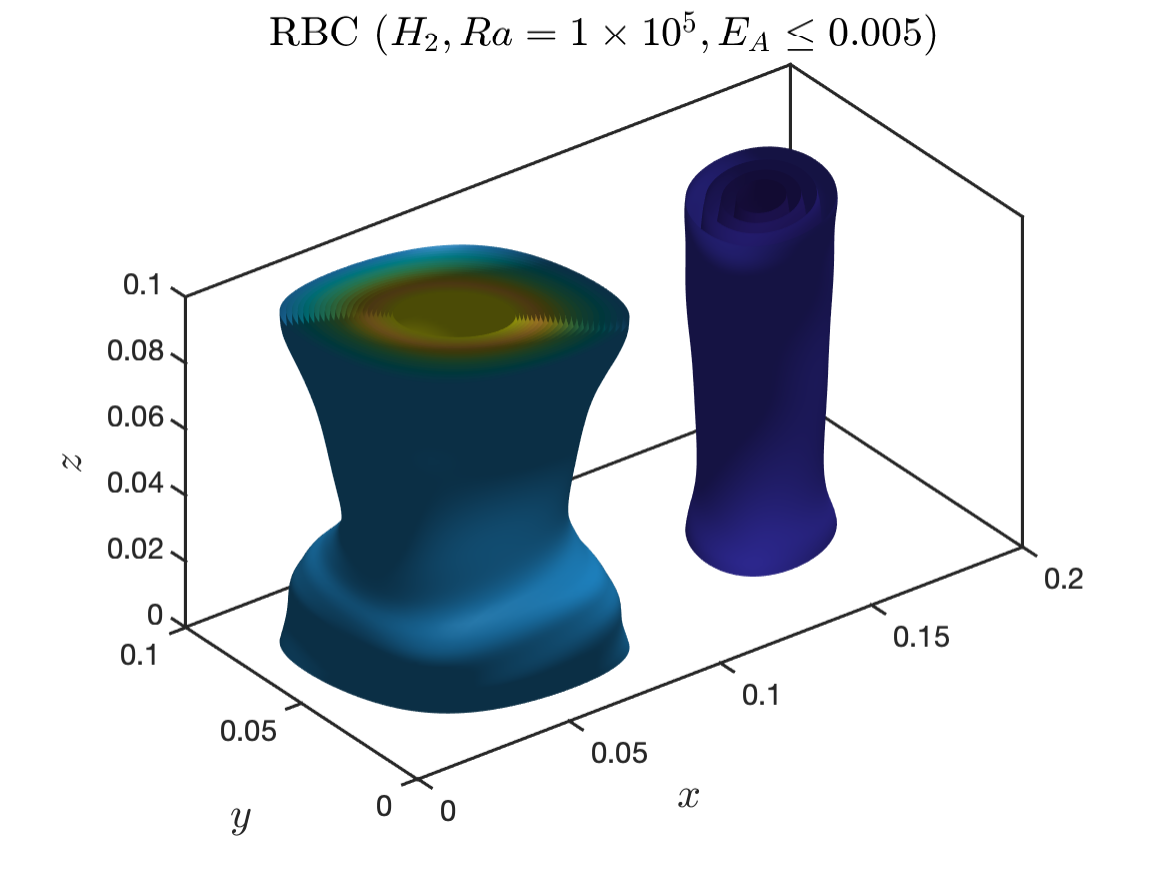}
\caption{Isosurfaces of ${H}_1$ (left panel) and $H_2$ (right panel) for a snapshot of the RBC at $Ra=1\times10^5$ obtained from finite-element methods with $N_x=60$ and $N_y=N_z=30$. Here we set a filter threshold $E_A=0.005$.}
\label{fig:RBC-1e5-isosurfaceH}
\end{figure*}

With $N_x=60$ and $N_y=N_z=30$ and quadratic interpolation, we obtain $\lambda_1=5.1\times10^{-6}$ along with a mean invariance error of $E_m=0.024$. The isosurfaces of the corresponding $H_1$ with a filter $E_A\leq0.005$ are presented in the left panel Fig.~\ref{fig:RBC-1e5-isosurfaceH}. A major vortex tube is observed for $x\geq0.1$, while a small vortex tube exists for $x\leq0.1$. We expect from Fig.~\ref{fig:RBC-1e5} that there should also be a comparable vortex tube for $x\leq0.1$ to the major one within $x\geq0.1$. Indeed, the isosurfaces of $H_2$ reveal the major vortex for $x\leq0.1$, as seen in the right panel of Fig.~\ref{fig:RBC-1e5-isosurfaceH}. Here we have $\lambda_2=6.9\times10^{-6}$, which is comparable to $\lambda_1$. Therefore, one may also check whether higher-order modes extract different structures than the first mode, provided that the eigenvalues of the higher-order modes are comparable to those of the first mode.

\begin{figure*}[!ht]
\centering
\includegraphics[width=0.3\textwidth]{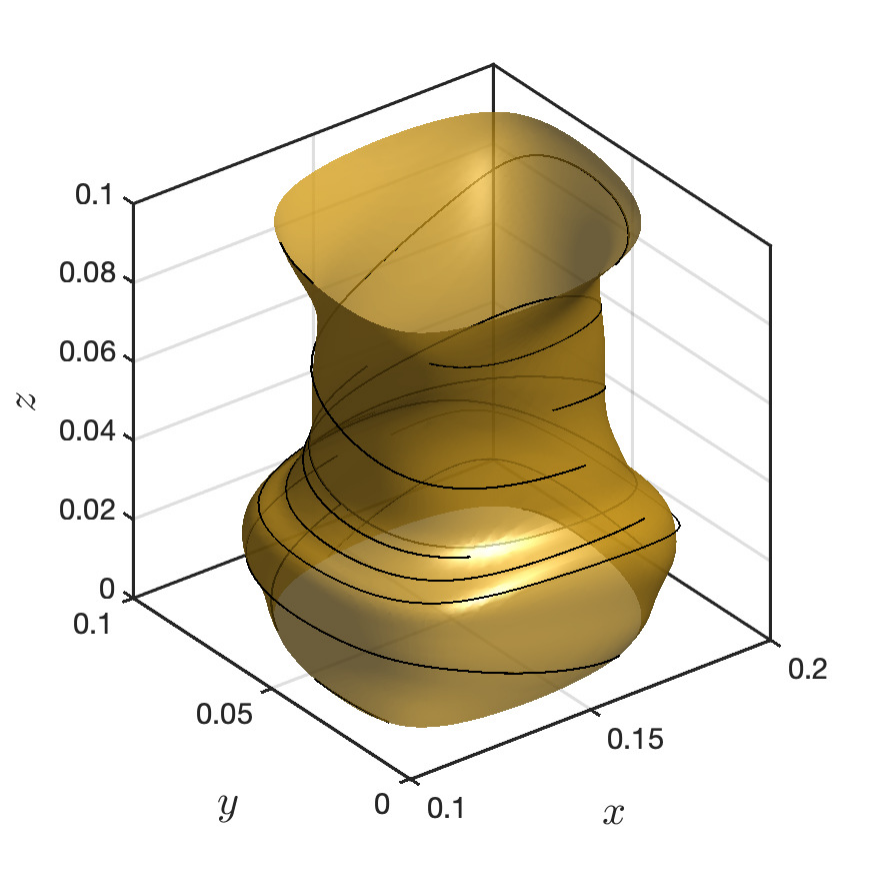}
\includegraphics[width=0.3\textwidth]{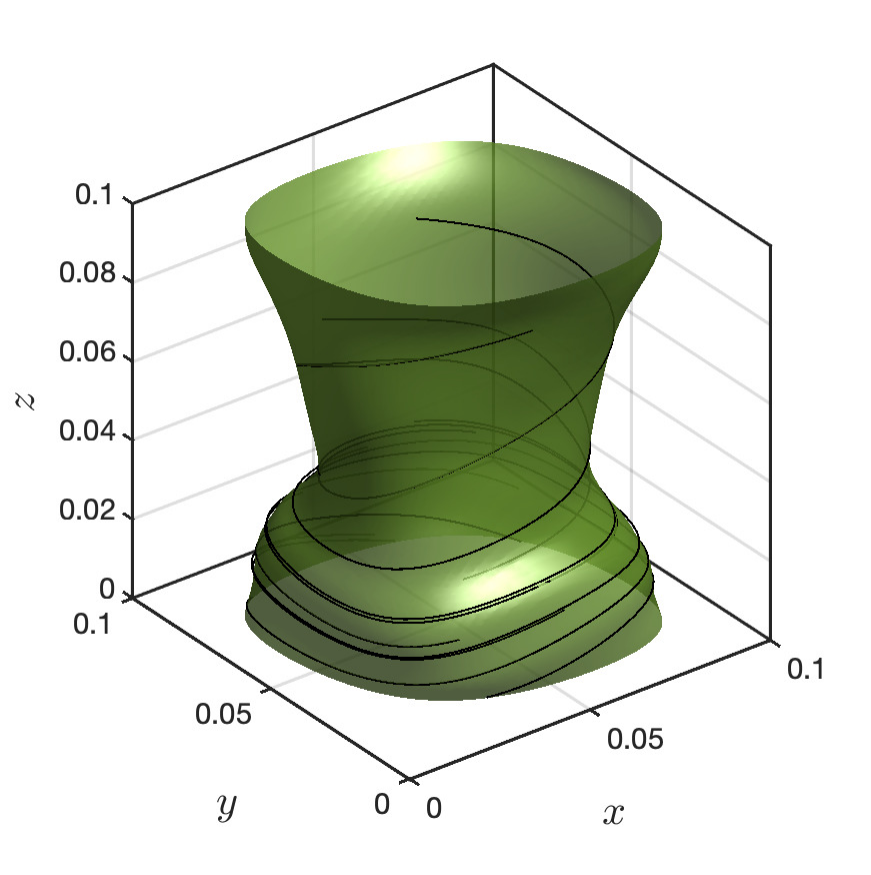}\\
\includegraphics[width=0.3\textwidth]{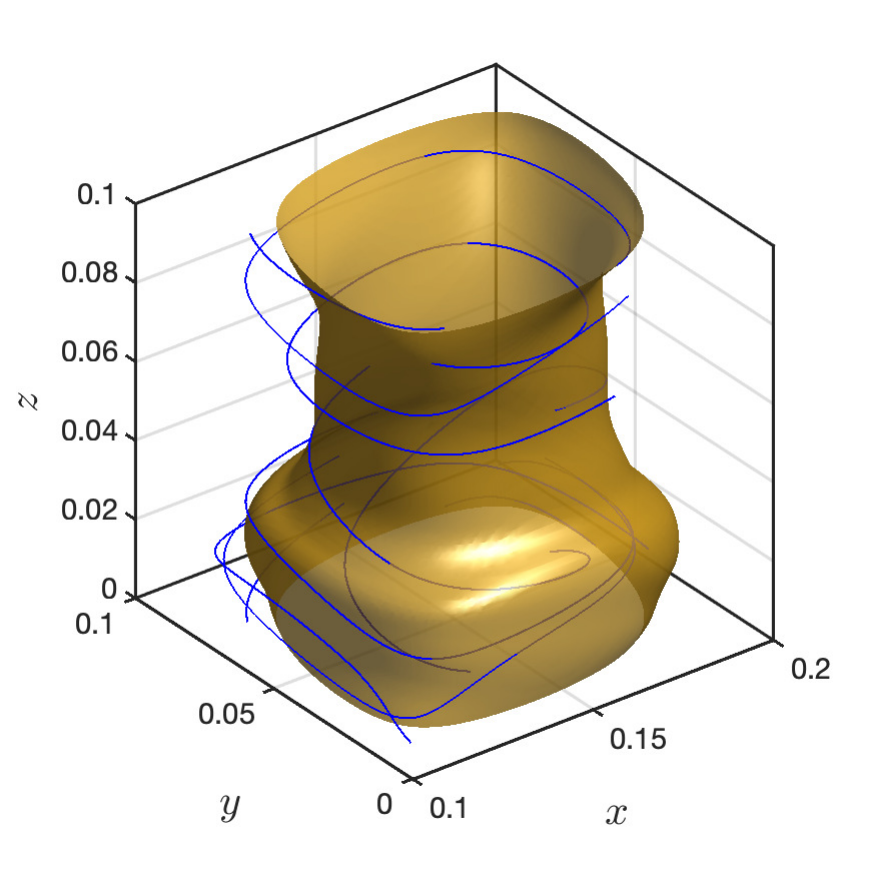}
\includegraphics[width=0.3\textwidth]{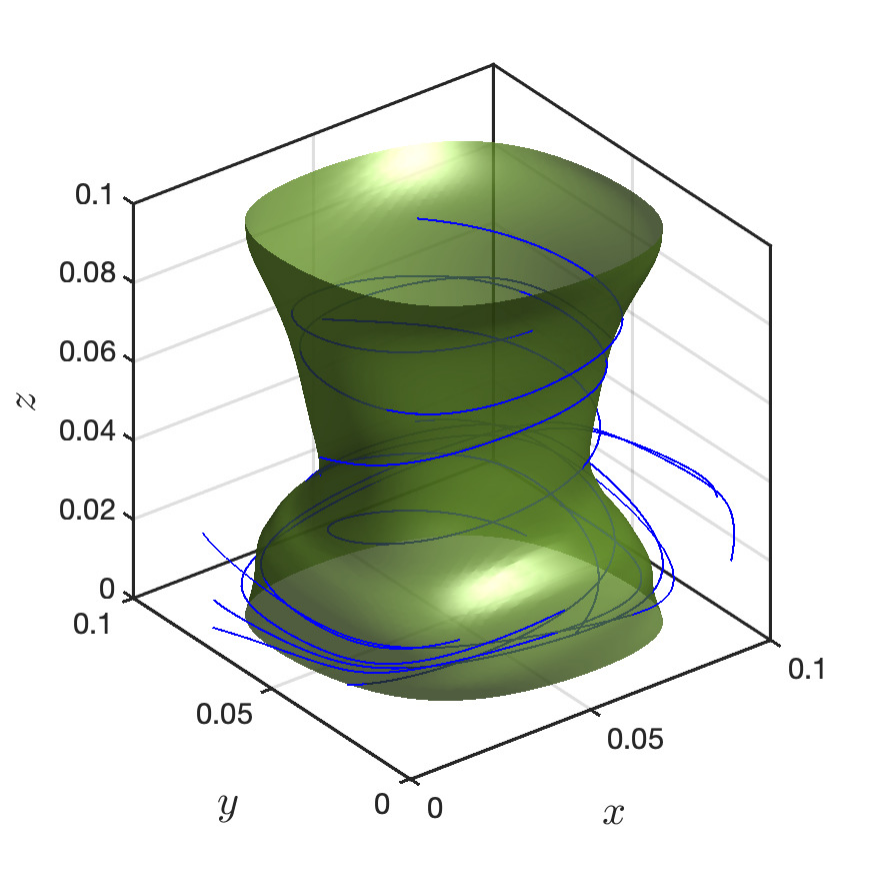}\\
\caption{Outermost isosurfaces of the two major vortex tubes in Fig.~\ref{fig:RBC-1e5-isosurfaceH}, along with streamlines of 10 randomly selected points in each of the surface. Trajectories from the forward simulations of the \emph{frozen} vector field are plotted with black lines and shown in the upper two panels. In contrast, the blue lines in the lower two panels denote streamlines of the unsteady flow started from the same initial points. Here the forward simulation time is 20 since $|\mathbf{u}|\sim\mathcal{O}(0.01)$.}
\label{fig:RBC-3d-streamline-steady}
\end{figure*}

Now we launch streamlines to validate the obtained results. We take 10 random points on the outermost layer of each major vortex tube in Fig.~\ref{fig:RBC-1e5-isosurfaceH} as initial conditions for forward simulation. Note that the extraction of the approximate first integral is an inherently Eulerian procedure. By performing the computation on a single snapshot of an unsteady flow we essentially freeze time. So, for validation, we also freeze time when we perform the time integration. As seen in the upper two panels of Fig.~\ref{fig:RBC-3d-streamline-steady}, these pseudo-streamlines (streamlines of the frozen flow) stay close to the extracted approximate streamsurfaces. We also launch streamlines for the unsteady flow field with the same initial conditions. As seen in the lower two panels of Fig.~\ref{fig:RBC-3d-streamline-steady}, the streamlines stay around the approximate streamsurfaces, which indicates that the unsteady flow field indeed admits two vortex tubes. This also indicates that the approximate stream surfaces obtained from the single snapshot of the velocity field are close to the real, time dependent Eulerian vortex tubes.

\begin{figure*}[!ht]
\centering
\includegraphics[width=0.32\textwidth]{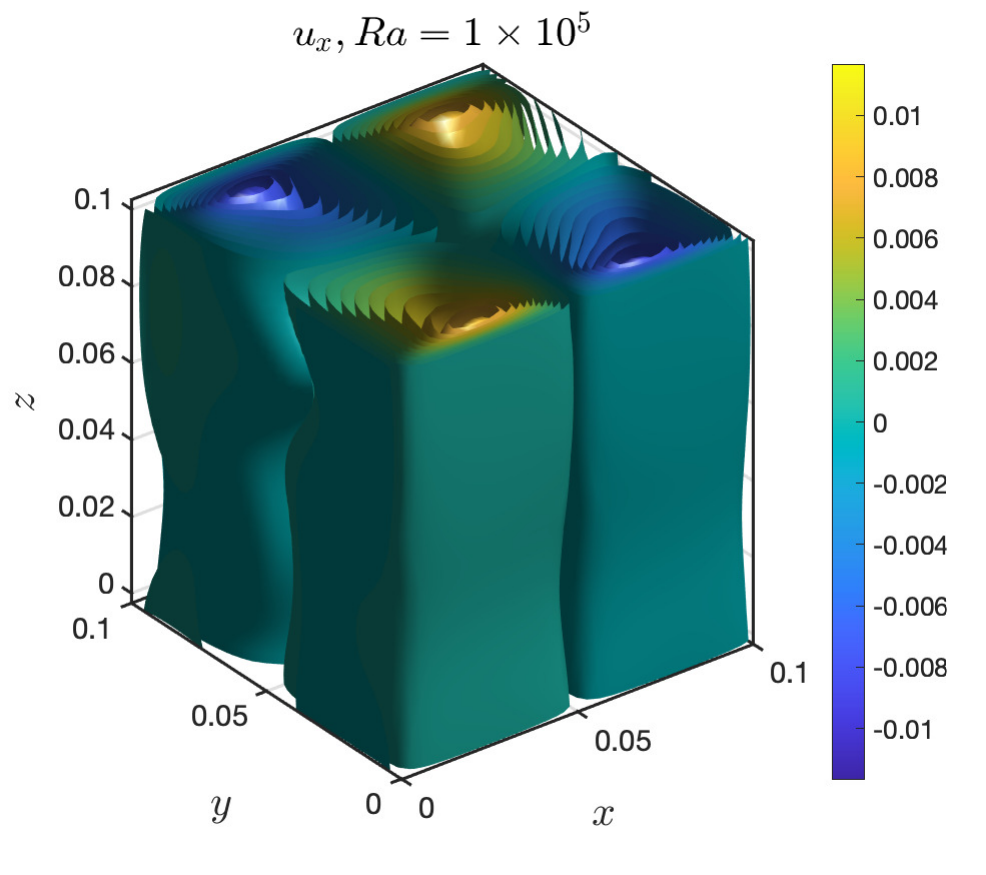}
\includegraphics[width=0.32\textwidth]{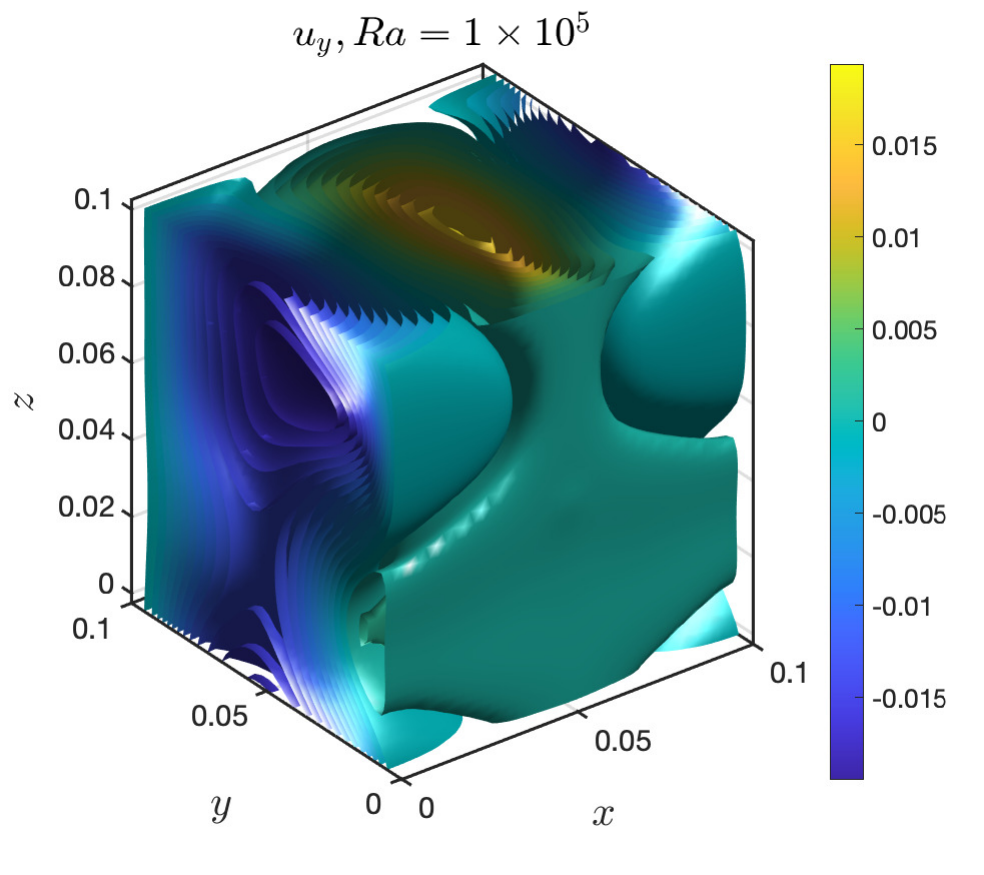}
\includegraphics[width=0.32\textwidth]{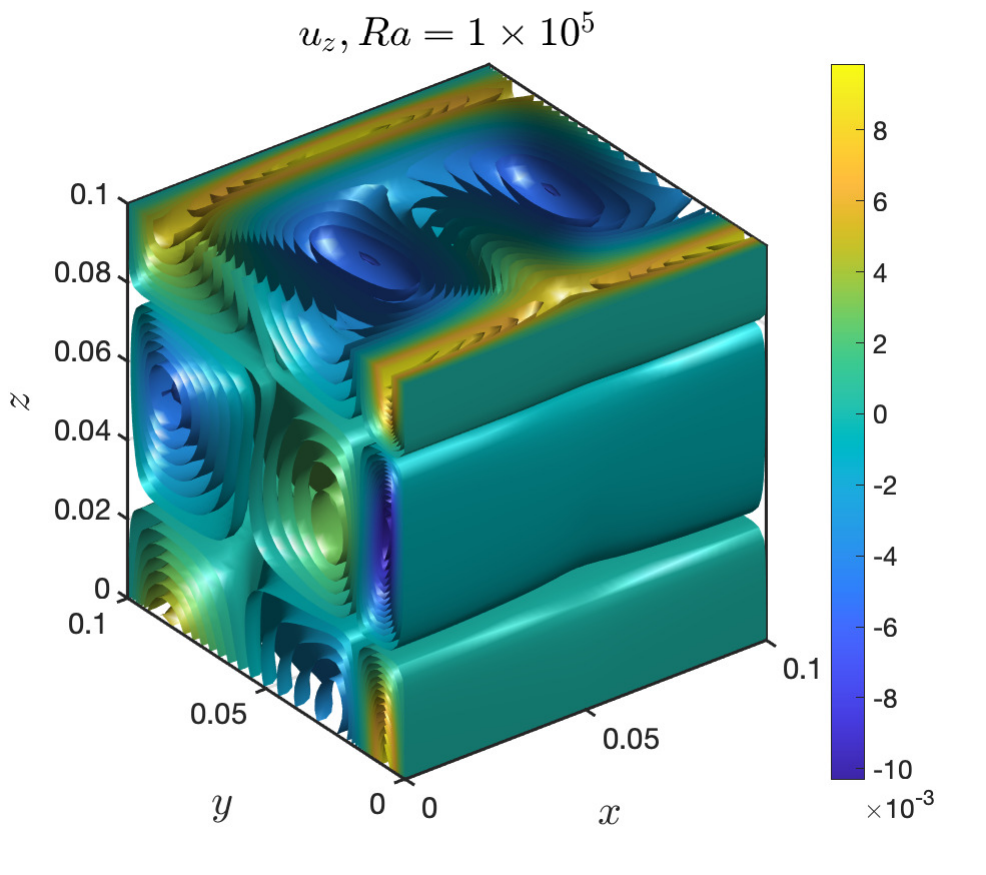}
\caption{Isosurfaces of the velocity components of a snapshot ($t=619.515$) of the RBC flow with $l_x=l_y=l_z=0.1$ and $Ra=1\times10^5$. The left, middle and right panels  show $u_x$, $u_y$ and $u_z$, respectively.}
\label{fig:RBCcube-1e5}
\end{figure*}

We conclude this section with a fully 3D unsteady flow in a cubic domain, where we extract vortex rings. Now we take $l_x=l_y=l_z=0.1$ and $Ra=1\times10^5$. This flow also converges to a limit cycle in steady state. We take a snapshot of the flow field and extract approximate first integrals. The contour plots of velocity components for this snapshot are shown in Fig.~\ref{fig:RBCcube-1e5}, which show that it is indeed a three-dimensional flow.

\begin{figure*}[!ht]
\centering
\includegraphics[width=0.32\textwidth]{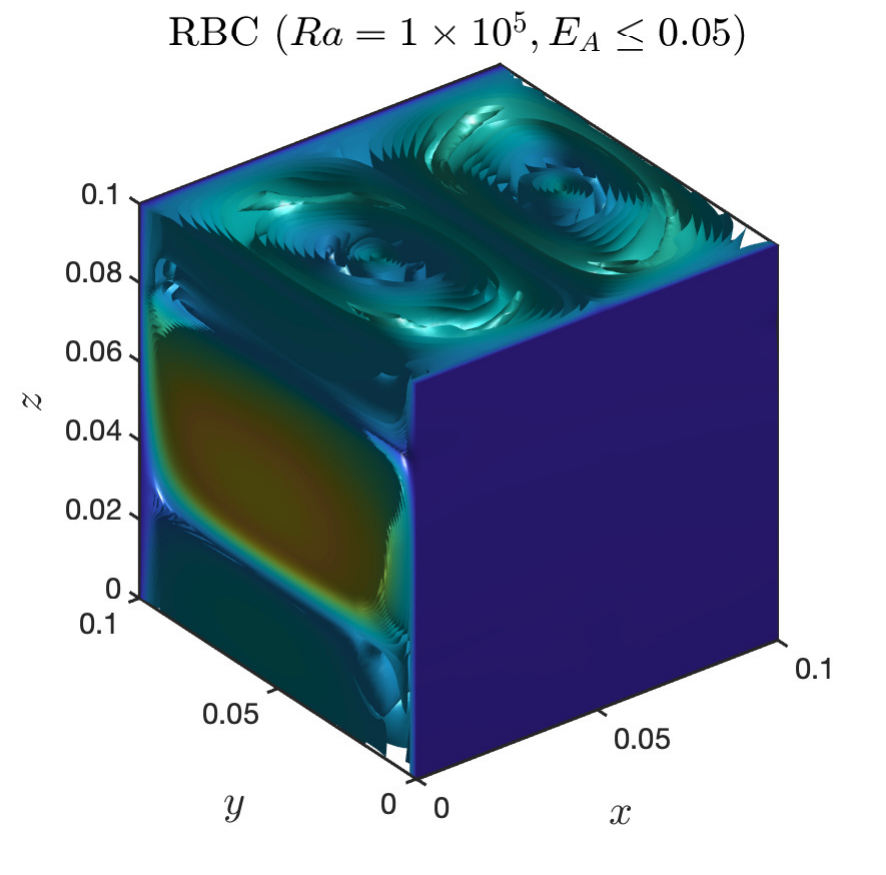}
\includegraphics[width=0.32\textwidth]{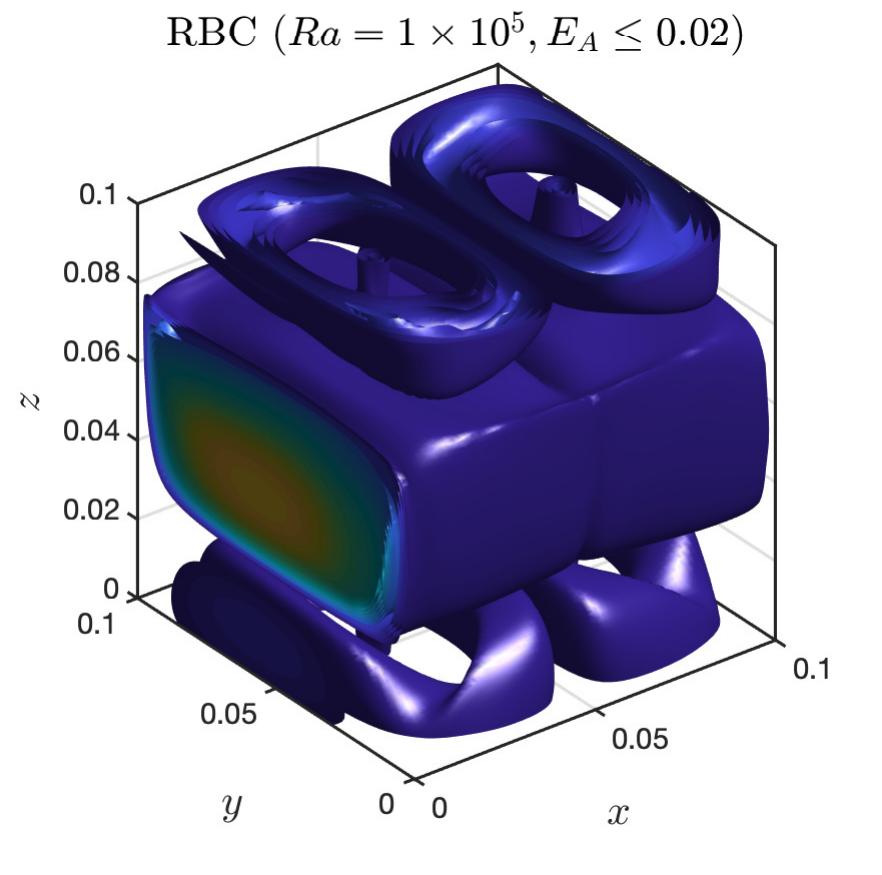}
\includegraphics[width=0.32\textwidth]{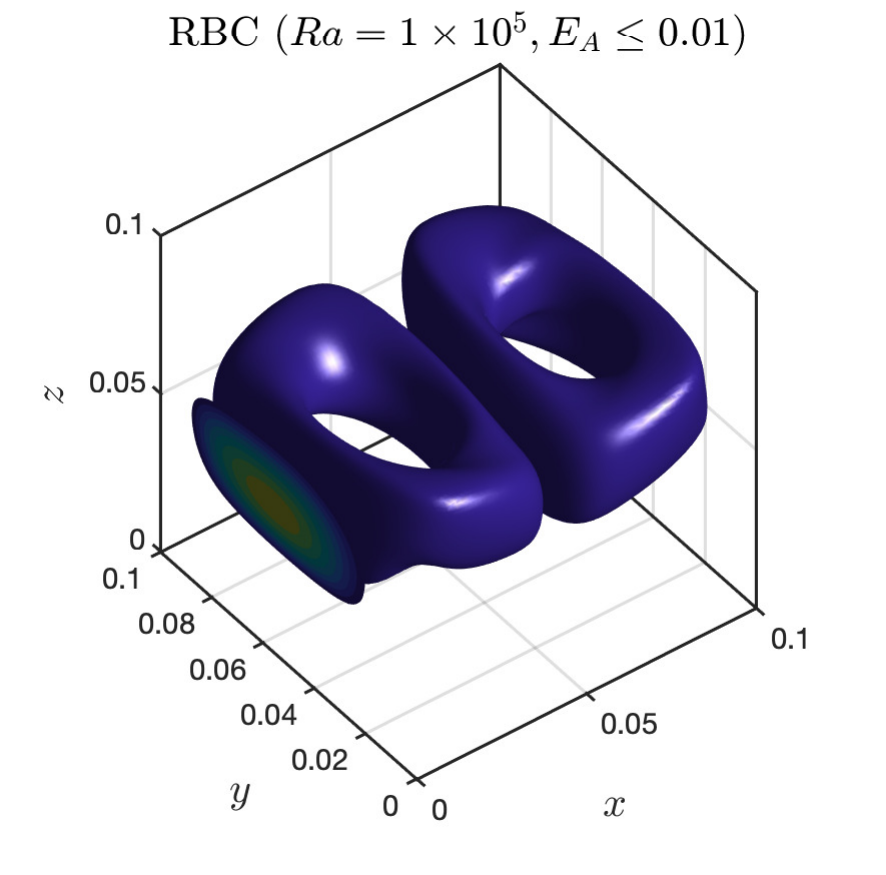}
\caption{Isosurfaces of $H_1$ for the RBC flow in a cube with various filter thresholds: $E_A\leq0.05$ (left panel), $E_A\leq0.02$ (middle panel), and $E_A\leq0.01$ (right panel).}
\label{fig:RBCcube-3d}
\end{figure*}

With $N_x=60$, $N_y=N_z=30$ and quadratic Lagrange elements, we obtain $\lambda_1=3.9\times10^{-5}$ along with mean invariance error $E_m=0.03$. By decreasing the filter threshold $E_A$, we are able to extract two vortex rings, as seen in the right panel of Fig.~\ref{fig:RBCcube-3d}. These vortex rings are  different from the vortex tubes that we extracted before.

\begin{figure*}[!ht]
\centering
\includegraphics[width=0.35\textwidth]{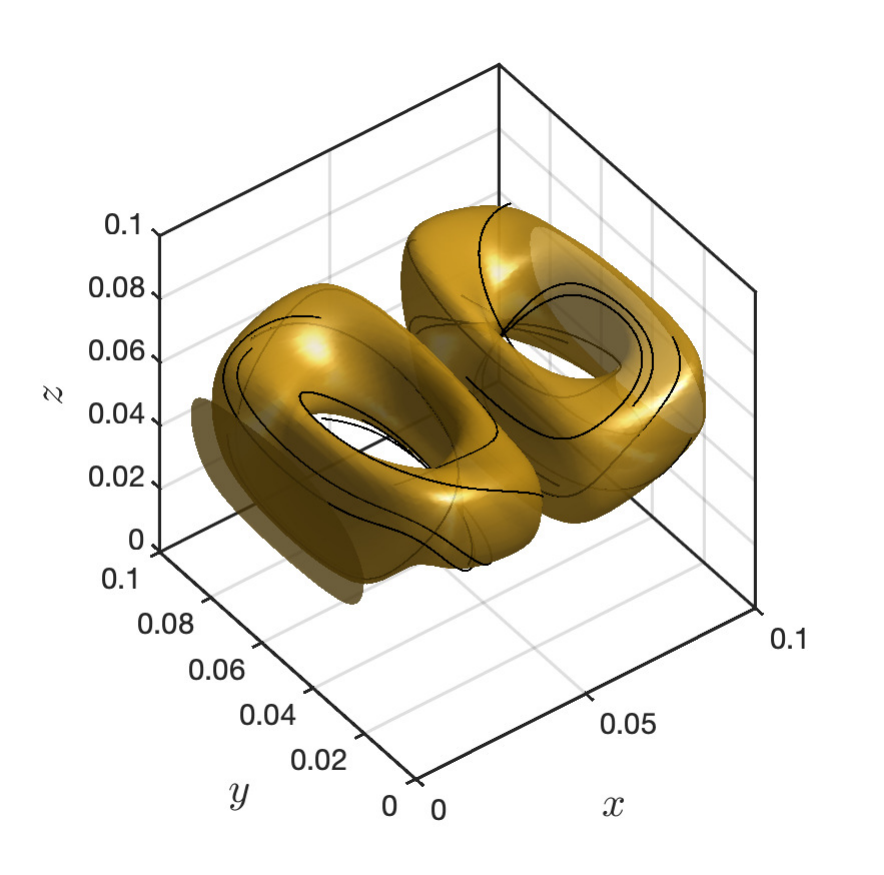}
\includegraphics[width=0.35\textwidth]{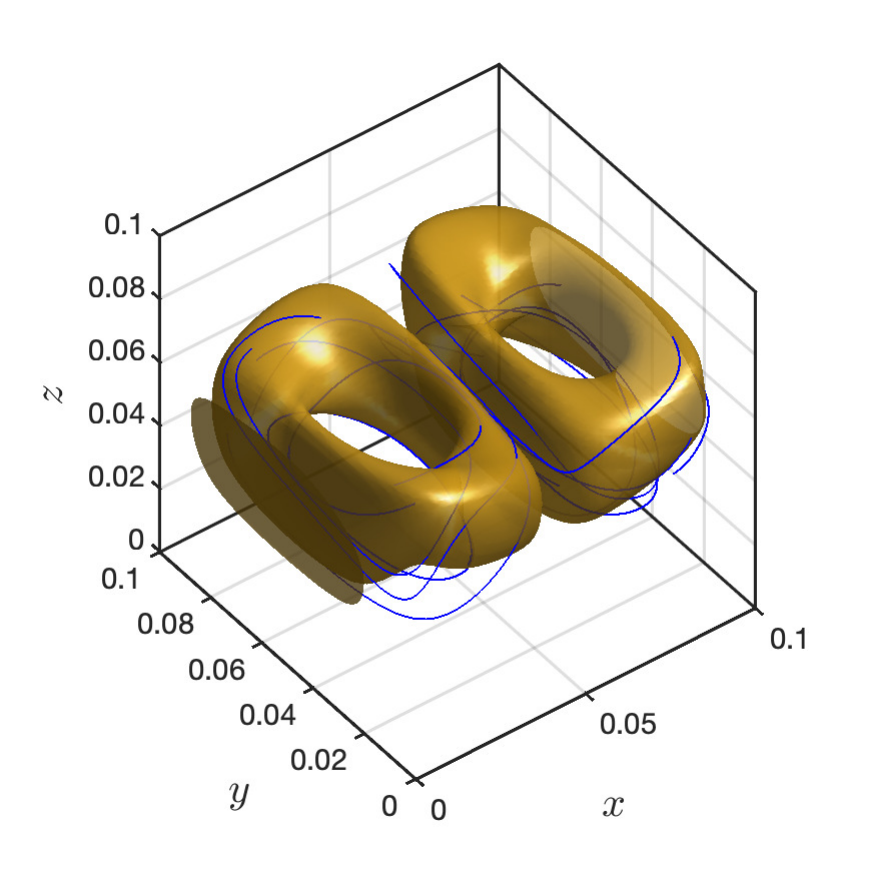}
\caption{Outermost isosurfaces of the two major vortex rings in Fig.~\ref{fig:RBCcube-3d}, along with streamlines of 10 randomly selected points in each  surface. Trajectories of the forward simulations of the \emph{frozen} vector field are plotted in black lines and shown in the left panel. The blue lines in the right panel denote streamlines of the unsteady flow started from the same initial points. Here the forward simulation time is 20, since $|\mathbf{u}|\sim\mathcal{O}(0.01)$.}
\label{fig:RBCcube-3d-streamline}
\end{figure*}

Repeating the procedure used to produce Fig.~\ref{fig:RBC-3d-streamline-steady}, we obtain results shown in Fig.~\ref{fig:RBCcube-3d-streamline}. The obtained pseudo-streamlines stay close to the extracted approximate streamsurfaces, and the streamlines of the unsteady flow field also stay around the approximate streamsurfaces, indicating the persistence of the vortex rings. 


\subsection{Momentum transport barriers}
Next we compute barriers to the transport of active vector fields as defined in~\cite{haller2020objective}. In particular, momentum transport barriers of a velocity field $\mathbf{u}(\mathbf{x},t)$ at time $t$ can be identified as streamsurfaces of the barrier equation 
\begin{equation}
    \mathbf{x}'(s)=\Delta\mathbf{u}\left(\mathbf{x}(s),t\right)
\end{equation}
where $s$ denotes a parameterization of streamlines forming the streamsurfaces. We will apply our FEM-based approach to the extract approximate streamsurfaces of the barrier field.

The flow near the top and bottom plates is contained in thin boundary layers. As a result $\Delta\mathbf{u}$ within the boundary layers has much larger magnitude than outside the boundary layers. Consequently, the solution $H$ will also exhibit boundary layers: $H$ is nearly constant outside the boundary layers given $\Delta\mathbf{u}$ is negligible, while $\nabla H$ is orthogonal to $\Delta\mathbf{u}$ inside the boundary layers. We are mainly interested in vortical structures outside the boundary layers because those boundary layers are very thin. To extract vortical structures outside boundary layers, we normalize the active velocity field as~\cite{aksamit2022objective}
\begin{equation}
    \mathbf{x}'(s)=\frac{\Delta\mathbf{u}\left(\mathbf{x}(s),t\right)}{|\Delta\mathbf{u}\left(\mathbf{x}(s),t\right)|}.
\end{equation}

\subsubsection{Quasi-two-dimensional flow}
\begin{figure*}[!ht]
\centering
\includegraphics[width=0.32\textwidth]{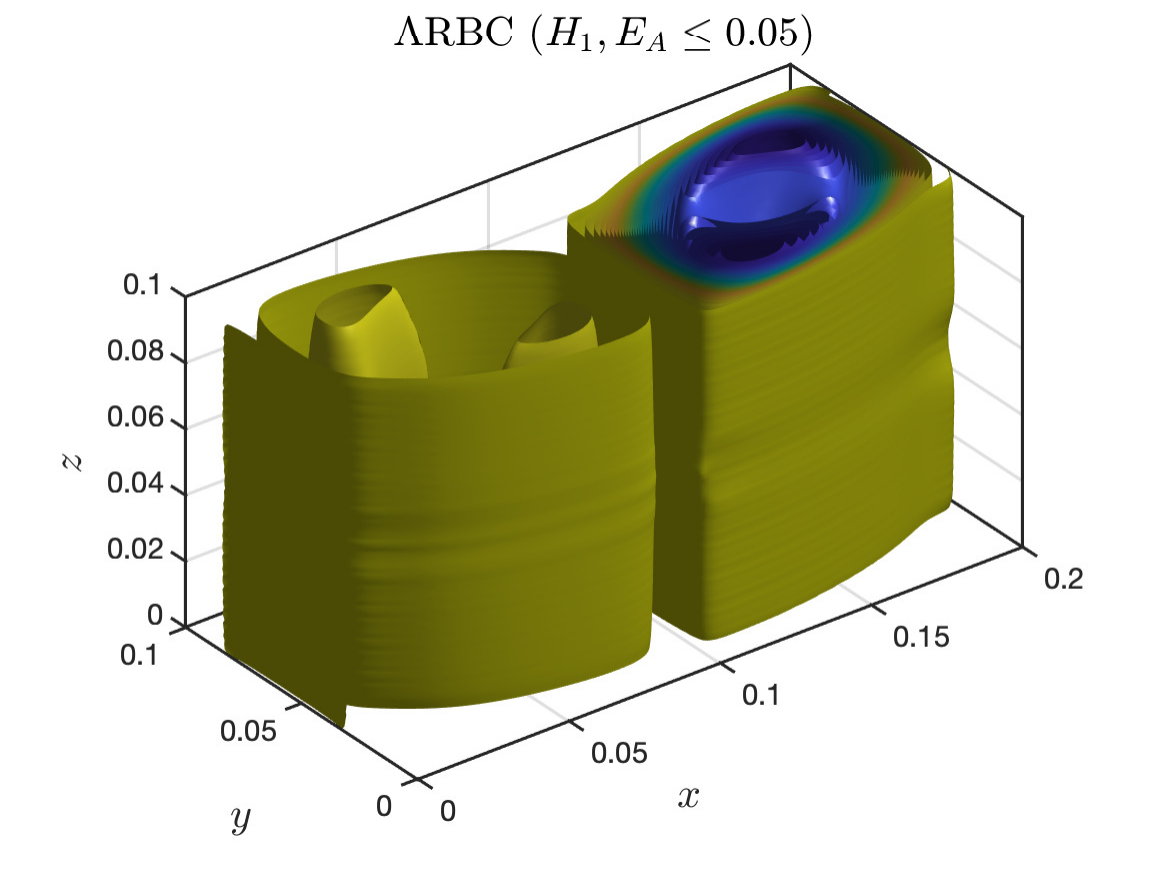}
\includegraphics[width=0.32\textwidth]{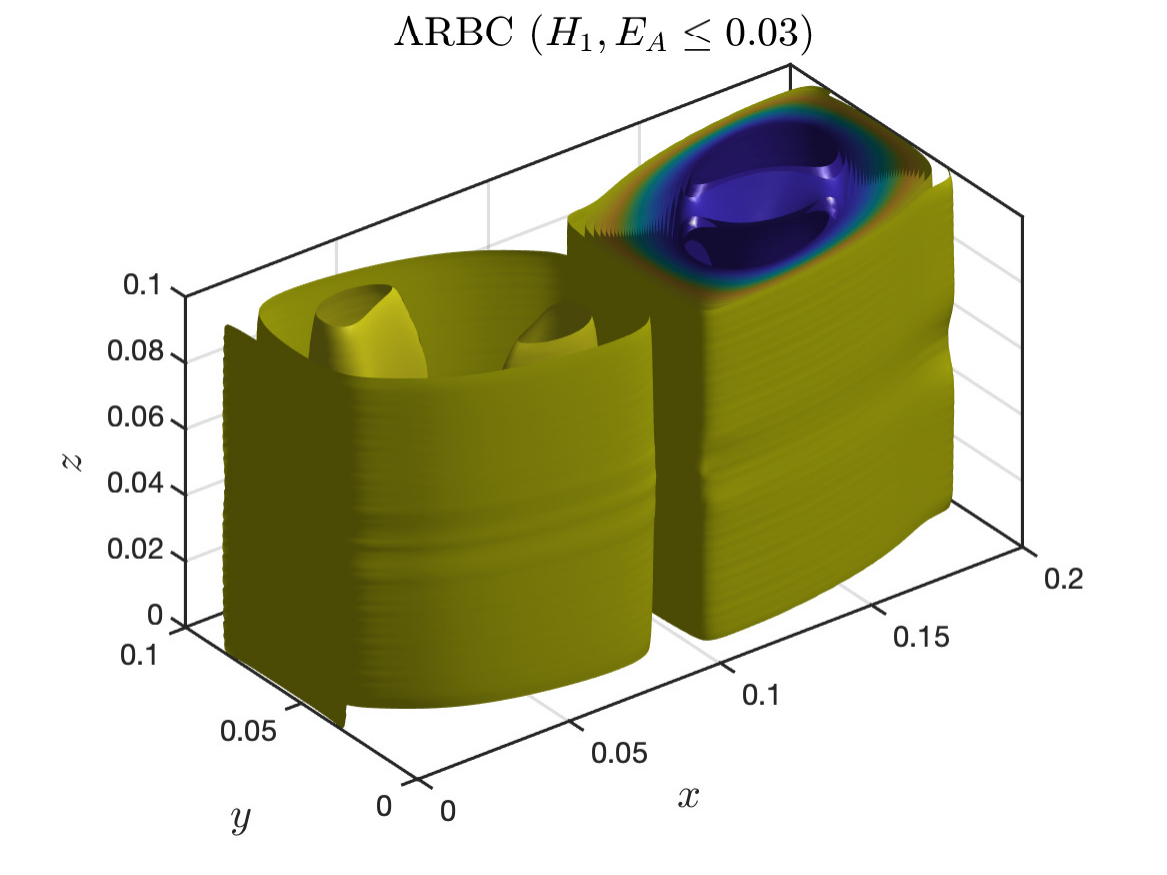}
\includegraphics[width=0.32\textwidth]{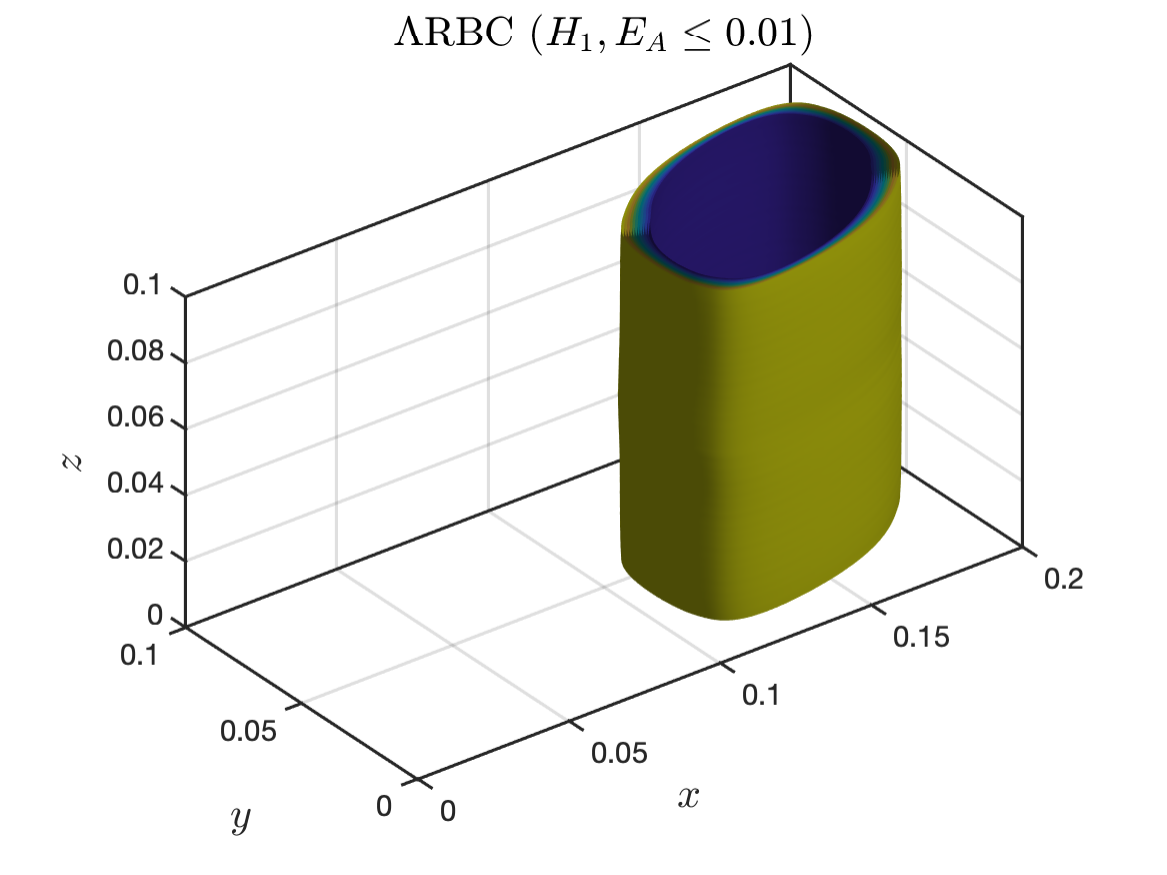}\\
\includegraphics[width=0.32\textwidth]{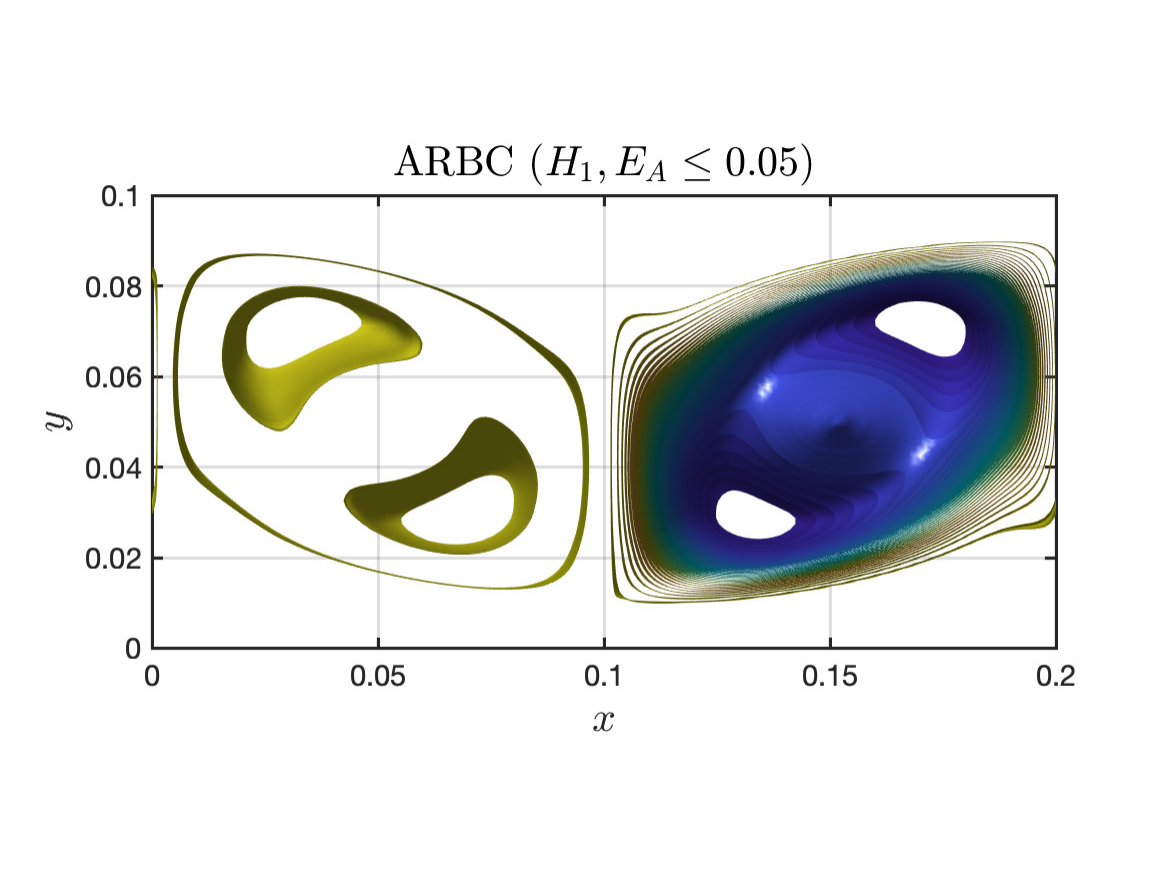}
\includegraphics[width=0.32\textwidth]{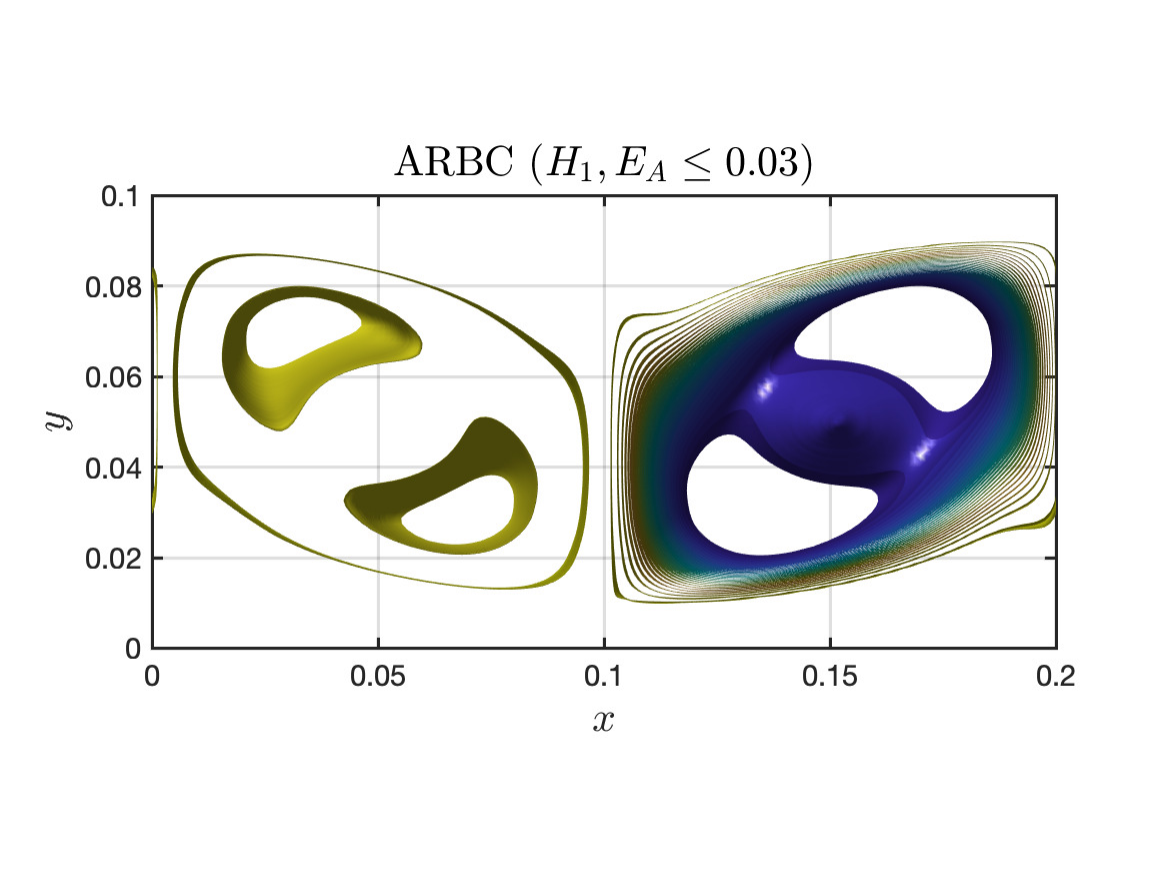}
\includegraphics[width=0.32\textwidth]{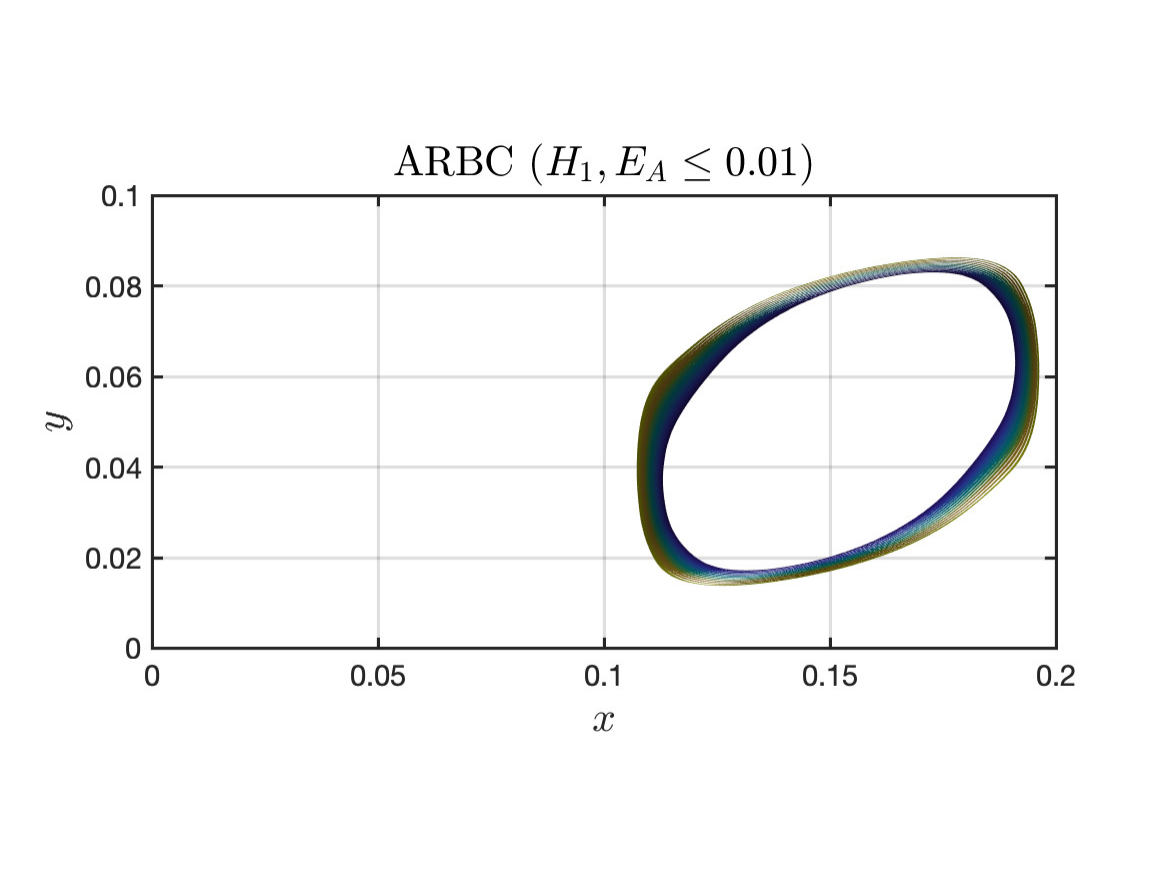}\\
\includegraphics[width=0.32\textwidth]{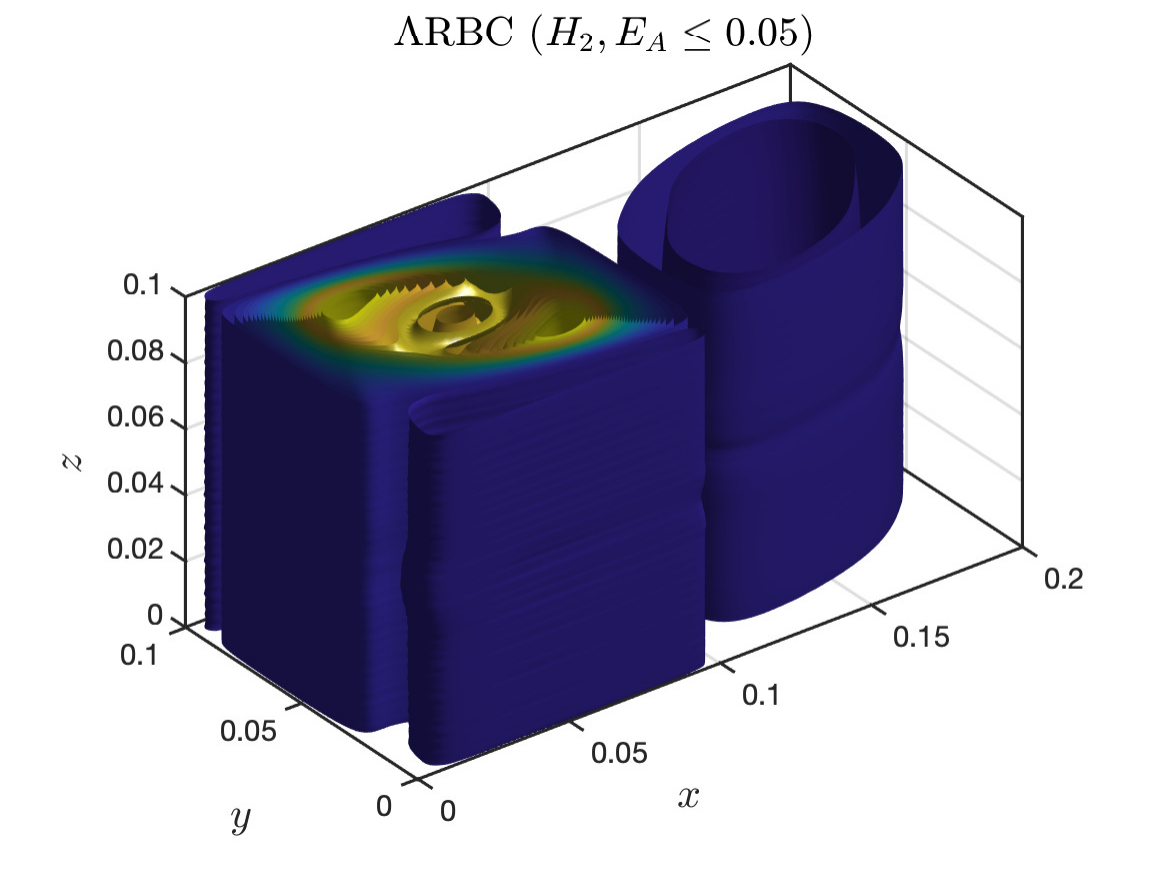}
\includegraphics[width=0.32\textwidth]{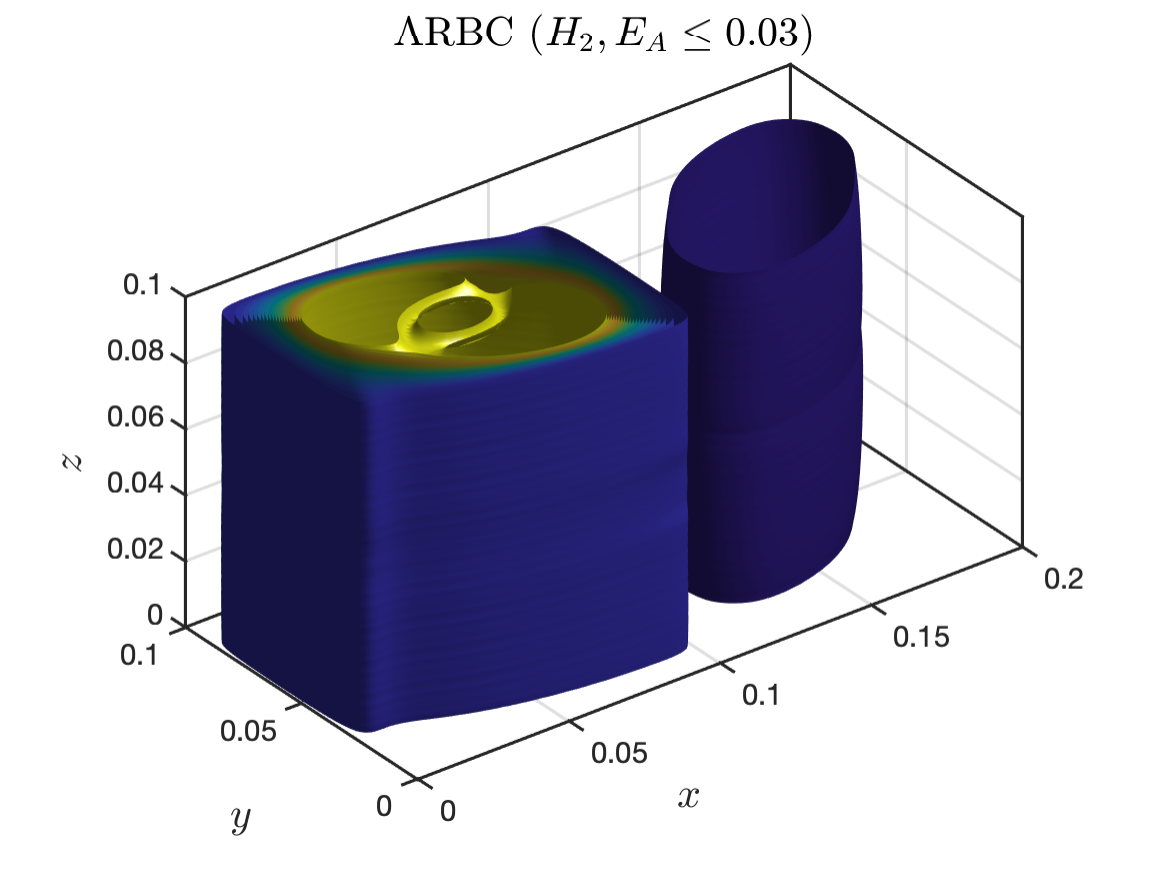}
\includegraphics[width=0.32\textwidth]{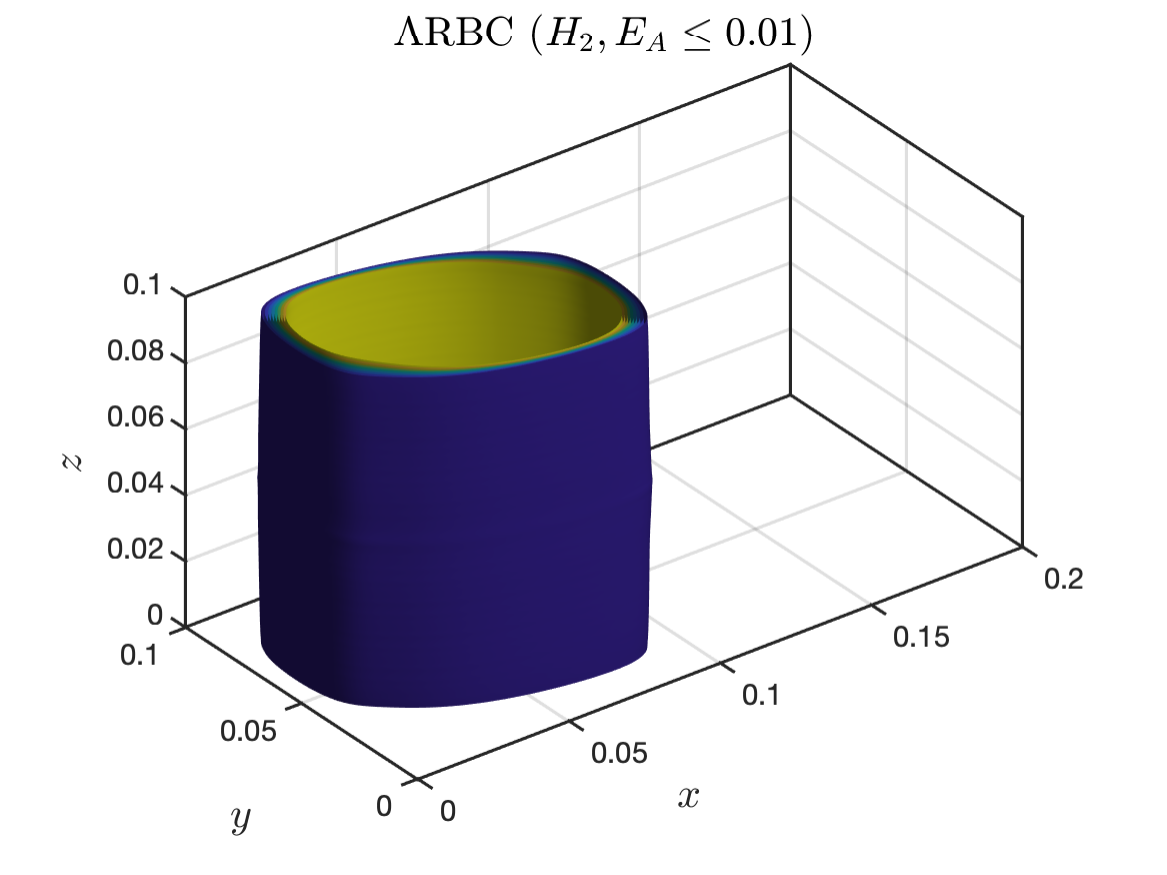}\\
\includegraphics[width=0.32\textwidth]{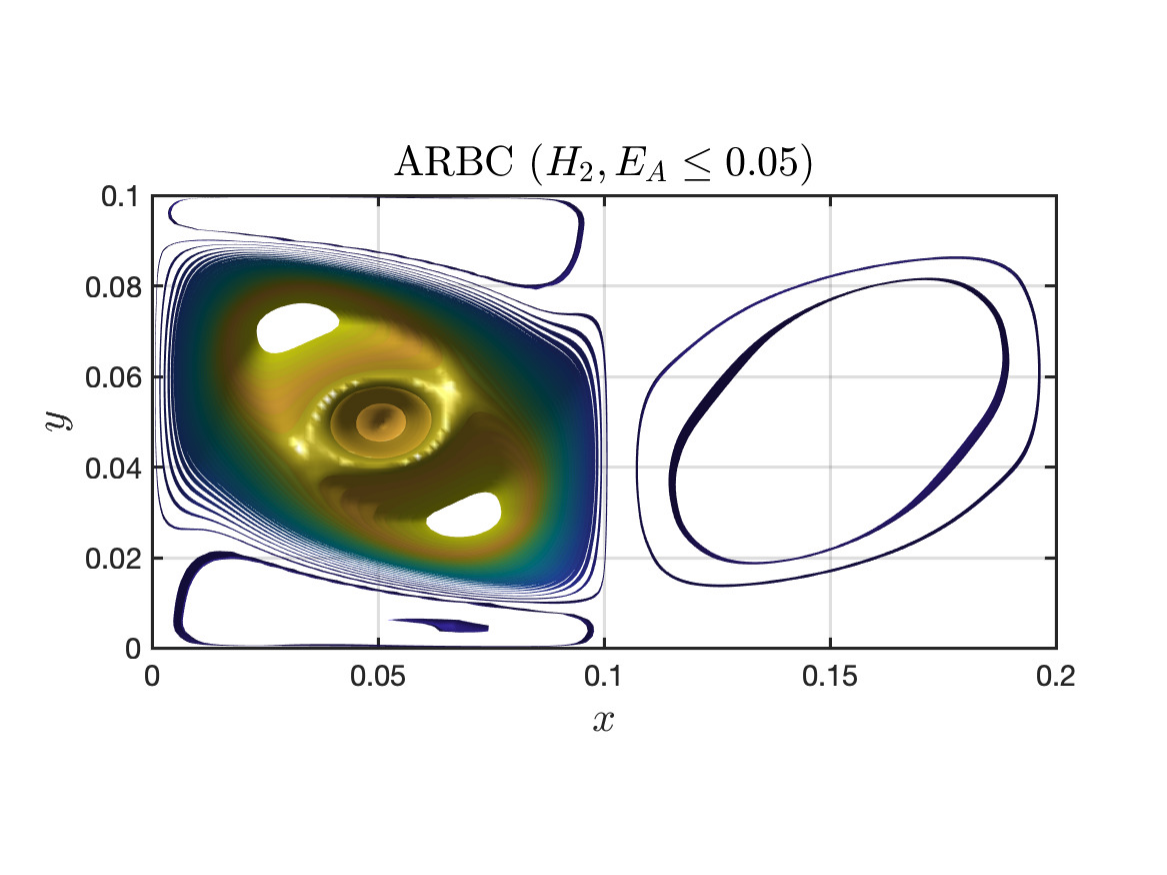}
\includegraphics[width=0.32\textwidth]{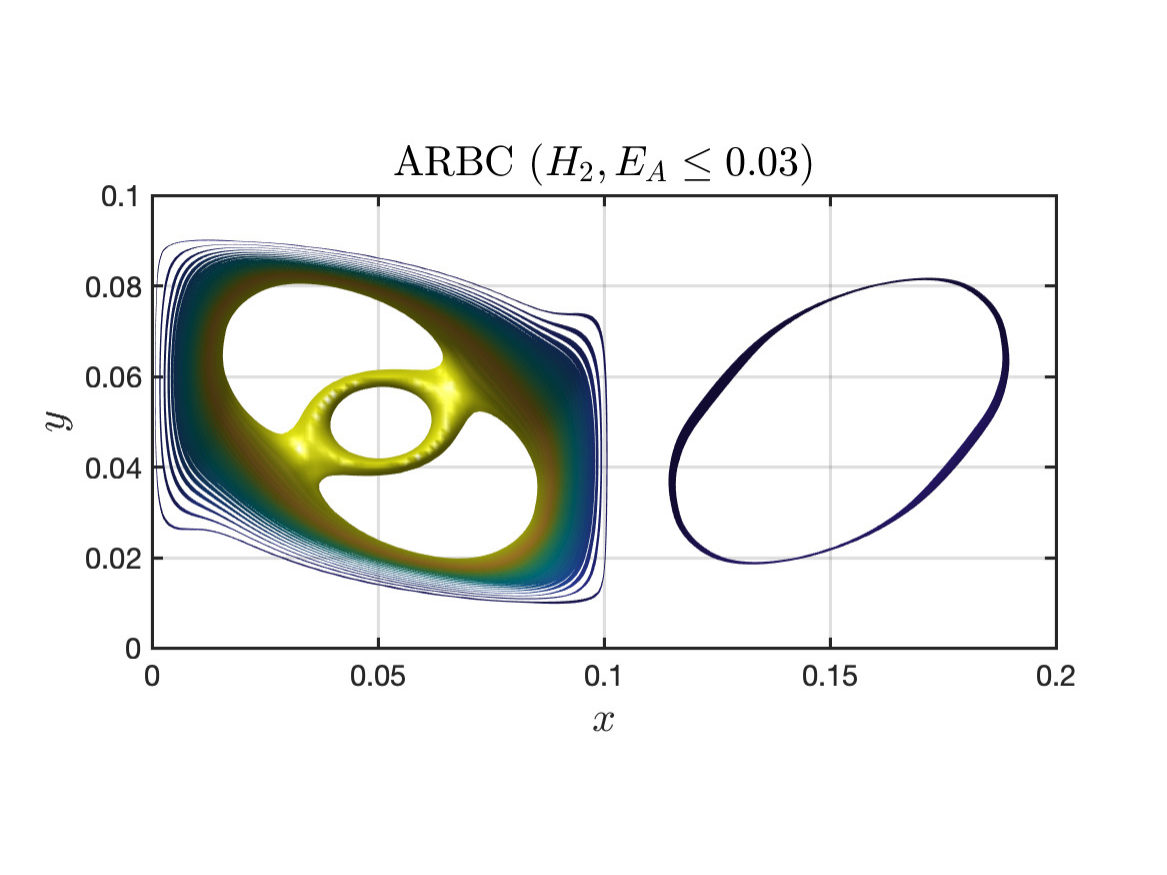}
\includegraphics[width=0.32\textwidth]{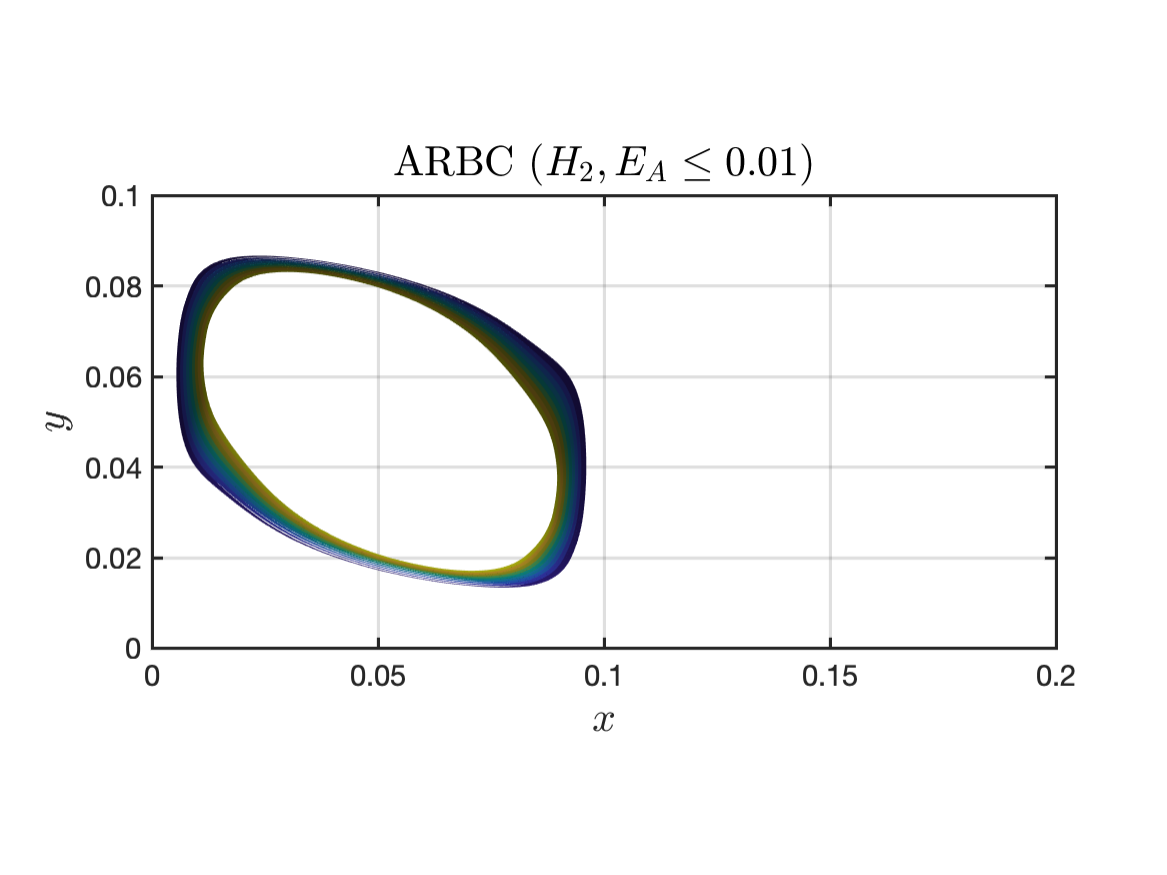}
\caption{Barriers to momentum transport in the RBC flow at $Ra=5\times10^4$ shown as isosurfaces of the approximate first integral. Here the results are based on various filter thresholds applied to $H_1$ (the first two rows) and $H_2$ (the last two rows). The panels in the second/fourth row are the projection of the panels in the first/third row onto $(x,y)$ plane.}
\label{fig:ARBC-5e4-H1}
\end{figure*}

We first consider the steady quasi-two-dimensional flow discussed in Sect~\ref{sec:quasi-2d}. We compute the Laplacian $\Delta\mathbf{u}$ at grid points using second-order finite difference~\cite{findiff}. With $N_x=60$, $N_y=N_z=30$ and quadratic interpolation, we obtain $\lambda_1=0.3420$ and $\lambda_2=0.5806$. With the filter~\eqref{eq:EA} applied, we obtain the isosurfaces for $H_1$ in Fig.~\ref{fig:ARBC-5e4-H1} (the first two rows), from which we see that the extracted barriers consist of a tube in the right half $(x\geq0.1)$ of the domain. We expect that there is another tube in the left half $(x\leq0.1)$ of the domain. Indeed, the isosurfaces for $H_2$ reveal the other tube, as shown in the last two rows in Fig.~\ref{fig:ARBC-5e4-H1}.

\begin{figure*}[!ht]
\centering
\includegraphics[width=0.245\textwidth]{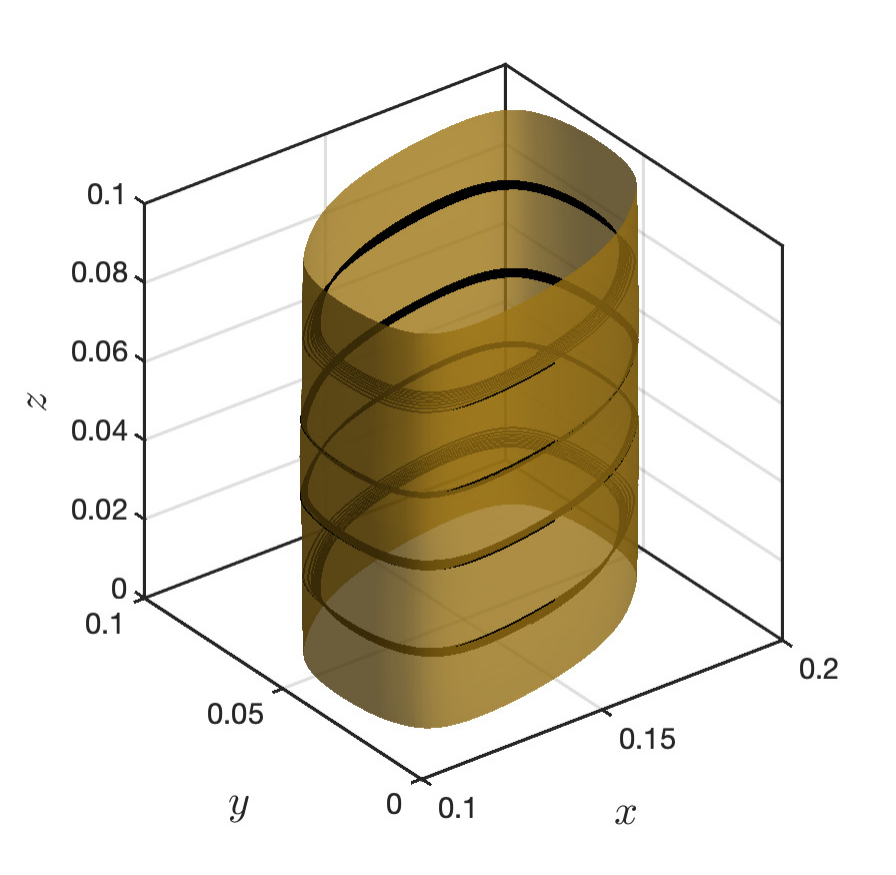}
\includegraphics[width=0.245\textwidth]{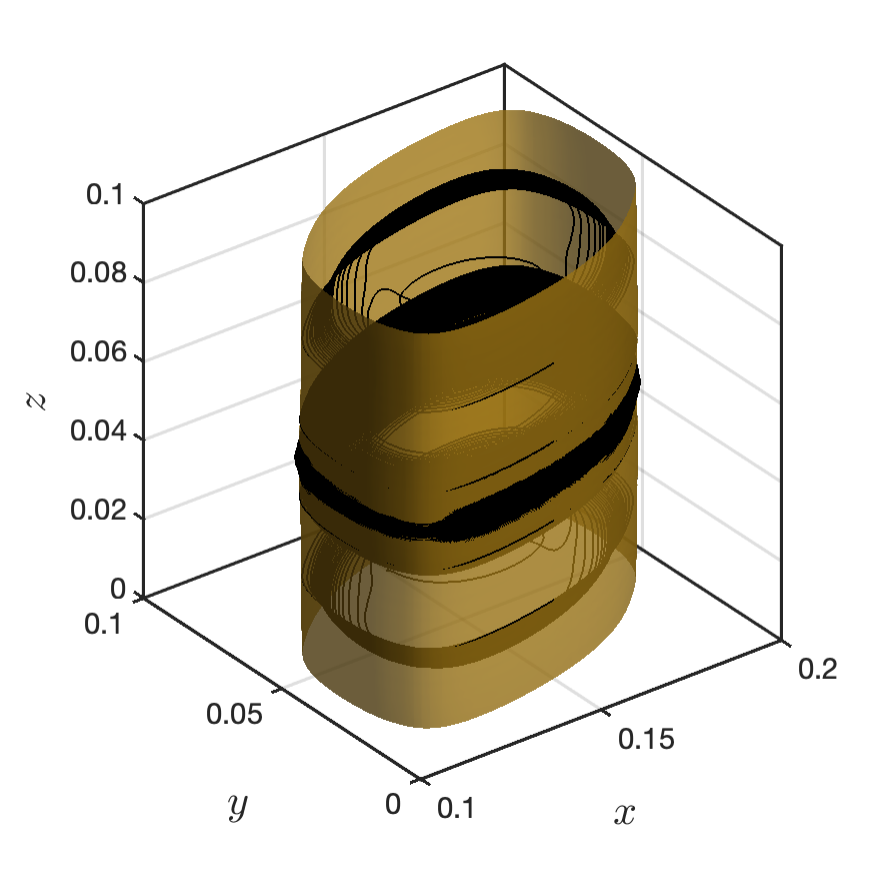}
\includegraphics[width=0.245\textwidth]{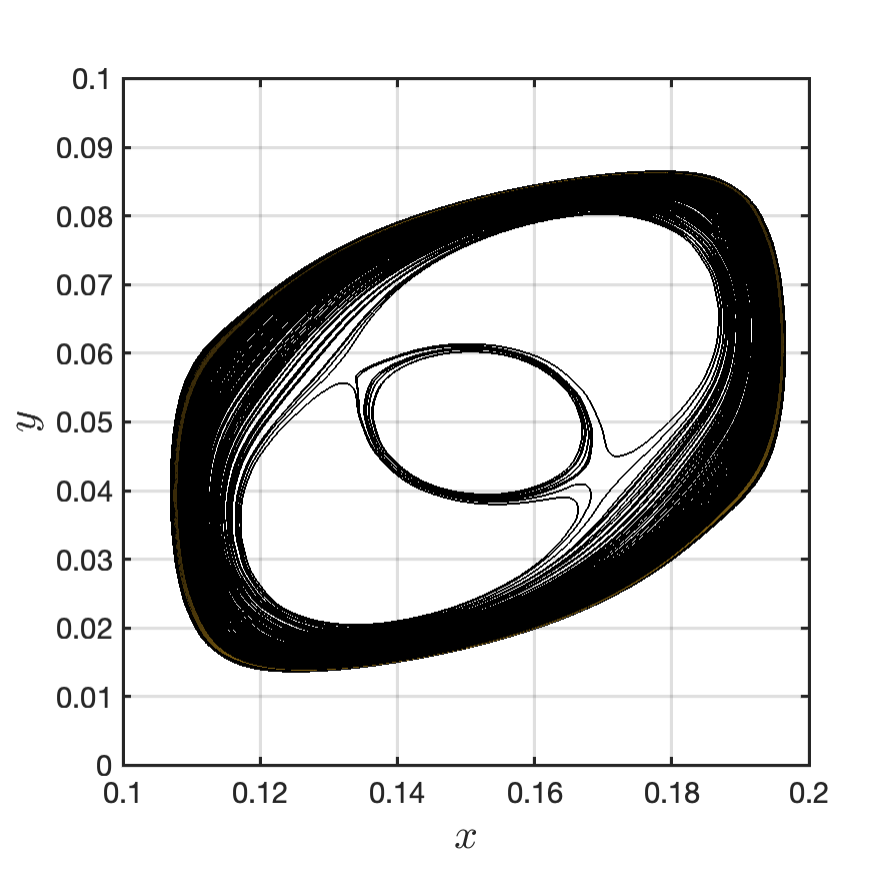}
\includegraphics[width=0.245\textwidth]{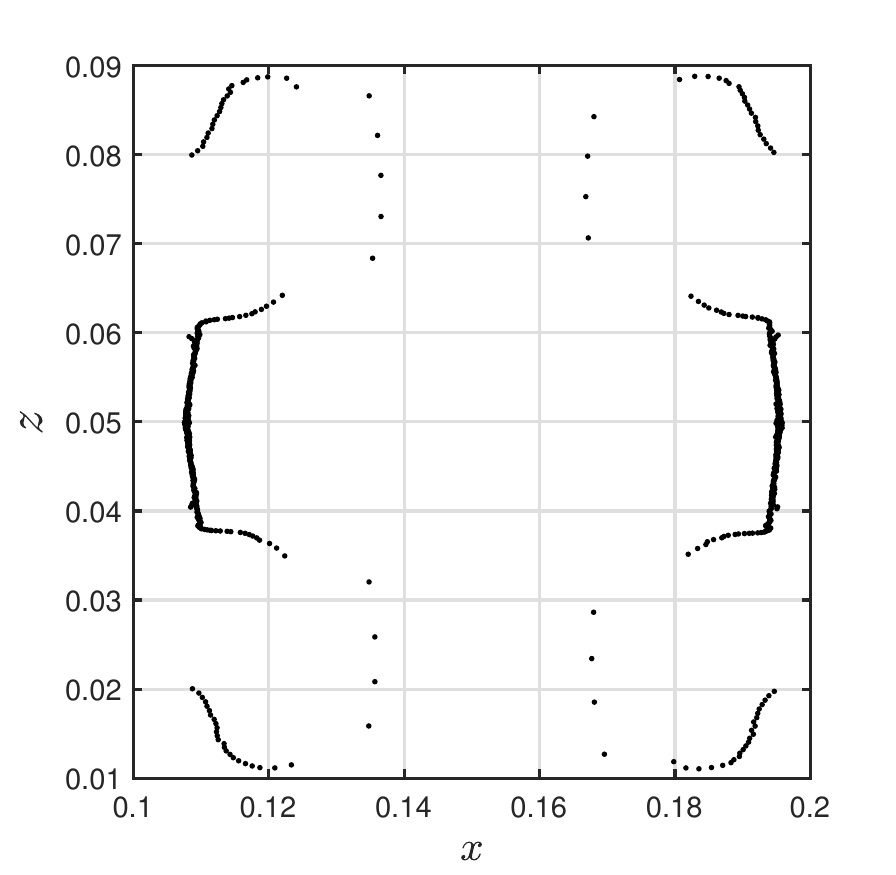}\\
\includegraphics[width=0.245\textwidth]{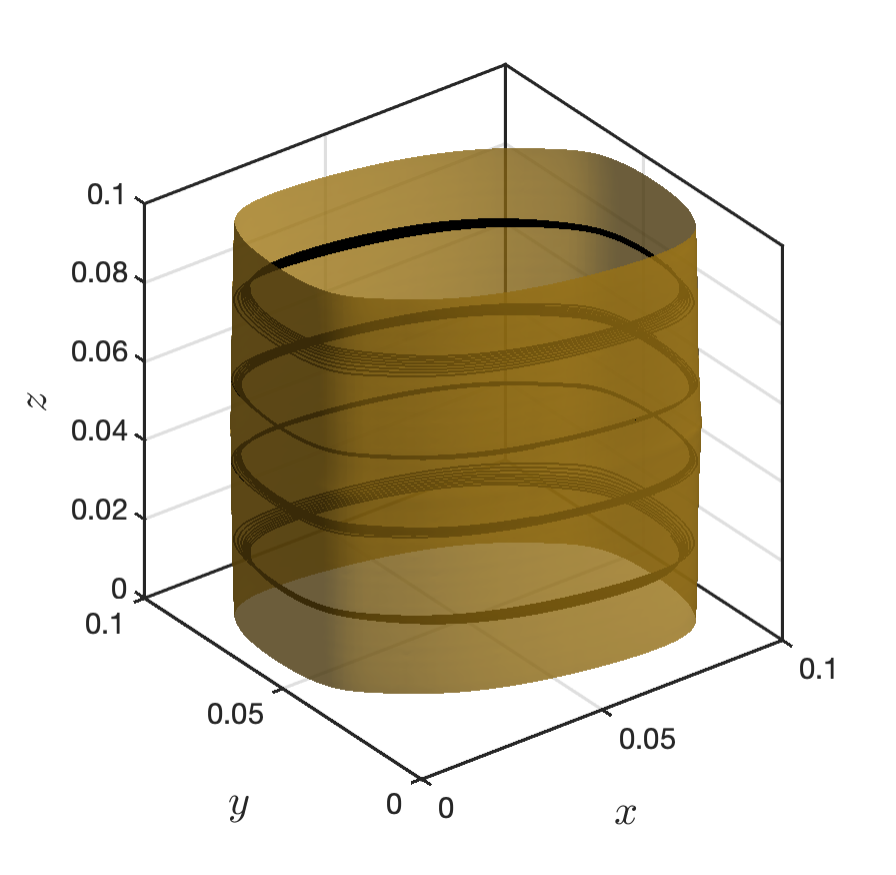}
\includegraphics[width=0.245\textwidth]{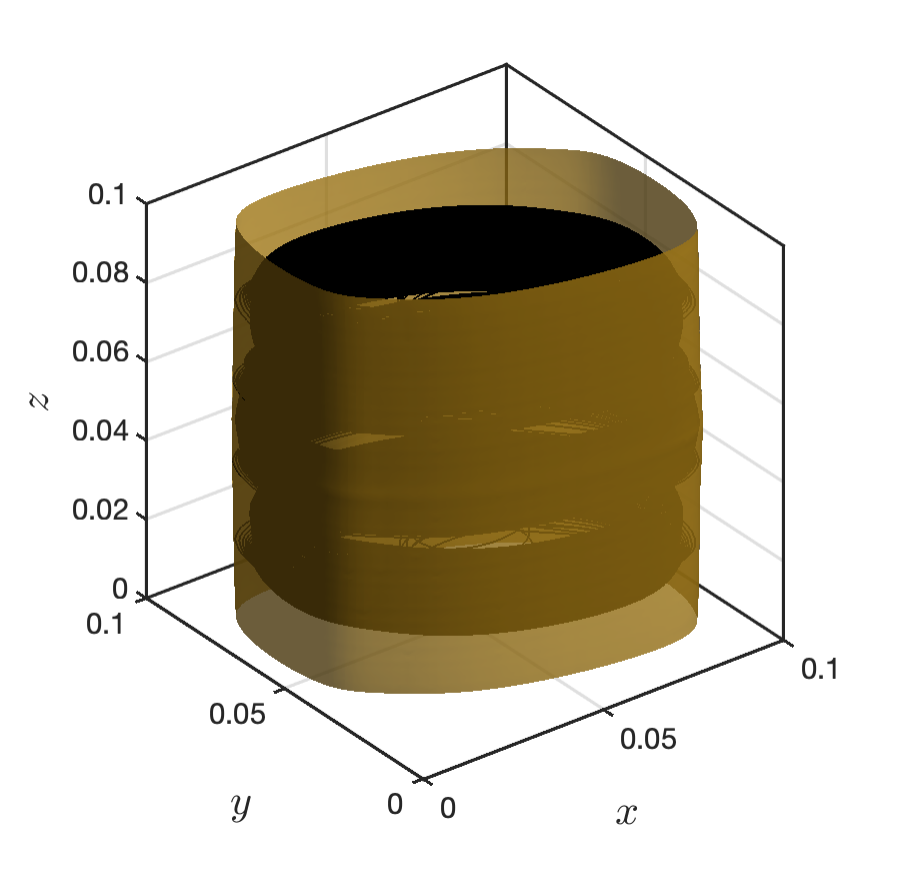}
\includegraphics[width=0.245\textwidth]{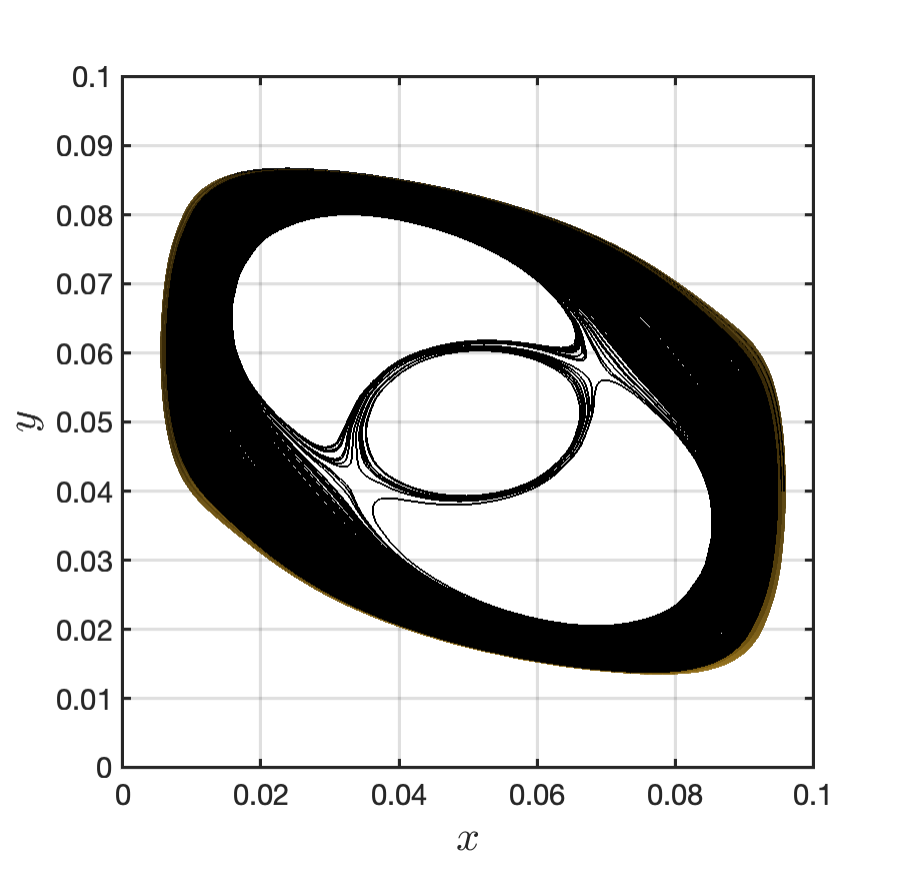}
\includegraphics[width=0.245\textwidth]{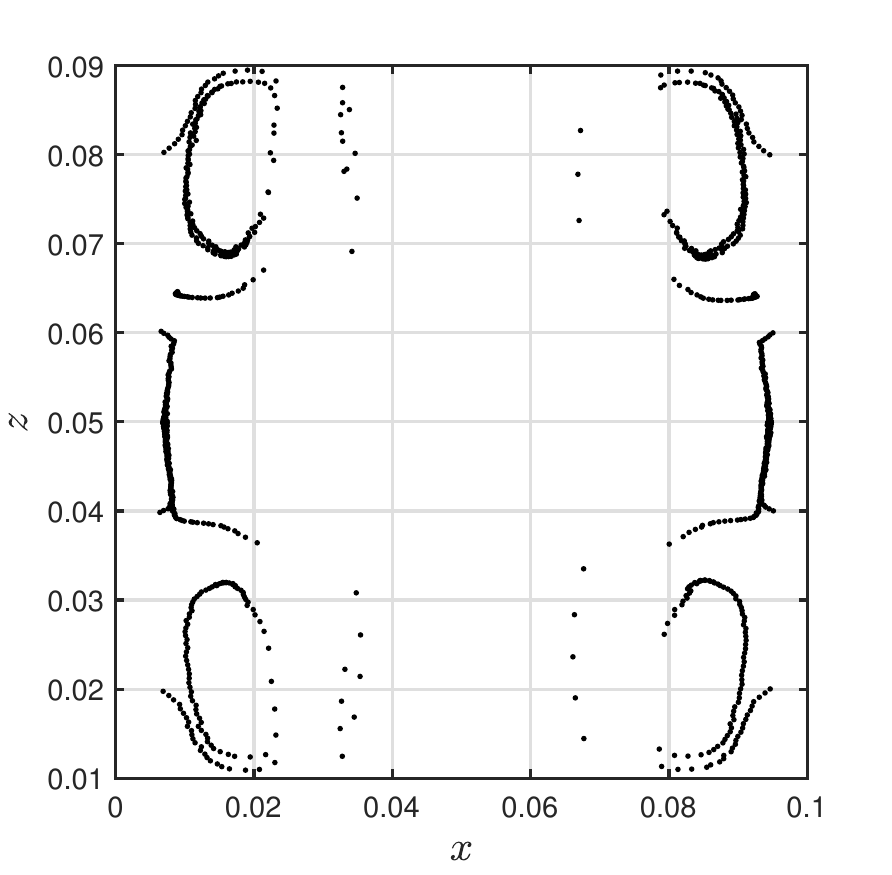}
\caption{Outermost isosurfaces of the two major vortex tubes in the right panel of Fig.~\ref{fig:RBC-1e5-isosurfaceH}, along with streamlines of 4 selected points in each of the surface. Trajectories from the forward simulations of active vector field are plotted in black lines. The integration time for the first and second columns is 2 and 50. The third column gives the projection of the panels in the middle plane onto $(x,y)$ plane, while the last column gives the intersection points of the simulated trajectories with the Poincar\'e section $y=0.05$.}
\label{fig:ARBC-5e4-streamline}
\end{figure*}

Next we launch streamlines on the outermost layers of the two tubes in the last column of Fig.~\ref{fig:ARBC-5e4-H1}. When the integration time is not too long, the obtained trajectories stay close to the extracted approximate streamsurfaces, as seen in the left column of Fig.~\ref{fig:ARBC-5e4-streamline}. However, given the streamsurfaces are not necessarily attracting, these trajectories may drift far away from the surfaces for longer time integration, as seen in the second and third columns of Fig.~\ref{fig:ARBC-5e4-streamline}, where panels in the third column are the projections of panels in the middle column onto the $(x,y)$ plane. We note that the drifted flow is nearly contained in the extracted barriers. Interestingly, the two panels in the third column are similar to the two projected plots in the middle column of Fig.~\ref{fig:ARBC-5e4-H1}. This again validates the obtained results. To further identify structures of the simulated trajectories, we present the intersection points of these trajectories along with a Poincar\'e section $y=0.05$. As seen in the last column, there exists invariant tori inside the vortex tube for $x\leq0.1$.

\subsubsection{Three-dimensional flow}
\begin{figure*}[!ht]
\centering
\includegraphics[width=0.45\textwidth]{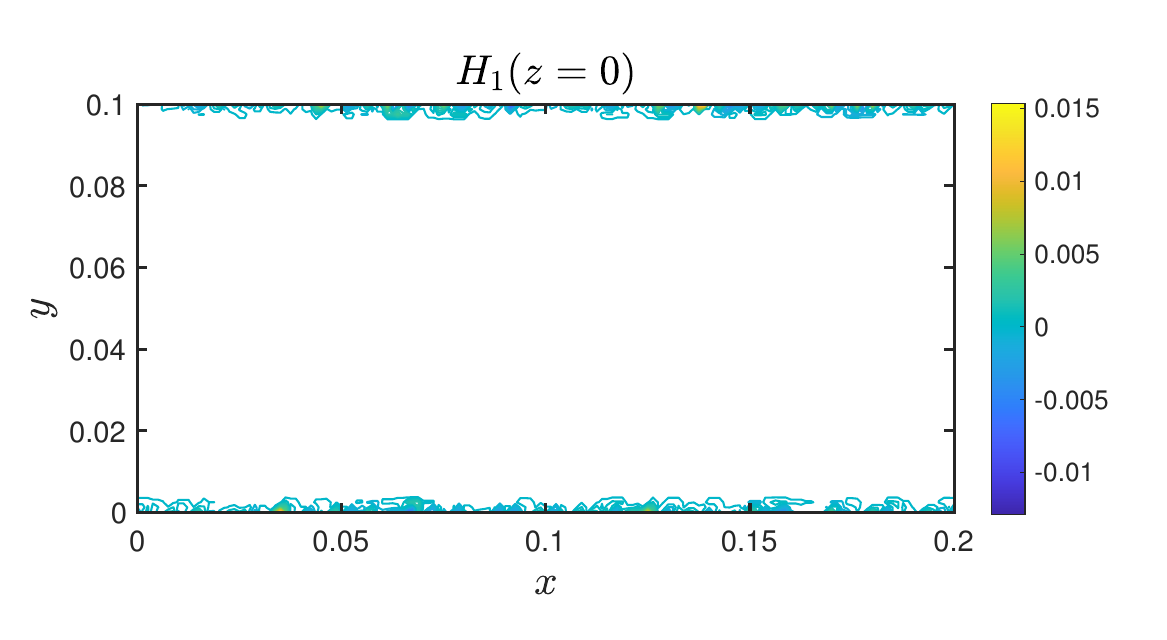}
\includegraphics[width=0.45\textwidth]{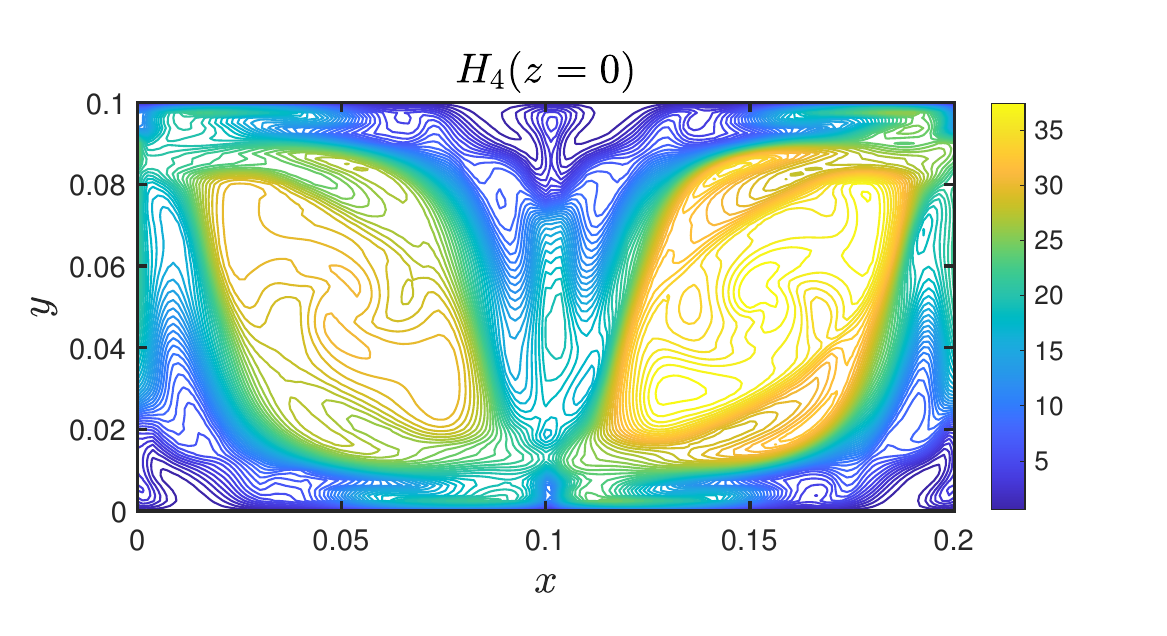}
\caption{Contour plots of $H_1$ and $H_4$ of the active vector field of a snapshot of the RBC flow with $Ra=1\times10^5$ at cross section $z=0$ (see Fig.~\ref{fig:RBC-1e5} for the snapshot).}
\label{fig:ARBC-1e5-2d}
\end{figure*}

Now we extract momentum barriers of the flow snapshot shown in Fig.~\ref{fig:RBC-1e5-isosurfaceH}. This is a snapshot of an unsteady three-dimensional flow. With $N_x=60$, $N_y=N_z=30$, we obtain $\lambda_1=\lambda_2=\lambda_3=1$ and $\lambda_4=19.2$. The first three modes correspond to boundary layer modes while the last one gives structures outside the boundary layer. As an illustration of the boundary layer modes, we present the contour plot of $H_1$ at the cross section $z=0$ in the left panel of Fig.~\ref{fig:ARBC-1e5-2d}. We see from the left panel that $H_1$ is barely changing outside the boundary layers. In contrast, the contour plot of $H_4$ at $z=0$ in the right panel of the figure reveals structures outside the boundary layers. So we should look for $H_4$. Note that $\lambda_4$ is large, which indicates that the barrier field does not admit any globally defined first integral. However, we can still apply the filter~\eqref{eq:EA} to extract approximate streamsurfaces.

\begin{figure*}[!ht]
\centering
\includegraphics[width=0.32\textwidth]{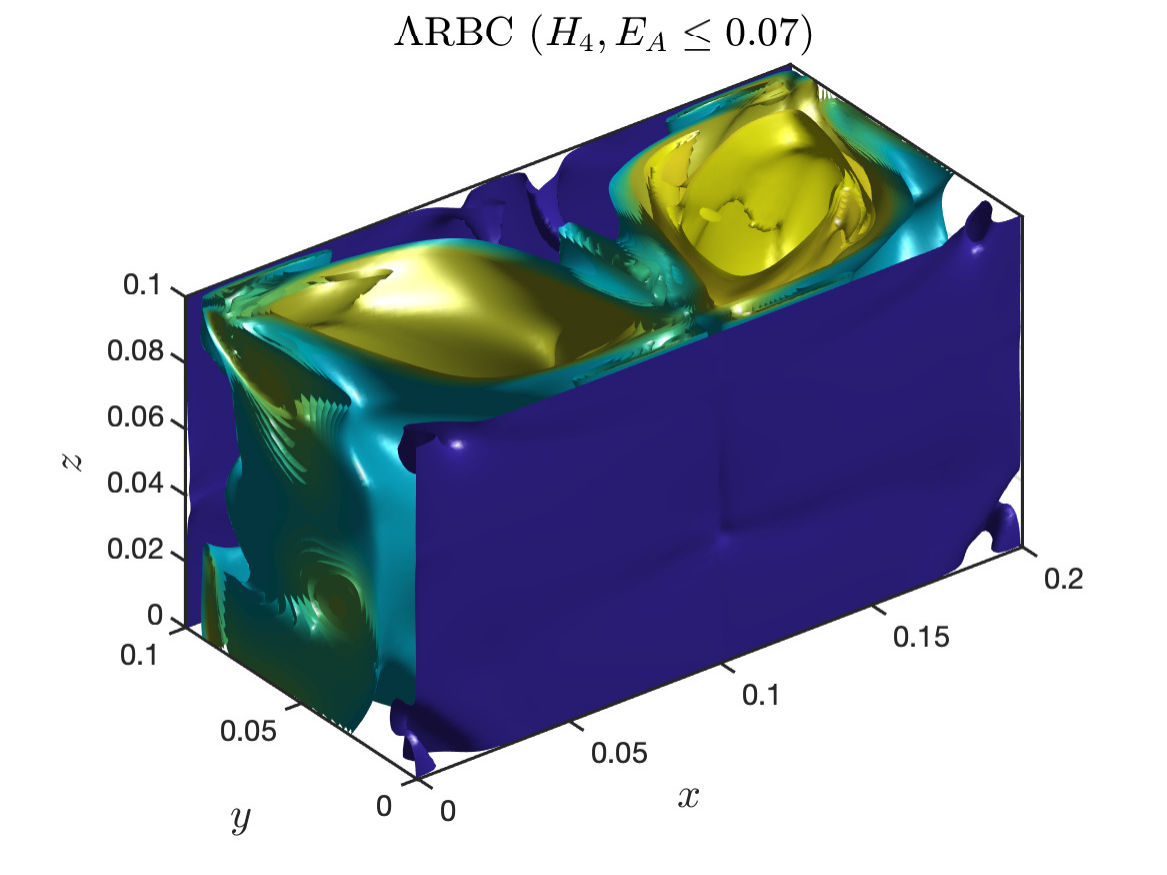}
\includegraphics[width=0.32\textwidth]{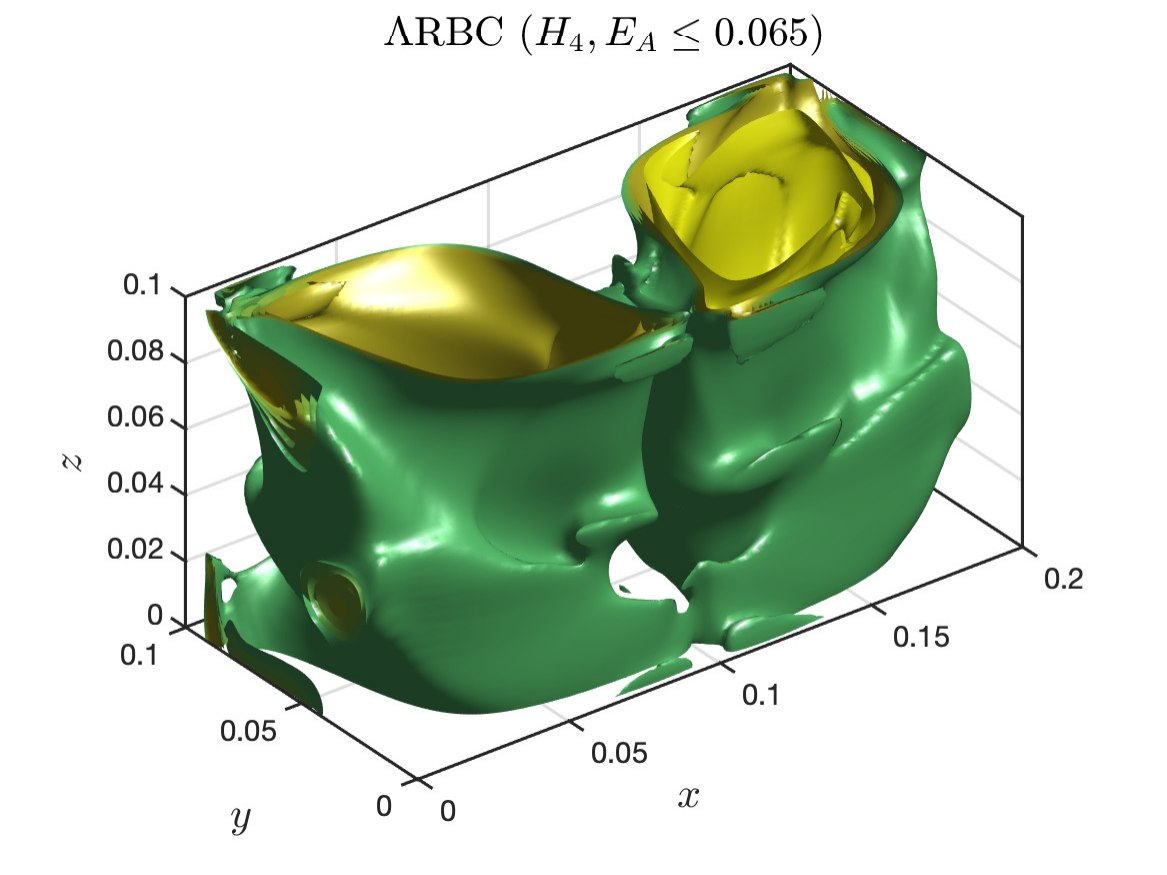}
\includegraphics[width=0.32\textwidth]{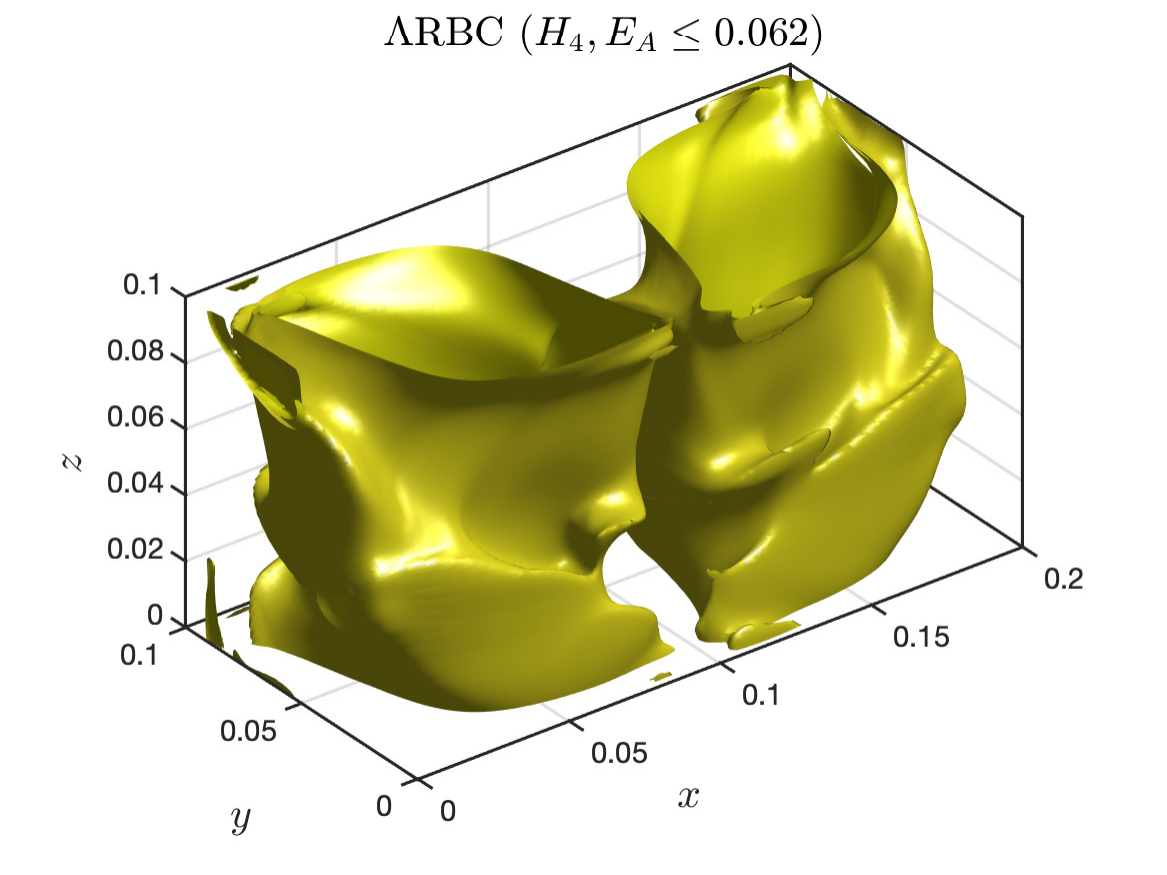}\\
\includegraphics[width=0.32\textwidth]{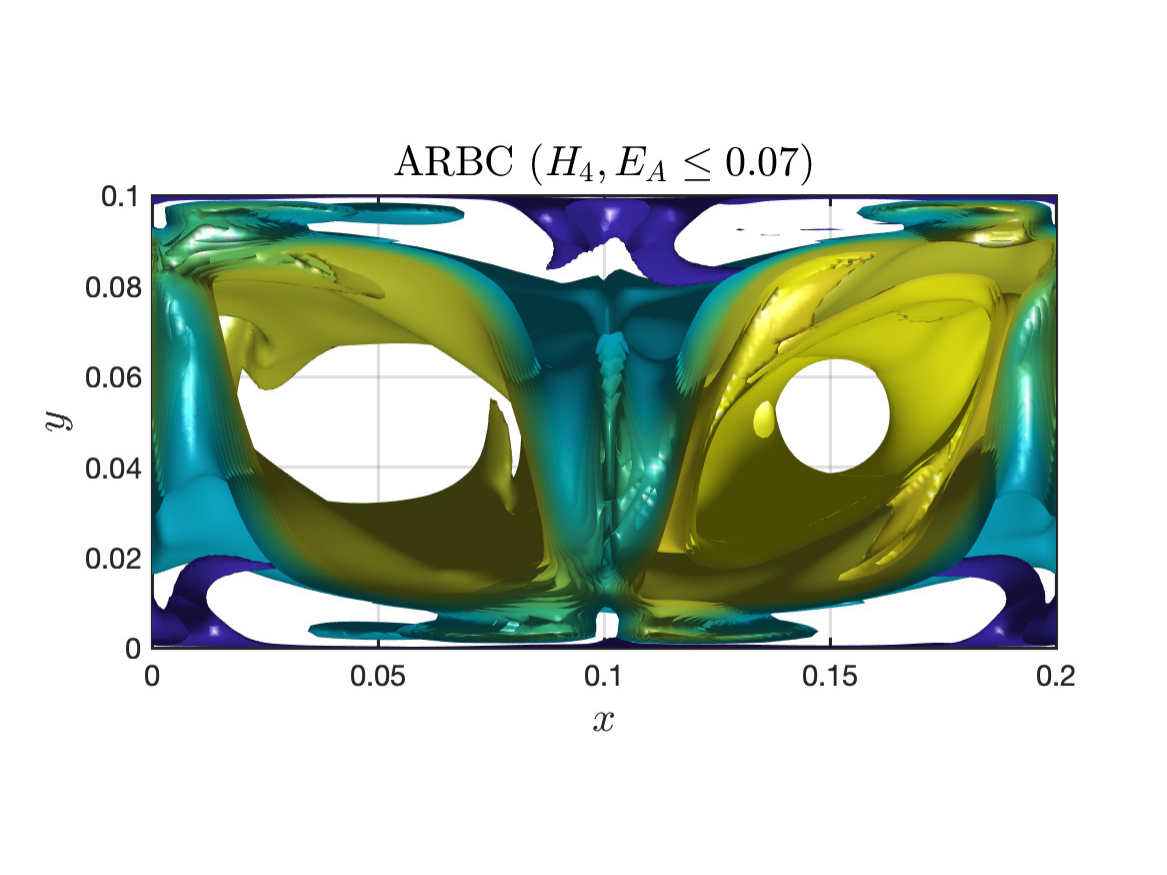}
\includegraphics[width=0.32\textwidth]{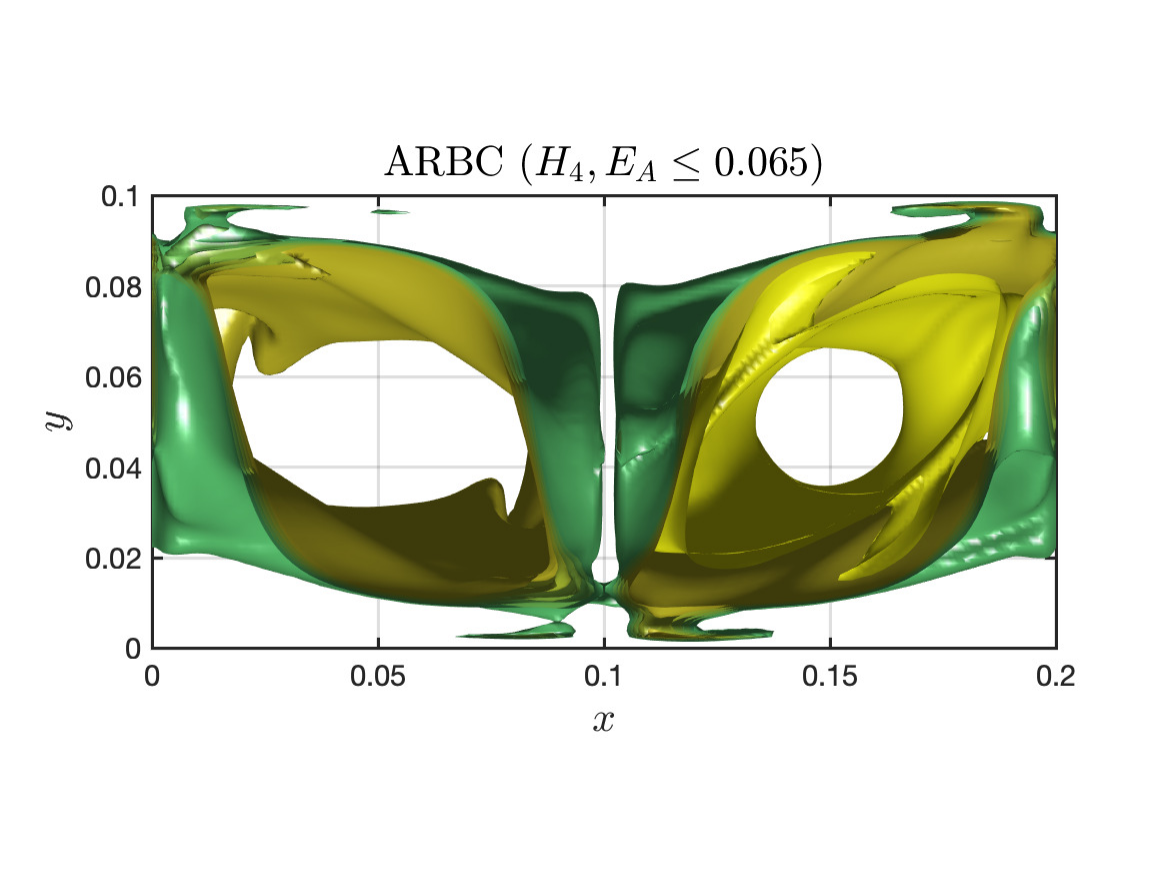}
\includegraphics[width=0.32\textwidth]{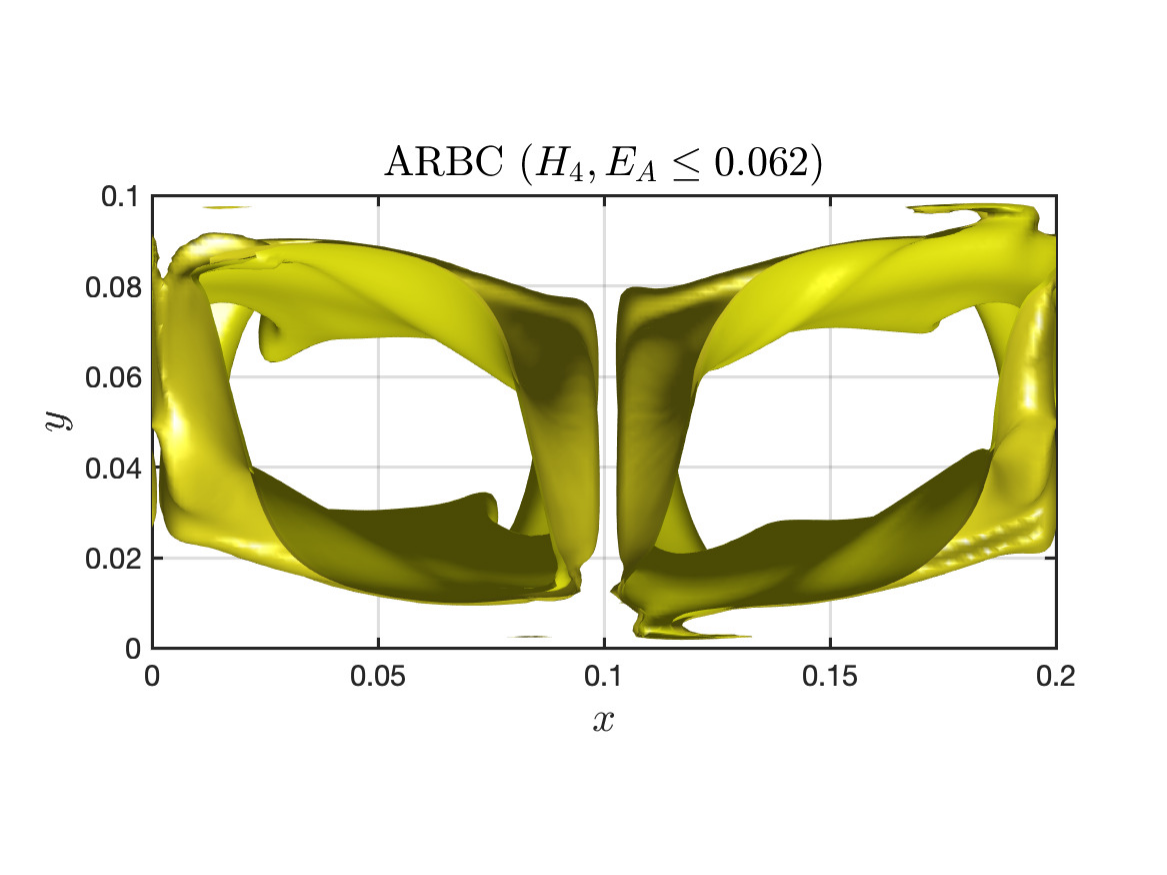}
\caption{Isosurfaces of $H_4$ for the active vector field of a snapshot of the RBC flow with $Ra=1\times10^5$ (see the right panel of Fig.~\ref{fig:ARBC-1e5-2d}) with various filter thresholds: $E_A\leq0.07$ (left panels), $E_A\leq0.065$ (middle panels), and $E_A\leq0.062$ (right panels). The lower panels are the projections of the upper panels onto $(x,y)$ plane.}
\label{fig:ARBCcube-3d}
\end{figure*}

\begin{figure*}[!ht]
\centering
\includegraphics[width=0.45\textwidth]{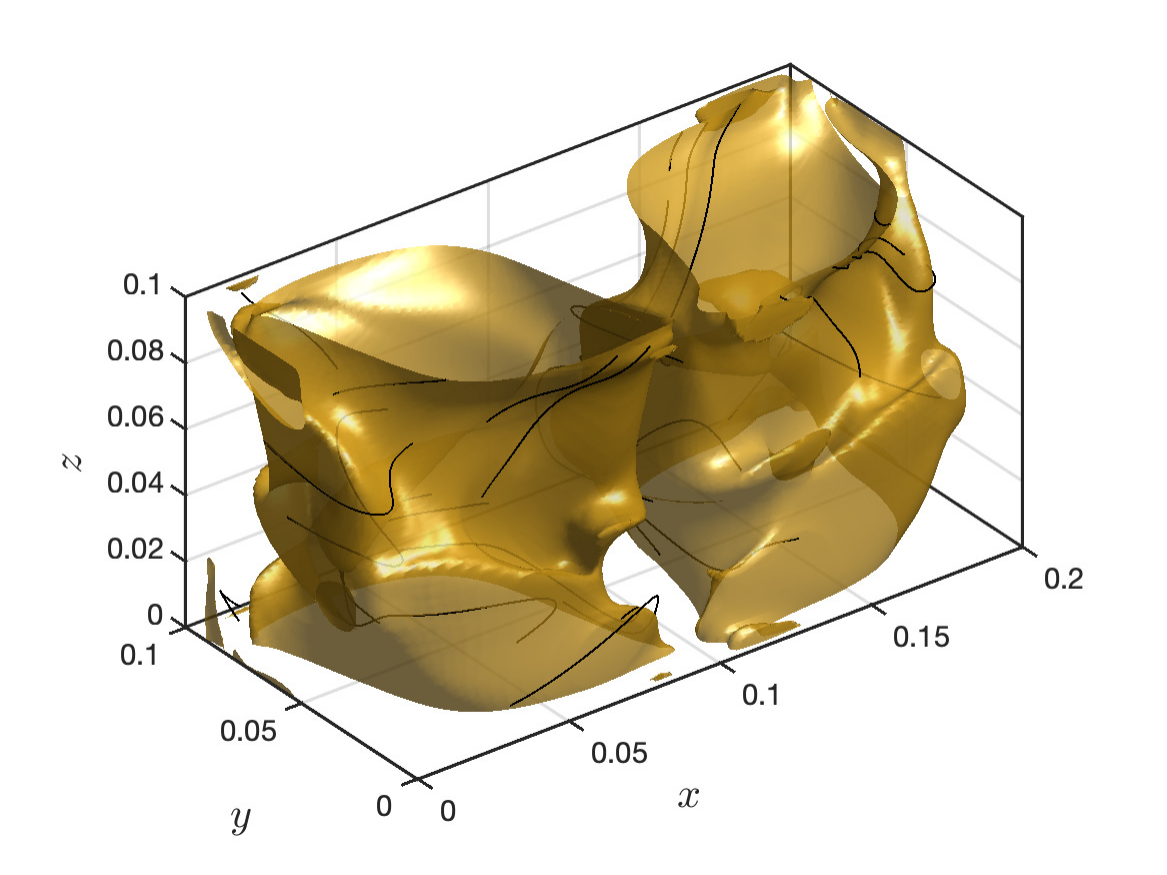}
\caption{Outermost isosurfaces of the two tubes in the right panel of Fig.~\ref{fig:ARBCcube-3d}, along with streamlines initialized from five randomly selected points on the surfaces.}
\label{fig:ARBC-1e5-streamline}
\end{figure*}

Isosurfaces of $H_4$ with different filter thresholds are plotted in Fig.~\ref{fig:ARBCcube-3d}. By decreasing the threshold properly, we are able to extract two disconnected tubes shown in the right panel of the figure. To validate the obtained approximate streamsurfaces, we launch streamlines started from 5 randomly selected points on the outermost layers of each of the two tubes in the right panel of Fig.~\ref{fig:ARBCcube-3d}. Here we set the integration time to be 0.1 given the velocity magnitude is of order 1 while the characteristic length scale for the tubes is of order 0.1. As seen in Fig.~\ref{fig:ARBC-1e5-streamline}, the trajectories from numerical integration stay close to the extracted streamsurfaces.

\section{Conclusion}

We have established a variational method for the construction of tubular and toroidal streamsurfaces for 3D flow visualization. This method is an extension of the Fourier series expansion proposed in~\cite{stergios} from spatially periodic domains to general spatial domains. We have formulated an optimization problem seeking the closest first integrals. The isosurfaces of these closest first integrals give approximate streamsurfaces in vortical regions of 3D flows. We have derived the first-order necessary conditions to the optimal solution which  gave rise to an eigenvalue problem of a set of linear partial-differential equations. We have used finite-element methods to solve the eigenvalue problem.

We have demonstrated the effectiveness of the proposed variational construction through a suite of examples. We started from simple benchmark studies including spherical and cylindrical vortex flows as well as Taylor-Couette flow to illustrate that the finite-element based implementation can handle flows in domains with arbitrary geometries. We have also applied the method to periodic flows such as ABC flows and Euler flows to show this method also works well for periodic flows. Finally, we have considered Rayleigh-Bénard convection flows to demonstrate the effectiveness of the proposed method for more complicated flows.

We have used regular mesh grids in the computations of this study. It is instructive to implement an adaptive mesh to enhance the performance of our variational construction. In particular, we can use the distribution of invariance error to conduct the adaptive change of mesh. This adaptation could play an important role in extracting tubular and toroidal streamfurfaces in complicated 3D flows, especially for turbulent flows.

We have implemented our variational construction using FEniCS. However, the variational method proposed here is generic and can be implemented in other finite-element packages or more specialized codes. In particular, one can use advanced eigensolvers that support high-performance computing to speed up the computation of eigensolutions.

\section{Appendix}
\label{sec:comp-fea-fourier}

\textcolor{black}{We compare the performance of our finite-element implementation against the Fourier series approach~\cite{stergios} for the two periodic flows in Sect.~\ref{sec:periodicflows}. We use the error metric $E_m$ defined in~\eqref{eq:Em} to make comparisons. This metric $E_m$ gives the averaged normalized invariance error evaluated at a collection of grid points.}

\textcolor{black}{We recall that our finite-element implementation seeks the leading eigenvalue of a generalized eigenvalue problem. As seen in Sect.~\ref{sec:fenics}, the matrices $\mathbf{A}$ and $\mathbf{B}$ of the generalized eigenvalue problem are of size $N_\mathrm{dof}\times N_\mathrm{dof}$, where $N_\mathrm{dof}$ denotes the the number of degrees-of-freedom of the finite-element discretization. In the Fourier approach~\cite{stergios}, one seeks the leading singular value of a matrix $\mathbf{C}\in\mathbb{C}^{m\times N_\mathrm{mode}}$. Here $N_\mathrm{mode}$ is the number of Fourier modes and $m$ is the number of grid points.}

\textcolor{black}{We infer from the size and sparsity of the matrices $\mathbf{A}, \mathbf{B}, \mathbf{C}$ that the Fourier approach requires much more memory than that of our finite-element implementation. Indeed, the number of  nonzero entries of the matrices $\mathbf{A}$ and $\mathbf{B}$ is $dN_\mathrm{dof}$ because the two matrices are sparse. Here $d$ is the bandwidth of the two matrices. We found that $d\approx29$ when we use Lagrange elements of interpolation order two. In contrast, the number of entries of the full matrix $\mathbf{C}$ is $mN_\mathrm{mode}$. In~\cite{stergios}, $m=100^3$ was used and hence $N_\mathrm{mode}$ often was restricted to be less than $10^4$.}  \textcolor{black}{Indeed, we found that for $N_\mathrm{mode}=1.7\times10^4$, the memory required to compute the leading singular value has exceeded 200 GB.} \textcolor{black}{Since $m=100^3\gg d\approx29$, the finite-element method requires much less memory than the Fourier approach for the same degree of fidelity, i.e., when $N_\mathrm{mode}=N_\mathrm{dof}$.}

\begin{figure*}[!ht]
\centering
\includegraphics[width=0.45\textwidth]{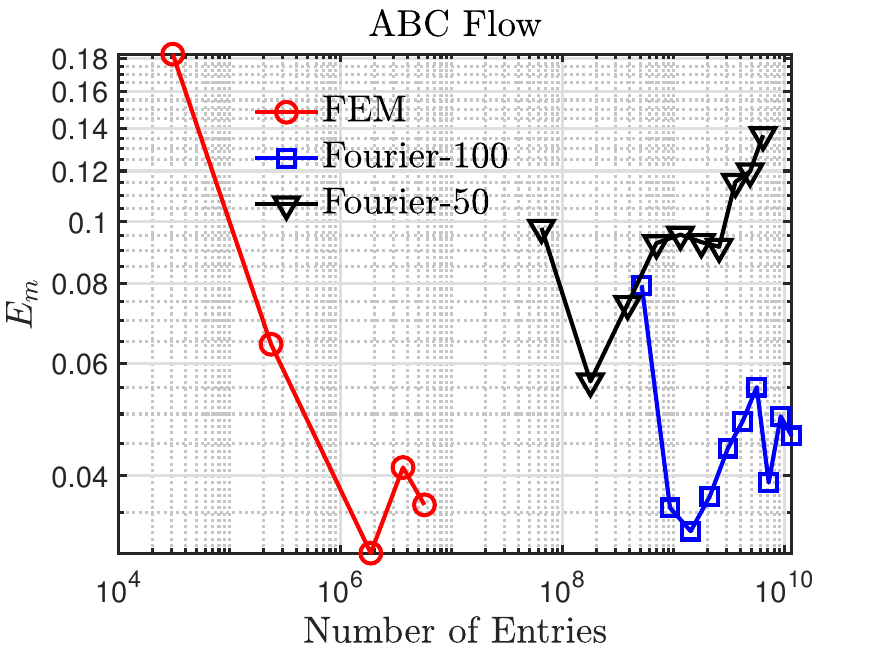}
\includegraphics[width=0.45\textwidth]{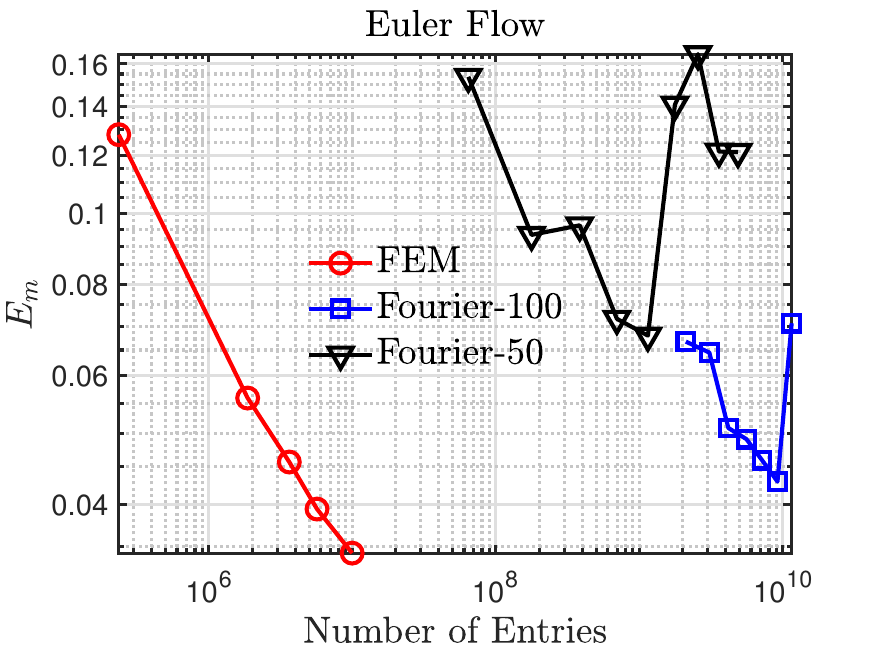}
\caption{Mean invariance error $E_m$ as a function of the number of entries ( $dN_\mathrm{dof}$ for the finite-element method and $mN_\mathrm{mode}$ for the Fourier series scheme) for ABC flow (left panel) and Euler flow (right panel) in Sect.~\ref{sec:periodicflows}. Here the legends `Fourier-100' and `Fourier-50' represent $m=100^3$ and $m=50^3$ respectively.}
\label{fig:fourier-fem}
\end{figure*}

\textcolor{black}{We plot the metric $E_m$ against the number of nonzero entries, namely, $dN_\mathrm{dof}$ or $mN_\mathrm{mode}$, to compare the performance of the two schemes. Indeed, the computational cost of leading eigenvalues or singular values is also directly related to these numbers of entries. As seen in Fig.~\ref{fig:fourier-fem}, in order to achieve the same level of error metric $E_m$, the number of entries needed for the finite-element method is much smaller than for the Fourier approach in both two periodic flows. In addition, the finite-element method can achieve smaller errors with increasing number of entries. Therefore, the finite-element implementation shows better scaling. We have performed the Fourier-based computations with both $m=100^3$ and $m = 50^3$, since decreasing the number of gridpoints allows us to use a Fourier series with higher number of modes. However, we have found that increasing the number of gridpoints, $m$, is more beneficial in terms of the error metric $E_m$.}

\enlargethispage{20pt}


\dataccess{The code and data used to generate the numerical results included in this paper are available at \url{https://github.com/mingwu-li/first_integral}.}

\aucontribute{M.L.: formal analysis, investigation, methodology, software, validation, visualization, writing-original draft, writing-review and editing; B.K: investigation, methodology, writing-review and editing; G.H: conceptualization, project administration, supervision, writing-review and editing.\\  All authors gave final approval for publication and agreed to be held for accountable for the work performed therein.}

\competing{We declare we have no competing interest.}

\funding{We received no funding for this study.}

\ack{Insert acknowledgment text here.}



\bibliography{RSPA_v2_kb.bbl} 
\bibliographystyle{ieeetr}

\end{document}